 \documentclass[final,3p,times]{elsarticle}

\usepackage{dsfont,amsmath,amssymb}
\usepackage{epsfig}

\def\bk{{\bf k}}

\def\br{{\bf r}}
\def\bu{{\bf u}}

\def\bx{{\bf x}}

\def\cF{{\cal F}}

\def\cB{{\cal B}}
\def\cJ{{\cal J}}

\def\cN{{\cal N}}
\def\cO{{\cal O}}
\def\cP{{\cal P}}

\def\cX{{\cal X}}
\def\cZ{{\cal Z}}

\def\mR{{\mathbb{R}}}
\def\mT{{\mathbb{T}}}
\def\mZ{{\mathbb{Z}}}

\def\ol{\overline}
\def\ul{\underline}
\def\wt{\widetilde}

\def\tJ{{\tilde{J}}}
\def\hn{\widehat{n}}
\def\hP{\widehat{P}}

\def\llangle{\left\langle}
\def\rrangle{\right\rangle}

\def\bdot{\hbox{\boldmath $\cdot$}}

\def\grad{\hbox{\boldmath $\nabla$}}

\def\bzed{\hbox{\boldmath $0$}}

\def\be{\begin{equation}}
\def\ee{\end{equation}}
\def\lb{\label}

\journal{Physica D: Nonlinear Phenomenon}

\begin{document}

\begin{frontmatter}

\title{Kinetic Wave Turbulence}

\author{Gregory L. Eyink${\,\!}^{1,2}$ and Yi-Kang Shi${\,\!}^{1}$}

\address{(1) Department of Applied Mathematics \& Statistics and (2) Department of Physics \& Astronomy,\\
The Johns Hopkins University, Baltimore, MD, USA}
 
\begin{abstract}
We consider a general model of Hamiltonian wave systems with triple resonances, in the
standard kinetic limit of a continuum of weakly interacting dispersive waves with random phases. 
We show in this limit that the leading-order, asymptotically valid dynamical  equation for multimode amplitude 
distributions is {\it not} the well-known equation of Peierls (also, Brout \& Prigogine and Zaslavskii \& Sagdeev), but 
is instead a reduced equation containing only a subset of the terms in that equation. Our equations are consistent
with the Peierls equation in that the additional terms in the latter vanish as inverse powers of volume in the
large-box limit.
The equations that
we derive are the direct analogue of the Boltzmann hierarchy obtained from the BBGKY hierarchy 
in the low-density limit for gases.  We show that the asymptotic multimode equations possess factorized 
solutions for factorized initial data, which correspond to preservation in time of the property of ``random 
phases \& amplitudes''. The factors satisfy the equations for the 1-mode probability density functions 
(PDF's) previously derived by Choi et al.  and Jakobsen \& Newell.  Analogous to the Klimontovich density 
in the kinetic theory of gases, we introduce the concepts of the  ``empirical spectrum'' and the ``empirical 1-mode 
PDF''. We show that the factorization of the hierarchy equations implies that these quantities are self-averaging: 
they satisfy the wave-kinetic closure equations of the spectrum and 1-mode PDF for almost any selection of     
phases and amplitudes from the initial ensemble. We show that both of these closure equations satisfy an 
$H$-theorem for an entropy defined by Boltzmann's prescription $S=k_B\log W.$ We also characterize 
the general solutions of our multimode distribution equations, for initial conditions with random phases but with no 
statistical assumptions on the amplitudes. Analogous to a result of Spohn for the Boltzmann hierarchy, these 
are ``super-statistical solutions'' that correspond to ensembles of solutions of the wave-kinetic closure equations 
with random initial conditions or random forces.  On the basis of our results, we discuss possible kinetic 
explanations of intermittency and non-Gaussian statistics in wave turbulence. In particular, we advance the 
explanation of a ``super-turbulence'' produced by stochastic or turbulent solutions of the wave kinetic equations
themselves.
 \end{abstract}


\end{frontmatter}

\section{Introduction and Outline of Main Results}\lb{intro}

The theory of wave turbulence has traditionally focused on the wavenumber spectrum, which is 
expected to satisfy the wave kinetic equation in the limit of a continuum of weakly interacting, 
phase-incoherent waves \cite{Zakharovetal92,NewellRumpf11}. However, recent very 
interesting works of  Choi et al. (2004,2005) \cite{LvovNazarenko04,Choietal04,Choietal05a,Choietal05b} 
and Jakobsen \& Newell (2004) \cite{JakobsenNewell04} have studied higher-order fluctuations of wave 
amplitudes by deriving equations for probability density functions (PDF's) in this same kinetic limit.
These works generalized the results of a pioneering study of Peierls (1929) \cite{Peierls29}, 
who obtained a set of evolution equations for multimode PDF's of energies of phonons (sound waves) in 
anharmonic crystals. Cf. also Brout \& Prigogine (1956) \cite{BroutPrigogine56}. Such equations were 
first discussed in the context of wave turbulence by Zaslavskii \& Sagdeev (1967) \cite{ZaslavskiiSagdeev67},
for some special 3-wave systems.The recent works cast new light on some outstanding problems of wave turbulence. 
On the one hand, the papers \cite{LvovNazarenko04,Choietal04,Choietal05a,Choietal05b,JakobsenNewell04} 
have made an important contribution to the foundations of wave kinetic theory, by showing that the 
``random phase'' and ``random phases \& amplitudes'' properties of initial wave fields are preserved 
in time by their multimode equations\footnote{Strictly speaking, such assumptions on initial 
data should be unnecessary for the validity of wave kinetics. RPA properties and, even more, Gaussian statistics 
should be dynamically established locally in physical space on the fast linear wave time-scale, due to the 
superposition of  dispersive waves trains arriving from distant, uncorrelated regions of space. See 
\cite{BenneySaffman66,NewellRumpf11} or, more rigorously, \cite{DudnikovaSpohn06}. Such properties 
are assumed valid at the initial time only to simplify the derivations.}\lb{fast}. These developments 
are surveyed in a recent monograph of Nazarenko \cite{Nazarenko11}.
Although there has been some recent progress \cite{LukkarinenSpohn11}, 
we are still far from a rigorous mathematical derivation either of the wave kinetic equation or of any 
of its  predictions.  Another important result of \cite{Choietal05a,Choietal05b,JakobsenNewell04} is a 
closed equation for the 1-mode PDF's of the independent wave amplitudes. These equations can 
have as solutions 1-mode PDFs that are far from Gaussian, especially when the equations are supplemented 
with boundary conditions and additional terms to represent strongly nonlinear process, such as wave-breaking,  
that lie outside the validity of weakly-interacting wave kinetics \cite{Choietal05a,Nazarenkoetal10}.
These results show promise to explain observations of intermittency and anomalous scaling in some 
recent experiments \cite{Falconetal07,Nazarenkoetal10} and simulations \cite{Yokoyama04} of gravity-capillary 
wave turbulence, especially for wavenumber regimes where such intermittency was not previously 
expected \cite{Connaughtonetal03} 


Unfortunately, the Peierls-Brout-Prigogine-Zaslavskii-Sagdeev (or simply, Peierls) equations for multimode
PDF's are not the leading-order, asymptotically correct equations in the standard limit of a continuum of weakly 
interacting, incoherent waves. It was not observed in \cite{Choietal05a,Choietal05b,JakobsenNewell04} 
that many of the terms in the Peierls equation vanish as inverse powers of volume in the infinite-box 
(continuous wavenumber) limit. Using the same methods as theirs but with a correct estimation 
of the terms, we shall obtain a novel set of multimode equations simpler than the Peierls equation
On the other hand, we shall show using these new multimode equations that the 1-mode PDF 
equations obtained in \cite{Choietal05a,Choietal05b,JakobsenNewell04} {\it are} correct, 
under the assumption that initial wave modes have statistically-independent random amplitudes as well as 
phases. More generally, we shall classify all possible realizable solutions of the new multimode 
equations, for initial conditions  with independent (and uniform) random phases but with no assumption 
on the statistics of initial wave amplitudes. We find that {\it the most general solutions correspond to 
``super-ensembles'' of solutions of the wave kinetic equations with random initial conditions for the spectrum 
or with random forcings.} This represents another possible mechanism for intermittency and non-Gaussian 
distributions by a ``super-turbulence'' of chaotic or stochastic solutions of the kinetic equations. As we discuss 
in detail, existing results on linear stability of Kolmogorov cascade solutions of the wave kinetic equations do not 
rule out transition to such ``super-turbulence.'' In fact, this possibility was anticipated in a review of the 
stability theory by Zakharov et al. (1992), (\cite{Zakharovetal92},  section 4.2.2), who referred to this possibility 
as ``secondary turbulence.'' The results of the present work show that this is the {\it only} possibility for 
explaining intermittency and anomalous scaling of wave turbulence strictly within the wave kinetics framework.
\footnote{We say ``strictly'' because the approach of \cite{Choietal05a,Choietal05b} invokes 
strongly nonlinear effects at sufficiently large wave amplitudes, outside the kinetic description. Our remark 
applies only to leading order in the small parameters $\epsilon$ and $1/L,$ whereas higher-order corrections 
could also in principle lead to (weakly) non-Gaussian statistics in either physical or Fourier space.} 

The invalidity of the Peierls equation as a leading-order description
will be shown below by specific, detailed calculations, but it can be 
understood on the basis of simple, general considerations. It will be useful to present such arguments 
here as a preliminary to the concrete calculations. This will also permit us to give an overview of the 
important new concepts, tools and results of our paper.  We must begin with a brief summary of the 
main results of \cite{JakobsenNewell04,Choietal05b} for a general Hamiltonian dynamics with 3-wave 
interactions:
\be \dot{a}_\bk^\sigma = \epsilon \sum_{\sigma_1,\sigma_2}\sum_{\bk_1,\bk_2}
L^{\sigma,\sigma_1,\sigma_2}_{\bk,\bk_1,\bk_2} a^{\sigma_1}_{\bk_1} a^{\sigma_2}_{\bk_2} 
e^{i(\sigma_1\omega(\bk_1)+\sigma_2\omega(\bk_2)-\sigma\omega(\bk))t}
\delta_{\sigma_1\bk_1+\sigma_2\bk_2,\sigma\bk} \ee
where the wave triplet nonlinear interaction $L^{\sigma_1,\sigma_2,\sigma_3}_{\bk_1,\bk_2,\bk_3}=
3i\sigma_1H^{-\sigma_1,\sigma_2,\sigma_3}_{\bk_1,\bk_2,\bk_3}$ in terms of the coefficient of the cubic 
term in the Hamiltonian and where the summations range over wavenumbers $\bk\in \frac{2\pi}{L}\mZ^d$ 
and a degeneracy index $\sigma=\pm 1,$ with $L$ the side-length of a periodic box containing the 
wave system. (For a systematic discussion of the notations used in this paper, see section \ref{notations} 
below.) The analysis of \cite{JakobsenNewell04,Choietal05b} is based on action-angle variables 
$(J_\bk,\varphi_\bk)$ for the linear wave dynamics, defined by $a_\bk^\sigma=\sqrt{J_\bk }
e^{i\sigma \varphi_\bk}.$ Note that the standard spectral density is related to the action variables as
\be n(\bk)=\lim_{L\rightarrow\infty}\left(\frac{L}{2\pi}\right)^d\langle J_\bk\rangle \lb{spectrum1} \ee 
in the infinite-volume limit, with total wave action per volume given by $N=\int d^dk\,\,n(\bk).$ 
Here the average $\langle\cdot\rangle$ is over the ensemble of initial amplitudes and phases.
In the large-box limit ($L\rightarrow \infty$) followed by the weak nonlinearity limit ($\epsilon\rightarrow 0$),
the spectrum is expected to satisfy the {\it wave kinetic equation}:
\begin{eqnarray}
\dot{n}(\bk,\tau)&=&36\pi\sum_{\ul{\sigma}=(-1,\sigma_2,\sigma_3)} \int d^dk_2 d^dk_3\,|H^{\ul{\sigma}}_{\ul{\bk}}|^2
\delta(\ul{\sigma}\cdot\omega(\ul{\bk}))\delta^d(\ul{\sigma}\cdot\ul{\bk})\cr
&&\times \Big\{n(\bk_2,\tau)n(\bk_3,\tau)-\sigma_2 n(\bk,\tau)n(\bk_3,\tau)-\sigma_3 n(\bk,\tau)n(\bk_2,\tau)\Big\},
\lb{wavekineq}
\end{eqnarray}
where $\tau=\epsilon^2 t$ is the nonlinear time. See \cite{Zakharovetal92,Nazarenko11}. To obtain information 
about the statistical distribution of amplitudes and phases beyond the spectrum (\ref{spectrum}), the authors 
of \cite{Choietal05a,Choietal05b} exploit a generating functional defined by  
\be \cZ_L(\lambda,\mu) =  \left\langle \exp\left(\sum_{\bk} 
i \lambda_\bk J_\bk + i\mu_\bk \varphi_\bk \right)\right\rangle.  \lb{Z-fun} \ee
In the same kinetic limit ($L\rightarrow \infty$, $\epsilon\rightarrow 0$), for an initial distribution with 
independent, uniform phases, this generating function is claimed to satisfy the Peierls equation in the form
\be
\frac{d\cZ}{d\tau}=-6\pi i \delta_{\mu,0} \sum_{\ul{\bk},\ul{\sigma}} |H^{\ul{\sigma}}_{\ul{\bk}}|^2 
\delta_{\ul{\sigma}\cdot\ul{\bk},\bzed}
\delta_T(\ul{\sigma}\cdot\omega(\ul{\bk})) (\ul{\sigma}\cdot\lambda_{\ul{\bk}})
 \partial_{\lambda_{\bk_1}}\partial_{\lambda_{\bk_2}}\partial_{\lambda_{\bk_3}}
  (\ul{\sigma}\cdot\lambda_{\ul{\bk}}) \cZ, \lb{Peierls} \ee
with $\ul{\sigma}=(\sigma_1,\sigma_2,\sigma_3)$,$\ul{\bk}=(\bk_1,\bk_2,\bk_3)$ and with $\delta_T(\omega)$
an approximate delta function of width $1/T$ in frequency space for some time $T$ intermediate between the 
wave period and the time $t=\epsilon^{-2}\tau$.
From this equation, \cite{Choietal05b,JakobsenNewell04}
derived results on the higher-order fluctuations and statistical distributions of the wave mode amplitudes. 
  

To see that the results, as stated above, cannot be asymptotically correct, consider the limiting behavior 
of the generating functional (\ref{Z-fun}) when the field $a_\bk$ is assumed to be ``RPA'', i.e. to have amplitudes and 
phases for distinct Fourier modes given by mutually independent random variables.  In that case, the 
generating functional must factorize for all independent variables, in particular for the amplitudes, as:
\be 
\cZ_L(\lambda,\mu=0) =  \left\langle \exp\left(\sum_\bk i \lambda_\bk J_\bk\right)\right\rangle 
= \prod_\bk\Big\langle\exp\left(i \lambda_\bk J_\bk\right)\Big\rangle, \lb{Z-prod} \ee
where each factor is a 1-mode generating function $\cZ^{(1)}(\lambda_\bk;\bk)=\Big\langle
\exp\left(i \lambda_\bk J_\bk\right)\Big\rangle.$ However, the definition of the spectrum implies that the 
rescaled variable
\be\tJ_\bk:=(L/2\pi)^dJ_\bk \ee
must be $O(1)$ in the limit as $L\rightarrow\infty.$  For the 1-mode generating function this gives,
employing the cumulant expansion, the result: 
\be\cZ^{(1)}(\lambda_\bk;\bk)=
\Big\langle\exp\left(i \lambda_\bk \left(\frac{2\pi}{L}\right)^d \tJ_\bk\right)\Big\rangle = 
\exp\left(\sum_{p=1}^\infty \frac{i^p}{p!} \left(\frac{2\pi}{L}\right)^{pd} \lambda^p_\bk\langle\tJ_\bk^p\rangle^c \right),
\ee
where $\langle\tJ_\bk^p\rangle^c$ is the $p$th-order cumulant. Inserting back into the product (\ref{Z-prod}), 
one obtains 
\be \cZ_L(\lambda,\mu=0) =\exp\left(\sum_{p=1}^\infty \frac{i^p}{p!}  \left(\frac{2\pi}{L}\right)^{pd}
 \sum_\bk \lambda^p_\bk\langle\tJ_\bk^p\rangle^c \right). \ee
Now assume that $\lambda_\bk=\lambda(\bk)$ for some smooth function $\lambda(\bk).$ Since for $L\rightarrow\infty$
\be   \left(\frac{2\pi}{L}\right)^{pd}\sum_\bk \lambda^p(\bk) \langle\tJ_\bk^p\rangle^c 
      \sim  \left(\frac{2\pi}{L}\right)^{(p-1)d}\int d^dk \,\,\lambda^p(\bk)\langle\tJ_\bk^p\rangle^c, \ee
the contributions of $p$th-order cumulants are $O(L^{-d(p-1)}).$  Only the $p=1$ contribution
$n(\bk)=\langle\tJ_\bk\rangle$  survives for very large $L$  
\be \cZ_L(\lambda,\mu=0) = \exp\left( i\int d^dk\,\, \lambda(\bk)n(\bk)  + O\left(L^{-d}\right)\right) 
\overrightarrow{\mbox{{\footnotesize$\,\,\,\,L\rightarrow\infty\,\,\,\,$}}} \exp\left( i\int d^dk\,\,  \lambda(\bk)n(\bk) \right).
\lb{1st-exp} \ee
Thus, the generating functional $\cZ_L(\lambda,\mu)$ of an RPA field is completely determined by its spectrum 
$n(\bk)$ in the limit as $L\rightarrow\infty$ and it then contains no information about higher-order fluctuations. 
This result, which we have derived here assuming existence of all higher-order cumulants, can be proved under
 much weaker assumptions (see section 2). 

There is a simple interpretation of the above limit in terms of the {\it empirical spectrum}, defined as
\be\hn_L(\bk) = \left(\frac{2\pi}{L}\right)^d \sum_{\bk_1\in \frac{2\pi}{L}\mZ^d} \tJ_{\bk_1}\delta^d(\bk-\bk_1). \ee
For an RPA field, this quantity is a sum of a large number of independent variables.\footnote{
The reader may worry that all terms appear to vanish in this summation except the single term for 
$\bk_1=\bk$! However, it is important to keep in mind that distributions (generalized functions) like the 
Dirac delta function are only meaningful when smeared with smooth test functions. In that case, 
$\int d^dk \,\varphi(\bk) \hn_L(\bk)= \left(\frac{2\pi}{L}\right)^d \sum_{\bk_1\in \frac{2\pi}{L}\mZ^d} \tJ_{\bk_1}\varphi(\bk_1)$
and this sum indeed contains a large number of terms in the limit $L\rightarrow\infty.$ More physically, the empirical 
spectrum must be ``coarse-grained'' by integration over cells $\Delta$ that partition the wavenumber space. This 
corresponds to replacing the smooth test function $\varphi(\bk)$ by the characteristic function $\chi_\Delta(\bk)$
which $=1$ for $\bk\in\Delta$ and $=0$ otherwise.} It is exactly analogous 
to the ``Klimontovich density'' or empirical 1-particle density in the kinetic theory of gases 
\cite{Klimontovich67}.  Unlike the usual spectrum $n(\bk),$ the empirical spectrum is a random variable 
that incorporates the information about amplitude fluctuations. The amplitude generating function $\cZ_L(\lambda)
=\cZ_L(\lambda,\mu=0)$ is just the characteristic functional of the empirical spectrum:
\be  \cZ_L(\lambda) = \left\langle \exp\left( i\int d^dk\,\, \lambda(\bk)\hn_L(\bk) \right) \right\rangle. 
\lb{Z-empspec} \ee
The previous limiting result for $\cZ_L(\lambda)$ is mathematically equivalent to the statement that 
the empirical spectrum has a {\it deterministic} limit for $L\rightarrow\infty$ which is the usual spectrum:   
 \be \lim_{L\rightarrow\infty} \hn_L(\bk) = n(\bk). \ee
This limit is a probabilistic law of large numbers for the empirical spectrum. There is a precisely analogous
law of large numbers for the empirical 1-particle distribution in the low-density limit for the kinetic theory
of gases, as first shown in 1975 by O. Lanford  \cite{Lanford75,Lanford76}. 
 
The above results should hold not only for RPA fields at initial times, but also for the evolved fields in the kinetic 
regime of wave turbulence, if the RPA property is propagated in time as expected. For example,  
\cite{Choietal05a},section 5, and \cite{Nazarenko11}, section 11.4, argue that the RPA property will be preserved
in a somewhat weaker form. As we shall discuss in detail below (see eq.(\ref{stosszahl})), these weaker forms 
still suffice to derive the limiting exponential expression (\ref{1st-exp}) for the generating functional. However, 
the Peierls equation does not have such exponential solutions! Thus, the current claims in the literature are not
consistent. We shall see that when carried out carefully, with due regard to the scaling of various terms 
with $\epsilon$ and $L,$ the standard analytical methods of wave turbulence yield {\it not}
the Peierls equation in the limit as first $L\rightarrow\infty$ then $\epsilon\rightarrow 0,$ but instead the following 
simpler equation:
\begin{align}
\begin{split}
\dot{\cZ}[\lambda,\mu]=&-36i\pi\delta_{\mu,\,0}\sum_{\ul{\sigma}=(-1,\sigma_2,\sigma_3)}
\int d^dk_1 d^dk_2 d^dk_3 \,\,\delta^d(\ul{\sigma}\cdot\ul{\bk})\delta(\ul{\sigma}\cdot\omega(\ul{\bk}))
|H^{\ul{\sigma}}_{\ul{\bk}}|^2\\
& \times\lambda(\bk_1)\left\{\frac{\delta^2\cZ}{\delta\lambda(\bk_2)\delta\lambda(\bk_3)}
-\sigma_2\frac{\delta^2\cZ}{\delta\lambda(\bk_1)\delta\lambda(\bk_3)}
-\sigma_3\frac{\delta^2\cZ}{\delta\lambda(\bk_1)\delta\lambda(\bk_2)}\right\}
\end{split}
\end{align}
This equation will be justified in detail in this work. Here we note only that it indeed differs from the Peierls equation.
It contains only a subset of the terms in the Peierls equation, lacking in particular all the terms involving third-order 
derivatives in $\lambda.$ The important property of the above corrected equation is that it has the 
expected solutions of exponential form 
\be {\cZ}[\lambda,\mu,\tau]=  \exp\left( i\int d^dk\,\,  \lambda(\bk) n(\bk,\tau)\right) \ee
where $n(\bk,\tau)$ solves the classical wave kinetic equation (\ref{wavekineq}), {\it if} the initial conditions $\cZ[\lambda,\mu,\tau=0]$ are 
also of this exponential form. 
The property of preserving 
exponential solutions implies a law of large-numbers for the empirical spectrum $\hn_L(\bk,\tau)$ at times $\tau>0$ 
and is equivalent to the ``propagation of chaos'' by the BBGKY hierarchy in the kinetic limit for gases. 

The analogy of kinetic wave turbulence with the kinetic theory of gases is in fact quite close. 
As we shall show below, the above equation for $\cZ$ is equivalent to a hierarchy of equations 
for $M$-point correlation functions of the empirical spectrum, which is exactly analogous to the ``Boltzmann 
hierarchy'' obtained from the BBGKY hierarchy for low-density gases \cite{Lanford75,Lanford76}.
Just as the Boltzmann hierarchy  has factorized solutions for factorized initial conditions, so does the kinetic
wave hierarchy for all multi-point spectral correlation functions.  The factors in both cases solve the relevant kinetic 
equation, which satisfies an $H$-theorem corresponding to positive entropy production. For both hierarchies 
the general solutions without assuming factorized initial data are ``super-statistical solutions'' which represent 
statistical mixtures of factorized solutions. This was first discussed for the Boltzmann hierarchy by Spohn \cite{Spohn84}. 
For wave turbulence these ``super-statistical solutions'' of the spectral hierarchy correspond to ensembles
of solutions $n(\bk,\tau)$ of the classical wave kinetic equation with random initial conditions $n_0(\bk)$. 
We shall discuss in this work both the mathematical derivation and the physical relevance of the
``super-statistical solutions''. These are a possible source of non-Gaussian statistics and intermittency 
not widely appreciated in the wave turbulence literature. 

Previous attempts to calculate intermittency effects within wave kinetic theory have employed the Peierls 
equation or its reduced forms for the PDFs of a finite number of modes \cite{Choietal05a, Choietal05b, Choietal09}. 
If the Peierls equation is not asymptotically valid in the usual wave kinetic limit (first $L\rightarrow\infty,$ then 
$\epsilon\rightarrow 0),$ then does this mean that the previously claimed results for  the finite-mode PDF's 
are also not leading-order asymptotic limits? The answer is yes, but fortunately only partially. The 
joint PDF of a set of $M$ wavenumber modes
$\tJ_{\bk_1},...,\tJ_{\bk_M}$ is characterized by its Fourier transform, the $M$-mode generating function:
\be \cZ^{(M)}(\lambda_1,...,\lambda_M,\mu=0;\bk_1,...,\bk_M)  = 
     \left\langle \exp\left(\sum_{m=1}^M i \lambda_m \tJ_{\bk_m} \right)\right\rangle. \ee
This is the same object which was introduced in the monograph of Nazarenko \cite{Nazarenko11},
section 5.4, except that we consider only a fixed number $M$ of the total number $N=O(L^d)$ of Fourier 
modes\footnote{It is easy to see that if all $N$ modes are considered, then $\cZ^{(N)}$ diverges 
exponentially in the limit $L\rightarrow\infty$ and, in particular, cannot converge to a solution of the Peierls equation. 
Only the quantity $\left(\frac{2\pi}{L}\right)^d\log \cZ^{(N)}$ can have a finite limit. We shall study this in detail 
in a following paper \cite{Eyink12}.}. We find that these objects do not individually satisfy closed equations 
but instead satisfy a hierarchy of equations that link $M$-mode functions to $(M+1)$- and $(M+2)$-mode functions: 
\begin{align}
\begin{split}
\dot{\cZ}^{(M)}=&-36i\pi\sum_{j=1}^M \sum_{\ol{\sigma}_2,\ol{\sigma}_3}\int d^d\ol{k}_2 d^d\ol{k}_3
\,\,\delta^d(\ul{\sigma}\cdot\ul{\bk}_j)\delta(\ul{\sigma}\cdot\omega(\ul{\bk}_j))|H^{\ul{\sigma}}_{\ul{\bk}_j}|^2\\
&\Big\{(\lambda_j+\lambda_j^2\frac{\partial}{\partial\lambda_j})\frac{\partial^2\cZ^{(M+2)}}{\partial\ol{\lambda}_2\partial\ol{\lambda}_3}\Big|_{\ol{\lambda}_2=\ol{\lambda}_3=0}-\ol{\sigma}_2\lambda_j\frac{\partial\cZ^{(M+1)}}{\partial\ol{\lambda}_3\partial\lambda_j}\Big|_{\ol{\lambda}_3=0}-\ol{\sigma}_3\lambda_j\frac{\partial\cZ^{(M+1)}}{\partial\ol{\lambda}_2\partial\lambda_j}\Big|_{\ol{\lambda}_2=0}\Big\}.
\end{split}
\end{align}
with the notations $\ul{\sigma}=(-1,\ol{\sigma}_2,\ol{\sigma}_3)$ and $\ul{\bk}_j=(\bk_j,\ol{\bk}_2,\ol{\bk}_3).$ 
The equations in this hierarchy can also be obtained from the Peierls equation, by setting all except $M$ of the 
$\lambda$'s equal to 0 and by retaining only a subset of terms. In particular, the terms in the Peierls 
equation which contain products $\lambda_\bk\lambda_{\bk'}$ for $\bk\neq \bk'$ are found to vanish in the limit 
$L\rightarrow\infty,$ when only a finite number $M$ of $\bk$-modes have $\lambda_\bk\neq 0.$
The above hierarchy for the $M$-mode generating functions is equivalent to a 
hierarchy for the joint PDF's $\cP^{(M)}(s_1,...,s_M;\bk_1,...,\bk_M)$ of $M$ Fourier modes $\bk_1,...,\bk_M$:
\begin{align}
\dot{\cP}^{(M)}+\sum_{j=1}^M\frac{\partial}{\partial s_j}\cF_j^{(M)}=0,
\end{align}
where, with the same notations as above, 
\begin{align}
\begin{split}
\cF_j^{(M)}=& -36\pi s_j\sum_{\ul{\sigma}=(-1,\ol{\sigma}_2,\ol{\sigma}_3)}\int d^d\ol{k}_2 d^d\ol{k}_3\delta^d(\ul{\sigma}\cdot\ul{\bk}_j)\delta(\ul{\sigma}\cdot\omega(\ul{\bk}_j))|H^{\ul{\sigma}}_{\ul{\bk}_j}|^2\\
&\Big[\int d\ol{s}_2d\ol{s}_3\frac{\partial\cP^{(M+2)}}{\partial s_j}(s_1,...,s_M,\ol{s}_2,\ol{s}_3)\ol{s}_2\ol{s}_3 \\
&+\ol{\sigma}_2\int d\ol{s}_2\cP^{(M+1)}(s_1,...,s_M,\ol{s}_3)\ol{s}_3
+\ol{\sigma}_3\int d\ol{s}_2\cP^{(M+1)}(s_1,...,s_M,\ol{s}_2)\ol{s}_2\Big].
\end{split}
\end{align}
Previously proposed closed equations for multi-mode equations, e.g. for $M=2$ in \cite{Choietal09}, are not 
asymptotically exact in the usual kinetic limit for wave turbulence but require additional ad hoc closure assumptions.

On the other hand, the one-mode equation that was derived in \cite{Choietal05b,JakobsenNewell04} 
for the PDF $P(s,\tau;\bk)=\cP^{(1)}(s,\tau;\bk)$ of a single amplitude $\tJ_\bk:$
\be \frac{\partial}{\partial\tau}P =\frac{\partial}{\partial s}\left[ s\left(\eta_\bk\frac{\partial P}{\partial s}
      +\gamma_\bk P\right)\right] \lb{1P-eq} \ee
is obtained from the above hierarchy as a leading-order asymptotic equation, for factorized solutions of the form: 
\be \cP^{(M)}(s_1,...,s_M,\tau;\bk_1,...,\bk_M)  = \prod_{m=1}^M
      P(s_m,\tau;\bk_m). \lb{P-factor} \ee
The coefficients $\eta_\bk,\, \gamma_\bk$ that appear also in the kinetic equation $\dot{n}_k=
\eta_\bk-\gamma_\bk n_\bk$ are obtained from the condition $n_\bk=\int ds\,s P(s;\bk)$  and the standard 
wavenumber integrals over products of $n.$ The one-mode 
equation is thus a nonlinear Fokker-Planck equation in the sense of McKean \cite{McKean66}.
Factorized solutions of the form (\ref{P-factor}) are obtained for factorized initial data, corresponding to RPA fields.
The hierarchy thus preserves in this sense the statistical independence of  amplitudes.  Just as for  
the spectral hierarchy, this factorization property of solutions implies a law of large numbers for 
{\it empirical 1-mode PDFs}
\be \hP_L(s,\Delta,t) = \frac{1}{N_L(\Delta)}\sum_{\bk\in \frac{2\pi}{L}\mZ^d\cap \Delta} \delta(s-\tJ_\bk) \ee
with $\Delta\subset \mR^d$ a subset of continuous wavenumbers and with $N_L(\Delta)$ the number of elements 
in $\frac{2\pi}{L}\mZ^d\cap \Delta.$ Also as for the spectral hierarchy, the general solutions of the PDF 
hierarchy are ``super-statistical solutions'' that correspond to ensembles of solutions of the 1-mode 
equation (\ref{1P-eq}) with random initial conditions $P_0(s;\bk).$ As we shall discuss, such ensembles 
may be physically relevant to explain non-Gaussian statistics in weakly nonlinear wave systems 
if the solutions of (\ref{1P-eq}) themselves become random, either through fluctuating driving forces or internal
chaos/turbulence. 

The detailed contents of this work are as follows. The introduction concludes with a brief section \ref{notations} 
summarizing our model assumptions and notations. The next section \ref{RPA} discusses the probabilistic 
properties of ``RP'' and ``RPA" fields.  The principal results of the paper are in section \ref{multimode}, which 
derives the multimode equations and discusses their basic properties. This is done first for the spectral generating 
functional and correlation functions (section \ref{spectral}) and next for the PDF generating functions and 
correlation functions (section \ref{PDF}). Many technical details are provided in \ref{appendix-spectral}
and \ref{appendix-pdf}. Finally, we discuss in section \ref{intermittent} possible kinetic theory mechanisms
of intermittency in wave turbulence.  The conclusion section \ref{conclusion} summarizes our results.

\subsection{Model and Notations}\lb{notations} 

We summarize here briefly the notations employed in our work.  We consider a system consisting 
of a complex wavefield $u(\bx,t)$ in a $d$-dimensional periodic cube with side $L$. 
As in \cite{JakobsenNewell04,Choietal05b}, we assume that there is a maximum wavenumber $k_{\max},$
to avoid ultraviolet divergences. This can be achieved by a lattice regularization with spacing $a=L/M,$ 
for some large integer $M,$ so that $k_{\max}=\pi/a.$ The location variable $\bx$ then ranges over the physical space
\be\Lambda_L=a \mZ_M^d, \ee
with the usual notation $\mZ_M$ for the field of integers modulo $M.$ This space has volume $V=L^d.$ 
The dual space of wavenumbers is 
\be\Lambda^*_L= \frac{2\pi}{L}\mZ_M^d \ee
with $k_{\min}=2\pi/L.$ The total number of modes is $N=M^d$, so that spatial volume $V=N a^d.$ We use 
the following index notation 
\be u^\sigma(\bx) = \left\{\begin{array}{ll}
                                            u(\bx)\, & \sigma=+1\cr
                                            u^*(\bx) & \sigma=-1
                                            \end{array} \right. \ee
for $u$ and its complex-conjugate $u^*,$ following \cite{JakobsenNewell04}. Likewise, we adopt their 
convention for (discrete) Fourier transform                                            
\be A^\sigma(\bk)= \frac{1}{N}\sum_{\bx\in\Lambda_L}u^\sigma(\bx,t)\exp(-i\sigma \bk\cdot\bx)\ee
so that $A^+(\bk)$ and $A^-(\bk)$ are complex conjugates. 
Notice that this quantity converges to the continuous Fourier transform $ \frac{1}{L^d} \int_{[0,L]^d}
d^dx\,\,u^\sigma(\bx,t)\exp(-i\sigma \bk\cdot\bx)$ in the limit $a\rightarrow 0.$ The discrete inverse transform is 
\be u^\sigma(\bx) = \sum_{\bk\in\Lambda_L^*} A^\sigma(\bk)\exp(i\sigma \bk\cdot\bx).\ee


The dynamics is assumed to be canonical Hamiltonian with a cubic Hamiltonian density (energy per volume) describing 3-wave 
interactions:
\be H=H_0+\delta H=\sum_{\bk}\omega(\bk)|A_{\bk}^{+}|^2
+\epsilon\sum_{\ul{\sigma},\ul{\bk}}H^{\ul{\sigma}}_{\ul{\bk}}
A^{\sigma_1}_{\bk_1} A_{\bk_2}^{\sigma_2} A_{\bk_3}^{\sigma_3} \delta_{\ul{\sigma}\cdot{\ul{\bk}},\,{\bf 0}}. \ee
As in \cite{JakobsenNewell04}, we denote triplets of variables by the notation $\ul{\sigma}=(\sigma_1,\sigma_2,\sigma_3)$, 
$\ul{\bk}=(\bk_1,\bk_2,\bk_3)$ and define the dot product $\ul{\sigma}\cdot\ul{\bk}=\sigma_1\bk_1+\sigma_2\bk_2+\sigma_3\bk_3.$
The Kronecker delta function in the triplet interaction term enforces the symmetry of the dynamics under the group of 
(discrete) space translations. The interaction coefficients $H^{\ul{\sigma}}_{\ul{\bk}}$ are furthermore assumed to satisfy 
\be H_{\ul{\bk}}^{\ul{\sigma}\,*}=H_{\ul{\bk}}^{-\ul{\sigma}} \,\,\,\, {\rm and} \,\,\,\, H_{\ul{\bk}}^{\ul{\sigma}}=H^{\pi(\ul{\sigma})}_{\pi(\bk)}
\ee
for any permutation $\pi\in S_3$. The first condition guarantees the reality of the Hamiltonian and the second can always be 
assumed without loss of generality. The Hamiltonian equations of motion for the Fourier coefficients are 
\be
\frac{\partial A_{\bk}^{\sigma}}{\partial t}=i\sigma\frac{\partial H}{\partial A_{\bk}^{-\sigma}}
=i\sigma\omega(\bk)A_{\bk}^{\sigma}+\epsilon\sum_{\sigma_1,\sigma_2}\sum_{\bk_1,\bk_2} L^{\sigma,\sigma_1,\sigma_2}_{\bk,\bk_1,\bk_2}
A^{\sigma_1}_{\bk_1}A^{\sigma_2}_{\bk_2}\delta_{\sigma_1\bk_1+\sigma_2\bk_2, \,\sigma\bk}.
\lb{A-eq} \ee
where $L^{\sigma,\sigma_1,\sigma_2}_{\bk,\bk_1,\bk_2}=3i\sigma H^{-\sigma,\sigma_1,\sigma_2}_{\bk,\bk_1,\bk_2}$. 
It is useful to introduce action-angle variables for the linear dynamics, $J_{\bk}=|A_{\bk}^\sigma|^2$ and $\varphi_\bk=\sigma \arg(A_\bk^\sigma),$
so that $A_{\bk}^\sigma=\sqrt{J_{\bk}}\psi_{\bk}^\sigma$, where $\psi_{\bk}=\exp(i\varphi_{\bk})$. In these variables,
the Liouville measure $\mu$ conserved by the Hamiltonian flow can be written variously as 
\be d\mu=\prod_\bk dq_\bk dp_\bk = \prod_\bk \frac{1}{i}dA^+_\bk dA^-_\bk = \prod_\bk dJ_\bk d\varphi_\bk \ee
where the canonical momenta and coordinates are given by real and imaginary parts of $A^\sigma_\bk=\frac{1}{\sqrt{2}}(p_\bk + i\sigma q_\bk).$

As usual in wave kinetic theory, we introduce the ``interaction representation'' 
\begin{align}
a^{\sigma}_{\bk}=A^{\sigma}_{\bk} e^{-i\sigma\omega(\bk) t}
\end{align}
which removes the rapid wave oscillations.  The Liouville measure is invariant under this transformation 
because of the rotation-invariance of the Haar measure on $\psi\in S^1.$ We shall often use the shorthand 
notation $a_1=a_{\bk_1}^{\sigma_1}$ together with $\sum_{1}:=\sum_{\sigma_1,\bk_1}$ and 
\begin{eqnarray}
&&\omega^{1}_{23}:=-\sigma_1\omega(\bk_1)+\sigma_2\omega(\bk_2)+\sigma_3\omega(\bk_3),\\
&&\delta^{1}_{23}:=\delta_{\sigma_1\bk_1,\,\sigma_2\bk_2+\sigma_3\bk_3}.
\end{eqnarray}
The dynamical equation of motion can then be written succinctly as 
\be
\dot{a}_1=\epsilon\sum_{2,3} L_{123} a_2 a_3 e^{i\omega^{1}_{23}t}\delta^1_{23}.
\ee

\section{Fields with Random Phases and Amplitudes}\lb{RPA}

It is often assumed in derivations of wave kinetic equations that initial fields have Fourier coefficients 
with random (and statistically independent) phases and amplitudes.  Furthermore, these properties are 
expected to be preserved in time, in some suitable sense, in the wave-kinetic limit. See the discussions in 
\cite{JakobsenNewell04} and, particularly, \cite{Choietal05a,Choietal05b}. It is therefore important to review 
here the definitions and properties of such ``RPA'' fields\footnote{More accurate acronyms would be ``IUP''
for ``independent uniform phases'' rather than ``RP'' and ``IPA'' for ``independent phases \& amplitudes'' 
rather than ``RPA''. However, we shall stick here to the abbreviations already employed in the literature.}. 
Our discussion shall be largely complementary to \cite{Choietal05b}, emphasizing the probabilistic 
characteristics of such fields. 

Consider then $N$ complex-valued random variables $a_\bk,$ $\bk\in \Lambda_L^*$ taken to be the Fourier 
coefficients of a random space field:
\be u(\bx) = \sum_{\bk\in\Lambda_L^*} a_\bk\exp(i\bk\cdot\bx).\ee
Here $a_\bk$ corresponds to $a^+_\bk=A^+_\bk$ in the previous section (no distinction need be made between 
the two at time $t=0$). It will be very important in what follows to work with normalized variables
\be \wt{a}_{\bk}= \left(\frac{L}{2\pi}\right)^{d/2} a_{\bk} \ee 
which are assumed to remain finite in the large-box limit $L\rightarrow\infty.$ As we shall remind the reader 
below, this normalization is required so that the spectrum of the random field is well defined in that limit. 
It is convenient to write the complex variables in polar coordinates $a_\bk=\sqrt{J_\bk}e^{i\varphi_\bk}$  
(action-angle variables) with the normalized action $\wt{J}_{\bk}= \left(\frac{L}{2\pi}\right)^{d} J_{\bk}.$
We use the notations $s_\bk$ and $\xi_\bk$ for possible values of the random variables $\wt{J}_\bk\in \mR^+$
and $\psi_\bk=e^{i\varphi_\bk}\in S^1.$ The Liouville measure of the previous section becomes
\be d\mu(s,\xi)  = \prod_{\bk\in \Lambda^*_L} ds_\bk  \frac{|d\xi_\bk|}{2\pi} \lb{liouville2} \ee
suitably normalized. We define the $N$-mode joint probability density function $\cP^{(N)}(s,\xi)$ 
with respect to Liouville measure, such that the average of the random variable $f(\tJ,\psi)$ is given by
\be\langle f(\tJ,\psi)\rangle=\int d\mu(s,\xi)\, \cP^{(N)}(s,\xi) f(s,\xi) \ee
where the integral is over $(s,\xi)$ in the product space $\left(\mR^+\right)^N\times \left(S^1\right)^N.$

Following \cite{JakobsenNewell04,Choietal05b} we define $u(\bx)$ to be a {\it random-phase field} (RP) if 
$\psi_\bk=e^{i\varphi_\bk}$ for all $\bk\in \Lambda_L^*$ are independent and identically distributed (i.i.d.)  
random variables, uniformly distributed over the unit circle $S^1$ in the complex plane. In terms of the joint 
PDF, this is equivalent to the condition that  
\be\cP^{(N)}(s,\xi)=\cP^{(N)}(s)\ee
independent of $\xi$. It is easy to see that an RP $u(\bx)$ is a homogeneous random field on $\Lambda_L$, 
statistically invariant under space-translations by the finite group $a\mZ_M^d$. This follows from the 
rotation invariance of the Haar measures $|d\xi_\bk|/2\pi$ on the phase variables and by the representation  
of the translations $u(\bx)\rightarrow \bu(\bx+\br)$ for $\br\in a\mZ_M^d$ as phase-rotations:
$\psi_\bk\rightarrow \psi_\bk e^{i\bk\bdot\br}.$
In the limit $L\rightarrow\infty$ the field $u_L(\bx)$ defined with appropriately chosen $\tJ_{\bk,L}$
will converge to a homogeneous random field $u(\bx)$ invariant under translations by $a\mZ^d$. The
standard definition of the spectrum $n(\bk)=\lim_{L\rightarrow\infty}(L/2\pi)^d\langle |a_{\bk, L}|^2\rangle$
implies that one must choose \\
\be \lim_{L\rightarrow 0}\langle \tJ_{\bk_L,L}\rangle=n(\bk), \lb{spectrum} \ee
for $\bk\in \Lambda^*=[-k_{\max},+k_{\max}]^d,$ where $\bk_L= \frac{\bk L}{2\pi} ({\rm mod}\,M)
\cdot \frac{2\pi}{L}\in \Lambda_L^*$ converges to $\bk$ as $L=aM\rightarrow\infty$ (for fixed $a$). 
In this case, $u_L(\bx)$ converges in distribution as $L\rightarrow\infty$ to a homogeneous field 
$u(\bx)$ with spectrum $n(\bk).$  

Again following \cite{Choietal05a,Choietal05b}, we define $u(\bx)$ to be a {\it random-phase and amplitude field}
(RPA) if $u(\bx)$ is RP and if also $\tJ_\bk$ are mutually independent random variables for all $\bk\in \Lambda_L^*.$ 
This is equivalent to the factorization of the $N$-mode PDF into a product of 1-mode PDF's:
\be\cP^{(N)}(s)=\prod_{\bk\in \Lambda_\bk^*} P(s_\bk;\bk). \ee
All homogeneous Gaussian random fields are RPA. Conversely, for any sequence of RPA fields satisfying condition
(\ref{spectrum}) the spatial field $u_L(\bx)$ converges in distribution to the homogeneous Gaussian field with mean 
zero and spectrum $n(\bk)$ as $L\rightarrow\infty.$  
See \cite{Kurbanmuradov95}  for related rigorous results. Here we note only that 
\be u_L(\bx) = \left(\frac{2\pi}{L}\right)^{d/2} \sum_{\bk\in\Lambda_L^*} \sqrt{\tJ_{\bk,L}} \exp(i\bk\cdot\bx+i\varphi_\bk) \ee
is a sum of $N$ independent variables scaled by $1/\sqrt{N}.$ It is important to emphasize that the Fourier coefficients 
$\wt{a}_{\bk,L}$ can remain far from Gaussian in this limit. In physical space also there are non-vanishing cumulants 
for large but finite $L.$  A curious property of RPA fields for finite $L$ is the slow decay of their cumulants in certain 
directions.  For example, it is easy to calculate the 4th-order cumulant as  
\begin{eqnarray} 
&& \langle u(\bx_1)u(\bx_2)u^*(\bx_3)u^*(\bx_4)\rangle
- [\langle u(\bx_1)u^*(\bx_3)\rangle\langle u(\bx_2)u^*(\bx_4)\rangle+\left(1\leftrightarrow 2\right)] \cr
&& \,\,\,\,\,\,\,\,\,\,\,\,\,\,\,\,\,\,\,\,\,\,\,\,\,\,\,\,\,\,\,\,\,\,\,=  \left(\frac{2\pi}{L}\right)^{2d} \sum_{\bk\in\Lambda_L^*} 
          \left[\langle \tJ_\bk^2\rangle-2\langle\tJ_\bk\rangle^2\right]
           e^{i\bk\cdot(\bx_1+\bx_2-\bx_3-\bx_4)} \cr
&& \,\,\,\,\,\,\,\,\,\,\,\,\,\,\,\,\,\,\,\,\,\,\,\,\,\,\,\,\,\,\,\,\,\,\, \sim \left(\frac{2\pi}{L}\right)^d\int d^dk\,\,  
         \left[\langle \tJ_\bk^2\rangle-2\langle\tJ_\bk\rangle^2\right]
           e^{i\bk\cdot(\bx_1+\bx_2-\bx_3-\bx_4)} 
\end{eqnarray}           
asymptotically as $L\rightarrow\infty.$ As expected, this goes to zero as $O(L^{-d})$ for $L\rightarrow\infty,$ but 
for finite $L$ the cumulant is constant on the hyperplane $\bx_1+\bx_2=\bx_3+\bx_4,$ even when the pair of 
points $\bx_1,\bx_3$ are separated very far from the pair $\bx_2,\bx_4,$ for example. This is in contrast to some 
more traditional derivations of wave kinetic equations, which assume rapid spatial decay of higher-order cumulants 
\cite{BenneyNewell69}.     

A most important result for RPA fields is that the {\it empirical spectrum}  
\be\hn_L(\bk) = \left(\frac{2\pi}{L}\right)^d \sum_{\bk_1\in \Lambda_L^*} \tJ_{\bk_1,L}\delta^d(\bk-\bk_1), 
\,\,\,\, \bk\in \Lambda^* \ee
converges under the condition (\ref{spectrum}) to the deterministic spectrum $n(\bk)$ with probability going to 1 
in the limit $L\rightarrow\infty$ (weak law of large numbers). It is worth sketching the simple proof.  Note for any 
continuous function $\lambda$ on $\Lambda^*$ that 
\be \int_{\Lambda^*} d^dk\,\, \lambda(\bk)\hn_L(\bk) = \left(\frac{2\pi}{L}\right)^d \sum_{\bk_1\in \Lambda_L^*} \lambda(\bk_1)
\tJ_{\bk_1,L}. \ee
Crucially,
\begin{eqnarray}
&&  \left\langle\left|\left(\frac{2\pi}{L}\right)^d \sum_{\bk_1\in \Lambda_L^*} \lambda(\bk_1)\tJ_{\bk_1}-
       \left(\frac{2\pi}{L}\right)^d \sum_{\bk_1\in \Lambda_L^*} \lambda(\bk_1) n(\bk_1)\right|^2\right\rangle \cr
&& \,\,\,\,\,\,\,\,\,\,\,\,\,\,\,\,\,\,\,\,\,\,\,\,\,\,\,\,\,\,\,\,\,\,\,
= \left(\frac{2\pi}{L}\right)^{2d} \sum_{\bk_1\in \Lambda_L^*} \lambda^2(\bk_1)\langle|\tJ_{\bk_1}-n(\bk_1)|^2\rangle \cr
&& \,\,\,\,\,\,\,\,\,\,\,\,\,\,\,\,\,\,\,\,\,\,\,\,\,\,\,\,\,\,\,\,\,\,\,
\sim \left(\frac{2\pi}{L}\right)^d \int d^dk\,\,\lambda^2(\bk)\langle|\tJ_{\bk}-n(\bk)|^2\rangle =O(L^{-d})
\end{eqnarray}       
under the modest assumption that $\int d^dk\,\,\langle|\tJ_{\bk}-n(\bk)|^2\rangle <\infty.$
Since also  $\lim_{L\rightarrow\infty} \left(\frac{2\pi}{L}\right)^d \sum_{\bk_1\in \Lambda_L^*} \lambda(\bk_1) n(\bk_1)
=\int d^dk\,\,\lambda(\bk)n(\bk),$ the $L^2$-convergence follows 
\be \lim_{L\rightarrow\infty} \left\langle\left| \int d^dk\,\,\lambda(\bk)\hn_L(\bk)
     -\int d^dk\,\,\lambda(\bk)n(\bk)\right|^2\right\rangle=0\ee                                
and thus $\int d^dk\,\,\lambda(\bk)\hn_L(\bk)$ converges in probability to $\int d^dk\,\,\lambda(\bk)n(\bk)$ 
for every bounded, continuous $\lambda.$\footnote{Technically, this implies convergence in the weak-$\star$ topology 
for $\hn_L(\bk),n(\bk)$ as bounded, positive measures on $\Lambda^*.$ More physically, the function $\lambda(\bk)$ 
may be taken to be a smooth kernel $G_\delta(\bk-\bk_0)$ of width $\delta$ centered around a particular wavenumber $\bk_0$.
The coarse-grained spectrum $\ol{n}_{L,\delta}(\bk_0)=\int d^dk\,\, G_\delta(\bk-\bk_0)\hn_L(\bk)$ can then be interpreted as
the result of measuring the spectrum at $\bk_0$ with a finite resolution $\delta$ in wavenumber. The weak-$\star$ topology 
implies the convergence of such coarse-grained spectra in the limit $L\rightarrow\infty,$ when arbitrarily many wavenumbers 
lie within distance $\delta$ of $\bk_0$.}
This is sufficient to infer that the amplitude generating function defined in the introduction satisfies
\footnote{This is another standard result in probability theory, but, for completeness, we here recall the proof. 
Let $X_L=\int_{\Lambda^*} d^dk\,\,\lambda(\bk)\hn_L(\bk)$ and $x=\int_{\Lambda^*} d^dk\,\,\lambda(\bk) n(\bk)$. 
For $\epsilon>0,$ write
\be
\left|\llangle e^{iX_L}-e^{ix}\rrangle\right|\leq \llangle |e^{i(X_L-x)}-1|\,:|X_L-x|<\epsilon\rrangle
        +\llangle |e^{i(X_L-x)}-1|:\,|X_L-x|>\epsilon\rrangle  \ee
For any $\delta>0,$ one can choose $\epsilon$ so that $|e^{iz}-1|<\delta$ when $|z|<\epsilon.$ The first term 
is thus bounded by $\delta.$ By Chebyshev inequality the second term is bounded as 
\be \llangle |e^{i(X_L-x)}-1|:\,|X_L-x|>\epsilon\rrangle \leq 
 2 P(|X_L-x|>\epsilon)\leq \frac{2}{\epsilon^2}\langle|X_L-x|^2\rangle\rightarrow 0 \ee
for fixed $\epsilon$ as $L\rightarrow\infty.$ Thus,
\be \limsup_{L\rightarrow\infty}\left|\llangle e^{iX_L}-e^{ix}\rrangle\right| \leq \delta. \ee
Since $\delta$ is arbitrary, it follows that $\lim_{L\rightarrow\infty} \llangle e^{iX_L}\rrangle=e^{ix}$.}
\be  \lim_{L\rightarrow\infty} \cZ_L(\lambda)=\exp\left(i\int d^dk\,\,\lambda(\bk)n(\bk)\right) \lb{Zexp0} \ee
with $n(\bk)$ the deterministic spectrum. The law of large numbers (LLN) derived above means that 
for RPA fields the empirical spectrum $\hn_L(\bk)$ coincides with $n(\bk)$ at large $L$ for almost 
every realization of the random phases and amplitudes, not just after averaging over these variables.  

Notice that for the above result one does not actually need the full independence assumption in RPA, but only 
uncorrelated amplitudes:
\be \langle\tJ_{\bk_1}\tJ_{\bk_2}\rangle=\langle\tJ_{\bk_1}\rangle\langle \tJ_{\bk_2}\rangle,
\,\,\,\,\, \bk_1\neq\bk_2.\ee 
An even weaker and more general condition can be stated in terms of the {\it $M$-mode correlation functions} 
of the empirical spectrum, defined as:
\be\cN^{(M)}_L(\bk_1,...,\bk_M)=\langle \hn_L(\bk_1)\cdots \hn_L(\bk_M)\rangle. \ee
Note that (\ref{spectrum}) implies that $\lim_{L\rightarrow\infty}\cN^{(1)}_L(\bk)=n(\bk).$ A careful examination of the 
previous proof shows that, in order to obtain the LLN for the empirical spectrum, it suffices that 
\be  \lim_{L\rightarrow \infty} [\cN^{(2)}_L(\bk_1,\bk_2)-\cN^{(1)}_L(\bk_1) \cN^{(1)}_L(\bk_2)]=0. \lb{stosszahl} \ee
This condition is the analogue of the {\it Stosszahlansatz} invoked by Boltzmann to derive his kinetic equation.  
Under this assumption, all of the $M$-th order correlations that exist will factorize in the large-box limit:
\be  \lim_{L\rightarrow\infty} \cN^{(M)}_L(\bk_1,...,\bk_M)=\prod_{m=1}^M n(\bk_m). \lb{spect-fac} \ee
These observations go back to Lanford in his derivation of the Boltzmann equation  \cite{Lanford75,Lanford76}.  
The results that we shall present below suggest that properties (\ref{spectrum}) and (\ref{stosszahl}) for the 
wave field at the initial time, together with the RP property, are sufficient for the validity of the wave kinetic 
equation and a LLN for the empirical spectrum at positive times.  

RPA fields whose Fourier amplitudes possess the full independence property satisfy an even
stronger law of large numbers for the {\it empirical 1-mode PDF}, which is defined as 
\be \hP_L(s; \bk) = \left(\frac{2\pi}{L}\right)^d \sum_{\bk_1\in \Lambda_L^*} 
     \delta(s-\tJ_{\bk_1})\delta^d(\bk-\bk_1).  \lb{emp-pdf-1} \ee
Assume that the limiting random variables $\tJ_\bk=\lim_{L\rightarrow\infty}\tJ_{\bk_L, L}$ 
of an RPA field exist and have PDF's $P(s;\bk)$ which are continuous in $\bk.$ Then an analogue 
of the previous argument implies that the random functions $\hP_L(s; \bk)$ converge to $P(s;\bk)$
with probability approaching 1 as $L\rightarrow\infty.$ This implies the previous LLN for the 
spectrum, since $\hn_L(\bk)=\int_0^\infty ds\, s \hP_L(s;\bk)$ and $n(\bk)=\int_0^\infty ds\, s P(s;\bk).$
Although the ``empirical PDF'' defined in (\ref{emp-pdf-1}) is mathematically very convenient, 
it is not a PDF for finite $L$. It is therefore more intuitive to use an alternative definition    
\be  \hP_L(s; \Delta) = \frac{1}{N_L(\Delta)}\sum_{\bk\in \Lambda_L^*\cap \Delta} \delta(s-\tJ_\bk), 
\lb{emp-PDF-2} \ee
for any open set $\Delta \subset \Lambda^*$ and with $N_L(\Delta)$ the number of elements 
in $\Lambda_L^*\cap \Delta$. This quantity is nearly the same as $\frac{1}{|\Delta|} \int_\Delta 
d^dk \,\, \hP_L(s; \bk)$ for large $L$ but it has the advantage that it defines a probability measure 
in $s$ for each fixed $\Delta$ and $L.$ Definition (\ref{emp-PDF-2}) also has a simple intuitive 
meaning, since it represents the instantaneous distribution of amplitudes of the large number of 
Fourier modes that reside in the set $\Delta$ for large box-size $L.$ Under the same assumptions 
as above, it follows with probability going to 1 that 
 \be \lim_{L\rightarrow\infty} \hP_L(s;\Delta) =\frac{1}{|\Delta|} \int_\Delta d^dk \,\,P(s;\bk)\equiv P(s; \Delta). \ee
 
As before, strict independence is not necessary for the above results to hold. The essential 
property that is required is a factorization property of {\it multimode PDF'}s, defined for 
$\bk_1,...,\bk_M\in \Lambda^*$ by
\be \cP^{(M)}_L(s_1,...,s_M;\bk_1,...,\bk_M)
=\langle \delta(s_1-\tJ_{\bk_{1,L},L})\cdots \delta(s_M-\tJ_{\bk_{M,L},L})\rangle. \ee
The factorization property of the 2-mode PDF's for all pairs of distinct $\bk_1,\bk_2\in \Lambda^*$
\be  \lim_{L\rightarrow \infty} [\cP^{(2)}_L(s_1,s_2;\bk_1,\bk_2)-\cP_L^{(1)}(s_1; \bk_1)\cP_L^{(1)}(s_2;\bk_2)]=0
\lb{fact-pdf2} \ee
suffices\footnote{Since the proof is so similar to the one given previously, 
we give just a few details here. Because of our assumptions on $\tJ_\bk=\lim_{L\rightarrow\infty}\tJ_{\bk_L, L}$ 
it is easy to check that the average of the empirical PDF converges as $L\rightarrow\infty:$  
\be \langle \hP_L(s;\bk)\rangle= \left(\frac{2\pi}{L}\right)^d \sum_{\bk'\in \Lambda_L^*}
      \cP_L^{(1)}(s;\bk')\delta^d(\bk'-\bk) \rightarrow P(s;\bk). \ee
Therefore, in order to prove that $\hP_L(s;\bk)$ converges in probability to $P(s;\bk),$ with convergence in the 
weak-$\star$ topology on regular Borel measures, it is enough to show that       
\be \lim_{L\rightarrow\infty} \left\langle\left| \int ds \int_{\Lambda^*} d^dk \,\,\varphi(s,\bk)\hP_L(s; \bk)-
     \int ds \int_{\Lambda^*} d^dk\,\,\varphi(s,\bk)\langle \hP_L(s; \bk)\rangle \right|^2\right\rangle=0 \ee
for any continuous function $\varphi(s,\bk)$ on $\mR^+\times \Lambda^*$ which vanishes as $s\rightarrow\infty.$ 
Now a direct calculation of the above average shows that it equals
\be  \left(\frac{2\pi}{L}\right)^{2d} \sum_{\bk_1',\bk_2'\in \Lambda^*_L}  \int ds_1\int ds_2 
 \,\,\varphi(s_1,\bk_1')\varphi(s_2,\bk_2') \Big[P_L^{(2)}(s_1,s_2; \bk_1',\bk_2')-P_L^{(1)}(s_1; \bk_1')P_L^{(1)}(s_2;\bk_2')\Big].   \ee
The bracketed expression for $\bk_1'=\bk_2'$ is found to be
\be P_L^{(2)}(s_1,s_2;\bk_1',\bk_1')-P_L^{(1)}(s_1;\bk_1')P_L^{(1)}(s_2;\bk_1')=
   \delta(s_1-s_2)P_L^{(1)}(s_1;\bk_1')-P_L^{(1)}(s_1;\bk_1')P_L^{(1)}(s_2;\bk_1') \ee
so that its contribution to the average vanishes as $\sim \left(\frac{2\pi}{L}\right)^d \int_{\Lambda^*} 
d^d k\,\, \left\langle \left|\varphi(\tJ_\bk,\bk)-\langle\varphi(\tJ_\bk,\bk)\rangle\right|^2\right\rangle$  
for $L\rightarrow\infty.$ Thus,  to prove convergence, it must only be shown that the bracket term 
for $\bk_1'\neq \bk_2'$ vanishes as $L\rightarrow\infty.$ This is obviously true for RPA fields, when 
the bracket term for $\bk_1'\neq \bk_2'$ is zero!
} 
to derive the LLN for the empirical PDF and also the factorization of the multi-mode PDF's 
\be \lim_{L\rightarrow \infty}\cP^{(M)}_L(s_1,...,s_M;\bk_1,...,\bk_M)=
    \prod_{m=1}^M P(s_m;\bk_m) \ee
for all integers $M>2$ and distinct $\bk_1,...,\bk_M\in \Lambda^*.$  The ``asymptotic independence'' 
property (\ref{fact-pdf2}) is considerably weaker than strict RPA, permitting statistical dependence between 
Fourier modes at finite $L.$  We shall discuss natural ``microcanonical measures'' in section \ref{property-PDF} 
which satisfy the condition (\ref{fact-pdf2}) but not the more stringent RPA condition.
                     
In the following section we shall show that the above ``generalized RPA'' properties 
(\ref{spect-fac}), (\ref{fact-pdf2}) of initial conditions are preserved in time by the limiting kinetic hierarchies of 
wave turbulence.

\section{Multi-Mode Hierarchy Equations}\lb{multimode}


In this section we present our formal derivations of the multimode equations of wave kinetics and a 
discussion of their main properties. Our derivations are quite similar to those of Choi et al. \cite{Choietal05b},
but our model is somewhat more general in appearance. The major difference between our analysis and theirs, however, 
lies in a correct accounting of the size of the various terms in the limits as $L\rightarrow\infty$ and $\epsilon\rightarrow 0.$

We shall focus on the multimode generating function for finite box-size $L,$ defined as in \cite{Choietal05b} by  
\begin{align}
\cZ_L[\lambda,\mu,T]=\llangle\exp\left(\sum_{\bk\in\Lambda_L^*} \lambda_\bk J_\bk(T)\right)
\prod_{\bk\in\Lambda^*}\psi_{\bk}^{\mu_\bk}(T)\rrangle,
\lb{ChoiZ} \end{align}
where $\lambda_\bk\in \mR$ and $\mu_\bk\in \mZ$ for all $\bk\in \Lambda_\bk^*.$ The time $T$ 
is free for the moment but will later be chosen to be a time intermediate between the wave-period and
the nonlinear time-scale. This generating function is calculated perturbatively in the nonlinearity parameter $\epsilon$
at finite $L$, by expanding the solution of the dynamical equation 
\begin{align}
\dot{a}_1=\epsilon\sum_{2,3}L_{123} a_{2} a_{3} e^{i\omega^1_{23}t}\delta^1_{23}
\end{align}
into a power series
\begin{align}
a_1(T)=a_1(0)+\epsilon a_1^{(1)}(T)+\epsilon^2 a_1^{(2)}(T)+\cO(\epsilon^3).
\label{expansion}\end{align}
A straightforward calculation gives 
\begin{eqnarray}
&& a_1^{(0)}=a_1(0), \lb{a0} \\
&& a_1^{(1)}=\sum_{2,3}L_{123}a_2^{(0)} a_3^{(0)} \Delta_T(\omega^1_{23})\delta^1_{23}, \lb{a1} \\
&& a_1^{(2)}=\sum_{2345}L_{123}L_{245} a_3^{(0)}a_4^{(0)}a_5^{(0)}
E_T(\omega^1_{345},\omega^1_{23})\delta^1_{23}\delta^{2}_{45} + (2\leftrightarrow 3) \lb{a2}
\end{eqnarray}
We employ here the standard definitions \cite{BenneyNewell69}:  
\be \Delta_T(x)=\int_0^T \exp(ixt)dt, \,\,\,\,\, E_T(x,y)=\int_0^T \Delta_t(x-y)\exp(iyt)dt.  \ee
The terms in this perturbative solution of the equations of motion can be represented by a 
version of the Wyld diagram expansion \cite{ZakharovLvov75}. In this technique the
various contributions are represented by tree diagrams, as illustrated in Figs. 1-3 for the
zeroth-, first- and second-order terms. In our conventions, similar to those of \cite{Choietal05b}, 
a solid line labeled with an integer $j$ represents a factor $a^{(0)}_j$, whereas a 
dashed line indicates the absence of such a factor. An arrow is added to a solid line
to indicate $\sigma_j=+1$ (``source'') when the arrow is pointed away from $j$ and $\sigma_j=-1$ 
(``sink'') when the arrow is pointed toward $j$. The triple vertex labelled $1,2,3$ represents a factor 
$L^{\sigma_1,\sigma_2,\sigma_3}_{\bk_1,\bk_2,\bk_3}e^{\omega^1_{23}t}\delta^1_{23}$ 
with $\sigma_1=+1$ when the arrow points out of the vertex and $\sigma_1=-1$
when the arrow points into the vertex. The times at each vertex are ordered causally, with 
the latest times at the ``root'' of the tree, here labelled by $1$. When integrations 
are performed over all times from $0$ to $T$ consistent with this ordering, then the various
contributions  to the perturbative solution result.   

\begin{figure}[!ht]
\begin{center}
\includegraphics[width=200pt,height=35pt]{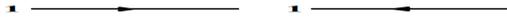}\\
\end{center}
\caption{Zeroth-order terms $a^{(0)+}_1$ and $a^{(0)-}_1$.} 
\end{figure}

\begin{figure}[!ht]
\begin{center}
\includegraphics[width=200pt,height=100pt]{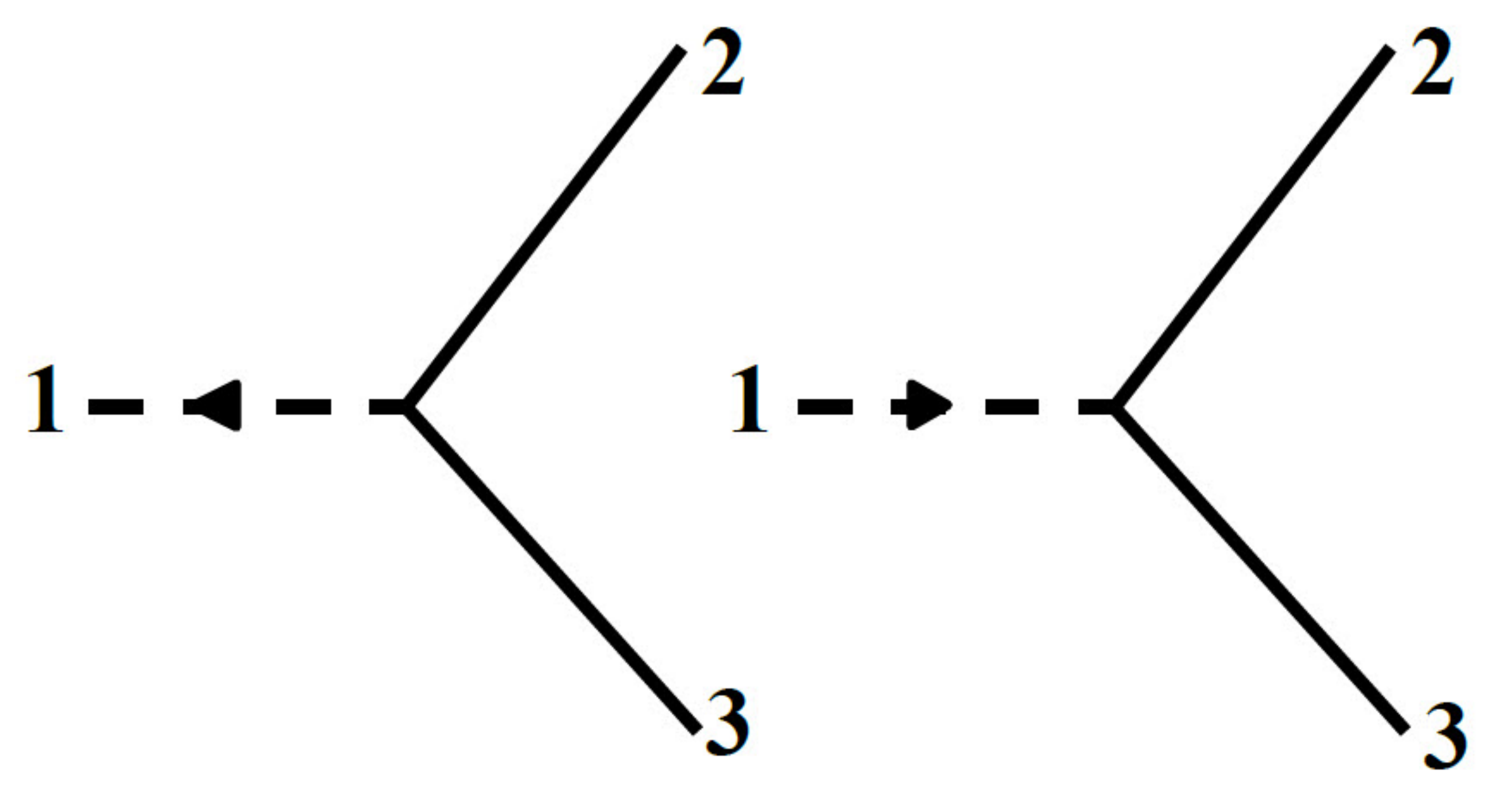}\\
\end{center}
\caption{First-order terms $a^{(1)+}_1$ and $a^{(1)-}_1$.} 
\end{figure}


\begin{figure}[!ht]
\begin{center}
\includegraphics[width=150pt,height=100pt]{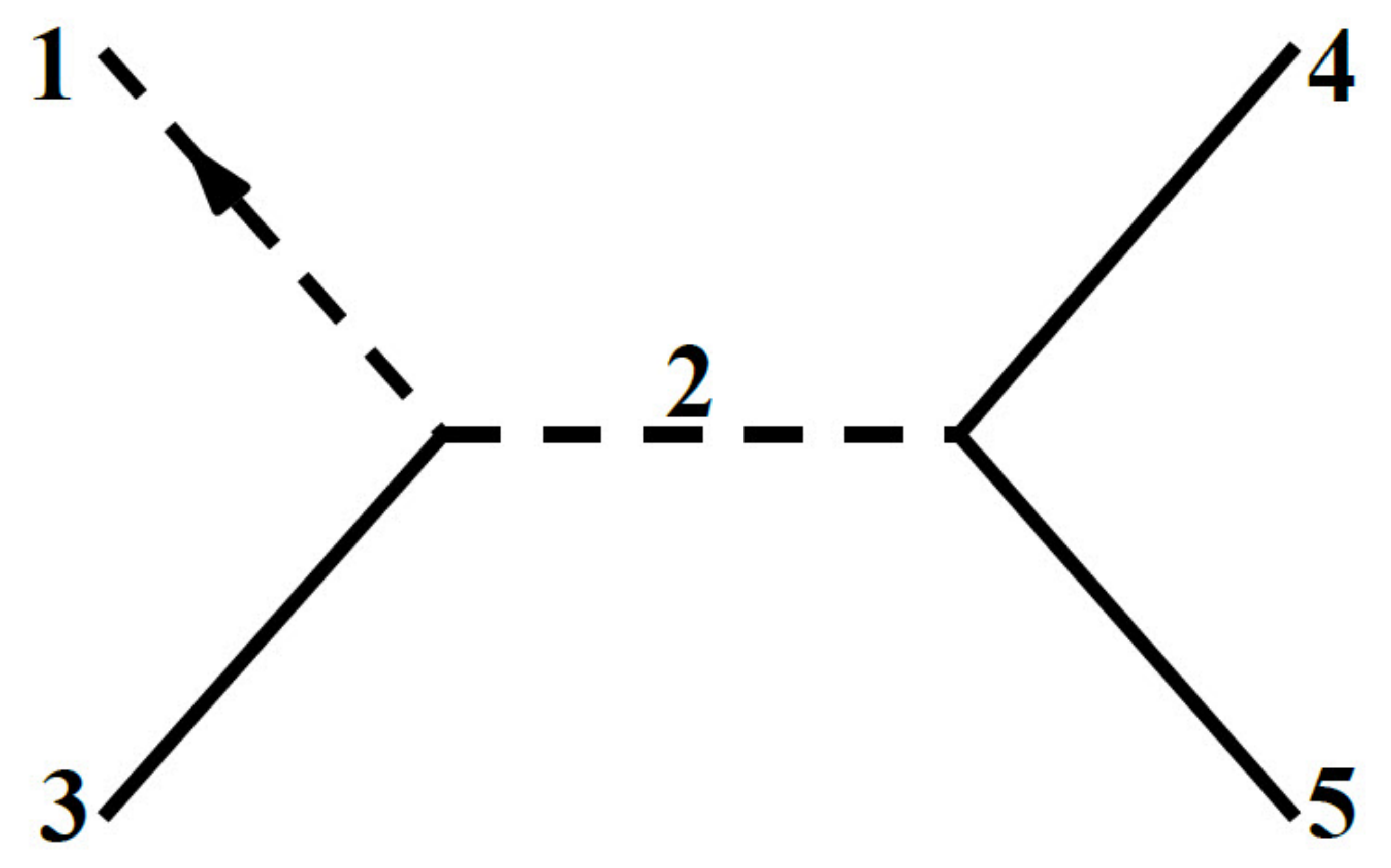}\\
\end{center}
\caption{Second-order term $a^{(2)+}_1$.} 
\end{figure}

The generating function is obtained perturbatively by substituting (\ref{expansion}) and expanding 
to obtain
\begin{equation}
\cZ_L[\lambda,\mu,T]=\cX_L\{\lambda,\mu,T\}+\cX_L^*\{\lambda,-\mu,T\} \label{ZXX}
\end{equation}
with
\begin{align}
\cX_L\{\lambda,\mu,T\}=\cX_L\{\lambda,\mu,0\}+\Big\langle\prod_{\bk\in \Lambda^*}e^{\lambda_\bk  J^{(0)}_\bk}[\epsilon \cJ_1+\epsilon^2(\cJ_2+\cJ_3+\cJ_4+\cJ_5)]\Big\rangle_J+\cO(\epsilon^3),
\lb{Xdef} \end{align}
where, as in \cite{Choietal05b}, 
\begin{eqnarray}
&& \cJ_1=\Big\langle\prod_{\bk}\psi_{\bk}^{(0) \mu_\bk}\sum_1 \Big(\lambda_1+\frac{\mu_1}{2J_1^{(0)}}\Big)a_1^{(1)}a_1^{(0)*}\Big\rangle_\psi, \lb{J1} \\
&& \cJ_2=\frac{1}{2}\Big\langle\prod_{\bk}\psi_{\bk}^{(0)\mu_\bk}\sum_1\Big(\lambda_1+\lambda_1^2 J_1^{(0)}-\frac{\mu_1^2}{4 J_1^{(0)}}\Big)|a_1^{(1)}|^2\Big\rangle_\psi, \lb{J2} \\
&& \cJ_3=\Big\langle\prod_{\bk}\psi_{\bk}^{(0)\mu_\bk}\sum_1 \Big(\lambda_1+\frac{\mu_1}{2J_1^{(0)}}\Big)a_1^{(2)}a_1^{(0)*}\Big\rangle_\psi, \lb{J3} \\
&& \cJ_4=\Big\langle\prod_{\bk}\psi_{\bk}^{(0)\mu_\bk}\sum_1\Big(\frac{1}{2}\lambda_1^2+\frac{\mu_1}{4J_1^{(0)2}}(\frac{\mu_1}{2}-1)+
\frac{\lambda_1\mu_1}{2J_1^{(0)}}\Big)(a_1^{(1)}a_1^{(0)*})^2\Big\rangle_\psi, \lb{J4} \\
&& \cJ_5=\frac{1}{2}\Big\langle\prod_{\bk}\psi_{\bk}^{(0)\mu_\bk}\sum_{1\neq 2}\Big(\lambda_1\lambda_2(a_1^{(1)}a_1^{(0)*}+a_1^{(1)*}a_1^{(0)})a_2^{(1)}a_2^{(0)*}\cr
&& \,\,\,\,\,\,\,\,\,\,\,\,\,\,\,\,\,\,\,\,\,\,\,\,\,\,\,\,\,\,\,\,\,\,\,\,\,\,\,\,\,\,\,\,\,\,\,\,\,\,\,\,\,\,\,\,\,\,\,\,\,\,\,\,\,\,
+(\lambda_1+\frac{\mu_1}{4J_1^{(0)}})\frac{\mu_2}{J_2^{(0)}}(a_2^{(1)}a_2^{(0)*}-a_2^{(1)*}a_2^{(0)})a_1^{(1)}a_1^{(0)*}\Big)\Big\rangle_\psi.
\lb{J5} 
\end{eqnarray}
The various contributions before averaging over phases can represented by diagrams, 
combining the tree diagrams for each of the factors by joining the trees with the same 
``root'' indices. Each of the integer labels indicates an index to be summed over independently
(except for the constraints imposed by delta-functions at the vertices).  We illustrate this representation 
in Figs.4-5 below for the first two contributions:

\begin{figure}[!ht]
\begin{center}
\includegraphics[width=150pt,height=100pt]{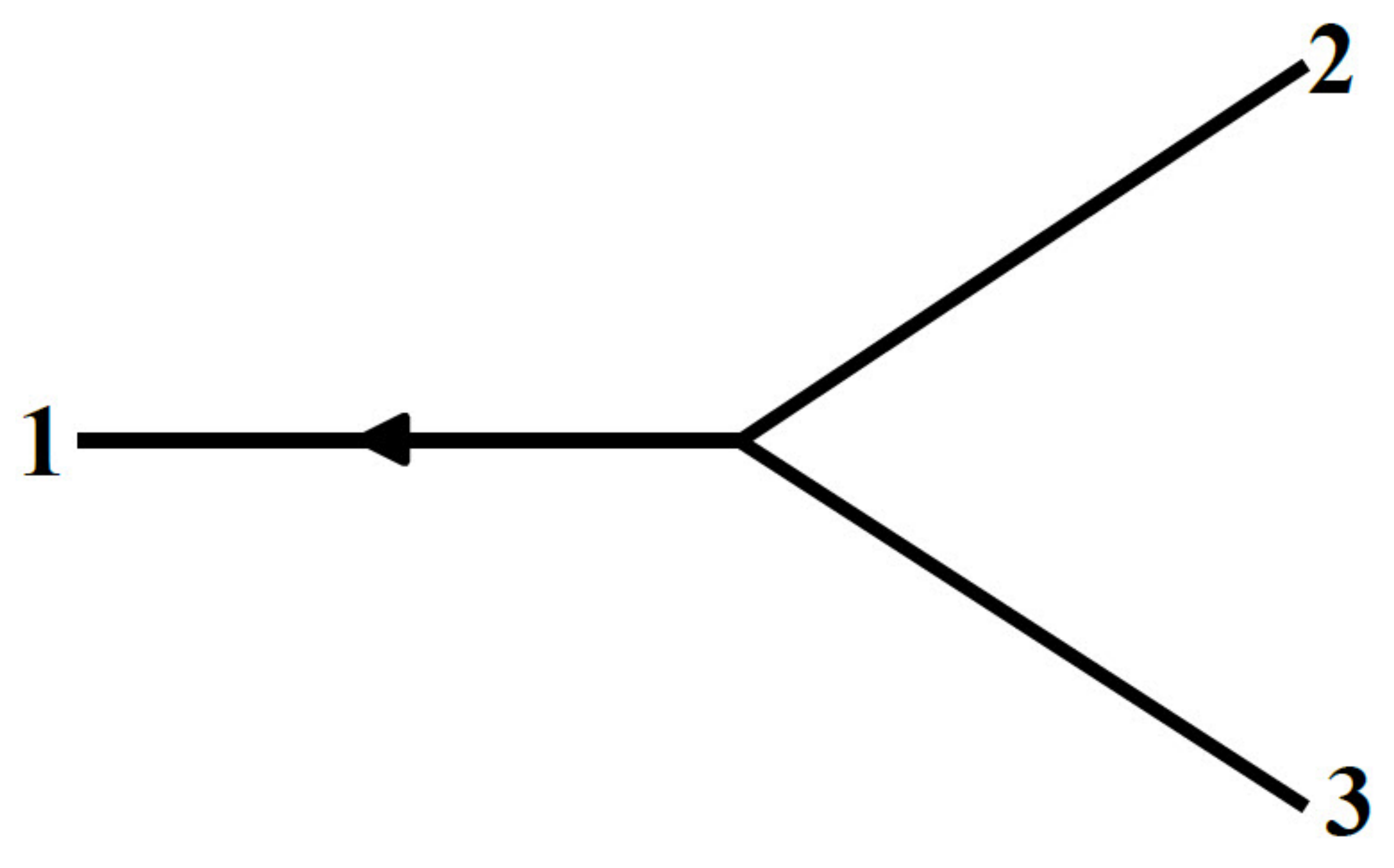}\\
\end{center}
\caption{First-order term $\sum_1(\lambda_1+\frac{\mu_1}{2J_1})a_1^{(1)}a_1^{(0)*}$.} 
\end{figure}

\begin{figure}[!ht]
\begin{center}
\includegraphics[width=150pt,height=100pt]{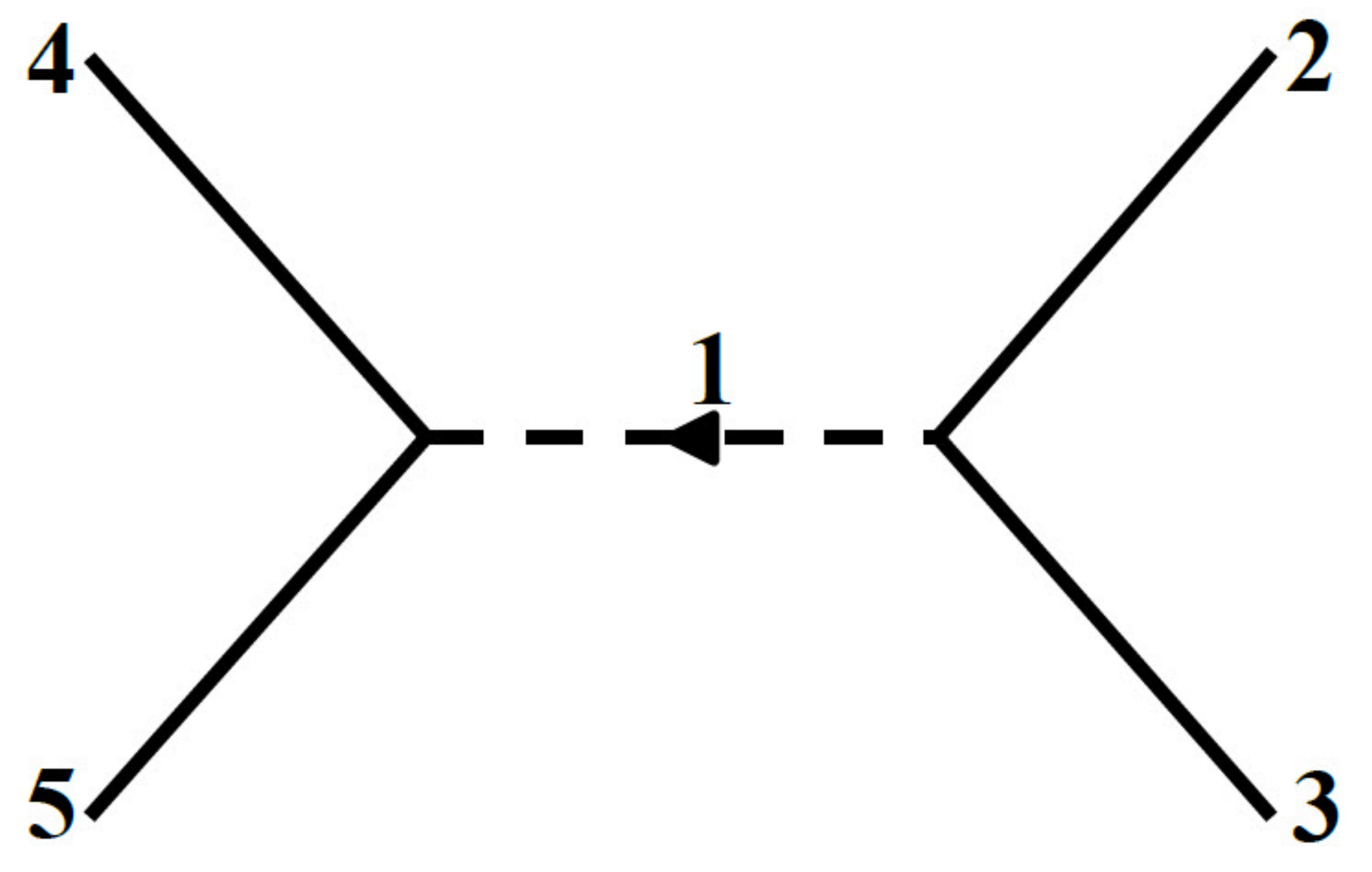}\\
\end{center}
\caption{Second-order term $\sum_1(\lambda_1+\lambda_1^2J_1-\frac{\mu_1^2}{4J_1})|a_1^{(1)}|^2$.} 
\end{figure}

The only contributions which survive the average over phases must have all phases summing to zero 
before averaging. This means that every $a^{(0)}$ factor must either pair with another factor  $a^{(0)}$ so that their 
phases sum to zero or belong to a set of $a^{(0)}$'s that pair with a $\psi_\bk^{(0)\mu_\bk}$ factor so that the sum 
of all their phases is zero. The first we call an ``internal coupling'', represented graphically by 
a solid line connecting the paired indices $i,j$ which contributes a factor $\delta_{\sigma_i+\sigma_j,\,0}\delta_{\bk_i,\bk_j}$
after phase averaging. The second we call an ``external coupling'', represented by joining all 
of the solid lines for indices $i_1,i_2,...,i_p$ to a blob $\bullet$ labelled $a$ that represents the 
phase $\psi_{\bk_a}^{(0)\mu_{\bk_a}}$ which contributes a factor $
\delta_{\sigma_{i_1}+\cdots\sigma_{i_p}+\mu_a,\,0}\prod_{j=1}^p \delta_{\bk_j,\bk_a}$
after phase averaging. This graphical representation is essentially the same as that employed
in \cite{Choietal05b}. We illustrate the representation in Figs.6 and 7 below for the terms in 
$\cJ_1$ and $\cJ_2$ which turn out to give the leading-order contributions to those quantities 
as $L\rightarrow\infty.$

\begin{figure}[!ht]
\begin{center}
\includegraphics[width=150pt,height=100pt]{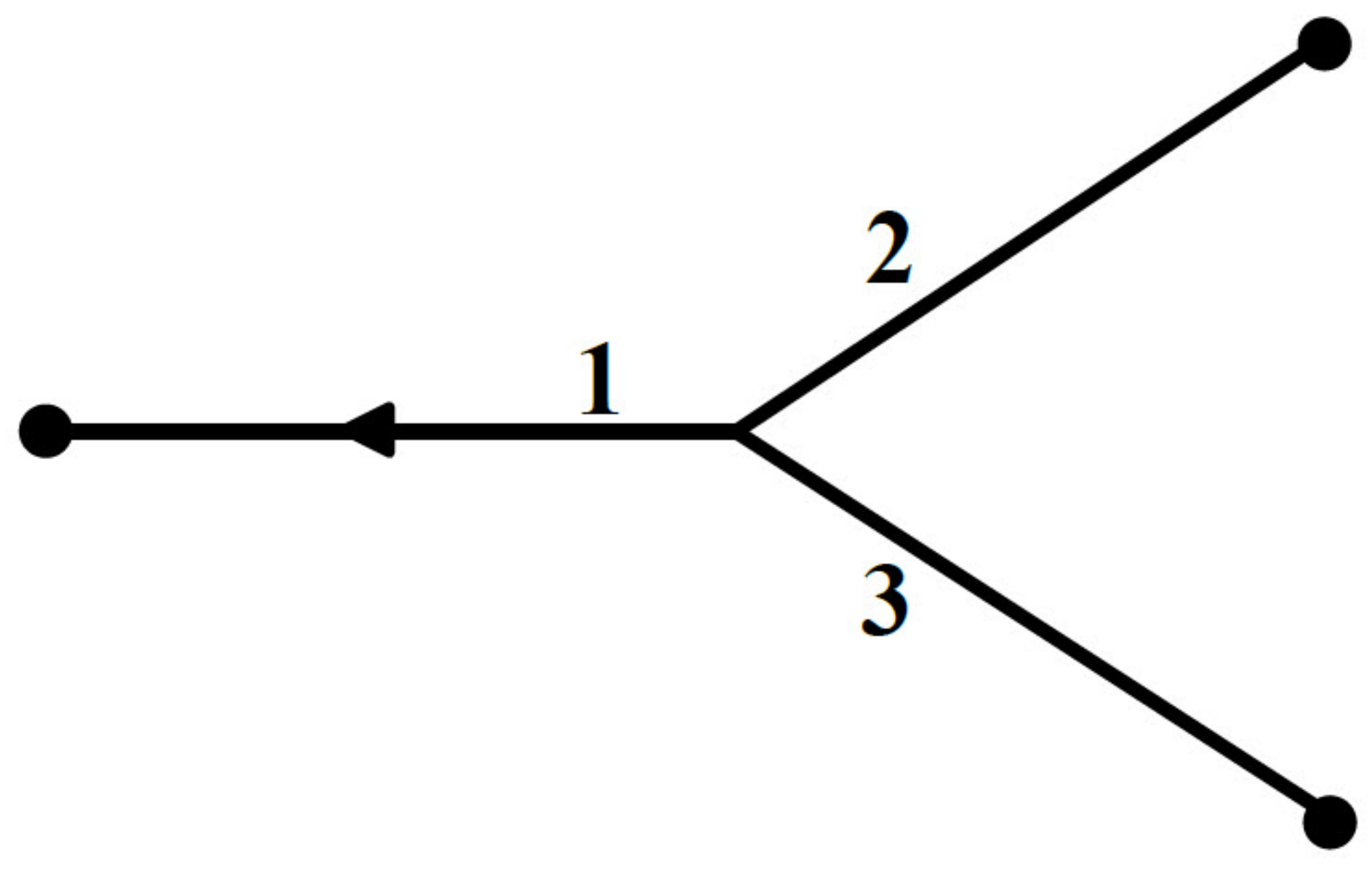}
\includegraphics[width=200pt,height=100pt]{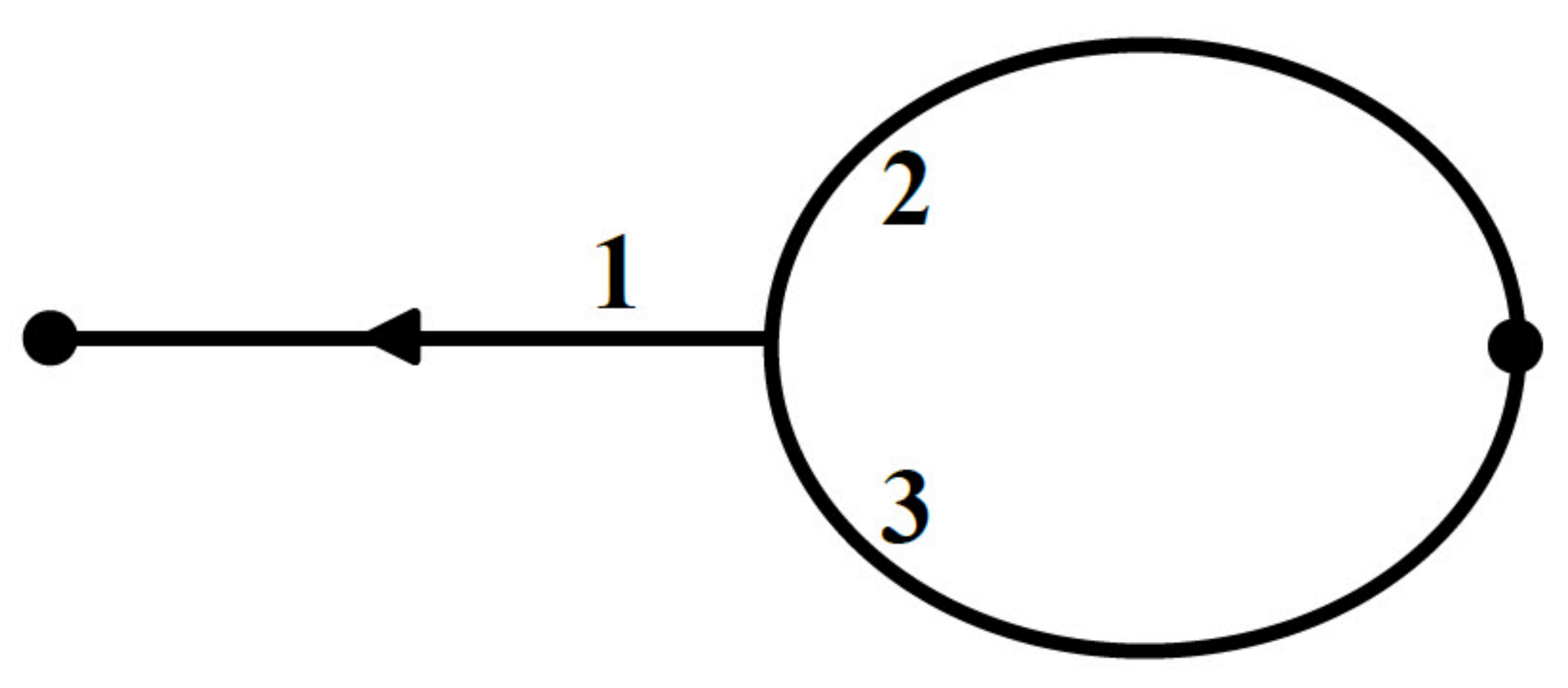}\\
\end{center}
\caption{Contributions to $\cJ_1$.} 
\end{figure}


\begin{figure}[!ht]
\begin{center}
\includegraphics[width=120pt,height=100pt]{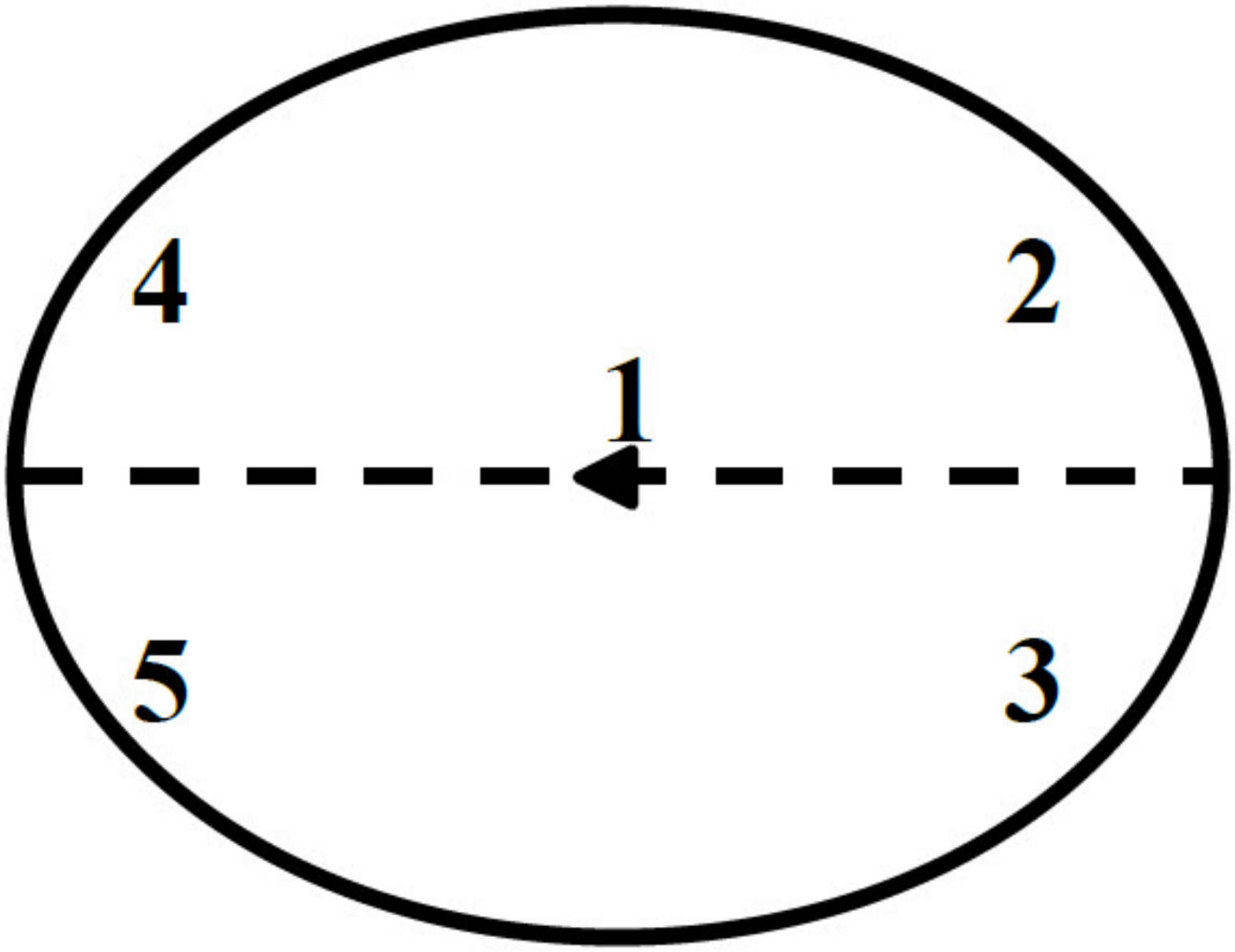}\\
\end{center}
\caption{Leading contribution to $\cJ_2$.} 
\end{figure}

\noindent Note that solid lines connected to external blobs have their wavenumbers ``pinned'' at the wavenumbers
of those blobs, so that those wavenumbers are no longer summed over.  For simplicity we often omit the labels
of the blobs, since those play no important role other than the ``pinning'' described above. 

\subsection{Spectral Hierarchy}\lb{spectral}

We now consider the first of the possible limits of $\cZ_L[\lambda,\mu]$ as $L\rightarrow\infty$ involving 
{\it all} of the $N$ modes. This leads to a set of equations for the spectral generating function and the 
spectral correlation functions. We first sketch the derivation of these equations, with more details in 
\ref{appendix-spectral}, and then analyze their basic properties.   

\subsubsection{Derivation}\lb{derive-spect}

The crucial observation which leads to our results differing from \cite{Choietal05b} is that one must keep 
$\tJ_\bk=O(1)$, {\it not} $J_\bk=O(1),$ in order to have a finite spectrum in the limit $L\rightarrow \infty.$
Thus, we take in the generating function of Choi et al.
\be J_\bk=\left(\frac{2\pi}{L}\right)^d \tJ_\bk, \,\,\,\lambda_\bk=i\lambda(\bk), \ee  
where $\lambda(\bk)$ is a smooth test function and, as before, $\mu_\bk$ are integers. This leads to 
\begin{align}
\cZ_L[\lambda,\mu]=\llangle\exp\left(i\sum_{\bk\in\Lambda_L^*}  \left(\frac{2\pi}{L}\right)^{d} \lambda(\bk)\tJ_\bk\right)\prod_{\bk\in\Lambda^*}\psi_{\bk}^{\mu_\bk}\rrangle,
\end{align}
We consider  the large-$L$ asymptotics of the various terms in the perturbation expansion of this quantity,
employing the following standard substitutions:
\begin{equation} 
\left(\frac{2\pi}{L}\right)^d\sum_\bk \Longrightarrow \int d^dk,\,\, 
      \left(\frac{L}{2\pi}\right)^d \delta_{\bk,\bk'} \Longrightarrow \delta^d(\bk-\bk'), \,\,
      \left(\frac{L}{2\pi}\right)^d \frac{\partial}{\partial\lambda_\bk} \Longrightarrow \frac{\delta}{\delta\lambda(\bk)} 
\label{largeL} \end{equation}      
From now on, we suppress superscript $(0)$ when there is no confusion. 

{\it{Calculation of $\cJ_1$}:} This quantity is represented by the graphs in Fig.6, or analytically:
\begin{eqnarray}
\cJ_1
&=& \sum_{1\neq 2\neq 3} L_{123} \Big(\lambda_1+\frac{\mu_1}{2J_1}\Big)\sqrt{J_1J_2J_3} \delta_{\mu_1,1}
                    \delta_{\mu_2+\sigma_2,\,0}\delta_{\mu_3+\sigma_3,\,0}\Delta(\omega^1_{23})\delta^1_{23} \cr
           && + \sum_{1\neq 2} L_{122} \Big(\lambda_1+\frac{\mu_1}{2J_1}\Big) \sqrt{J_1} J_2 \delta_{\mu_1,1}
                    \delta_{\mu_2+2\sigma_2,\,0}\Delta(\omega^1_{22})\delta^1_{22}
\end{eqnarray}   
Note that $\sigma_1=+1$ in coefficient $L_{123}$. Taking into account the wavenumber delta functions, 
there are two summations in the first term and one in the second.  However, as noted by Choi et al. (2005), 
these sums contain only a couple of non-zero terms, and then only for special choices of the $\mu$'s. 
In their terminology, the sums are ``pinned'' by these choices of $\mu$. Making the substitutions  
$J_i= (2\pi/L)^{d/2} \tJ_i$ and $\lambda_1=i\lambda(\bk_1)$ into the above gives the leading contribution
\begin{eqnarray}
\cJ_1
 &=&  \left(\frac{2\pi}{L}\right)^{d/2}\sum_{1\neq 2\neq 3} L_{123} \frac{1}{2} \mu_1 \sqrt{\frac{\tJ_2\tJ_3}{\tJ_1}} \delta_{\mu_1,1}
                    \delta_{\mu_2,-\sigma_2}\delta_{\mu_3,-\sigma_3}\Delta(\omega^1_{23})\delta^1_{23} \cr
           && +  \left(\frac{2\pi}{L}\right)^{d/2}\sum_{1\neq 2} L_{122} \frac{1}{2}\mu_1 \frac{\tJ_2}{\sqrt{\tJ_1}}\delta_{\mu_1,1}\delta_{\mu_2,-2\sigma_2}
                    \Delta(\omega^1_{22})\delta^1_{22}\propto L^{-d/2} 
\end{eqnarray}   
As we shall see, this term first-order in $\epsilon$ gives a subleading correction in the limit $L$ large but finite,
larger than many of the terms that Choi et al. retained in their evolution formula for $L\rightarrow\infty.$

{\it{Calculation of $\cJ_2$}}: 
Averaging over phases, the leading terms are contained in the contribution from Fig.7 and a similar contribution 
with 2 and 3 interchanged. Because there are no external couplings, all $\mu's$ must vanish. Because of the 
internal couplings, there are no sums over the wavenumbers 4,5 which are the same as wavenumbers 2,3. The final 
contribution to the generating function is:
\begin{eqnarray} 
\left\langle e^{\sum_\bk \lambda_\bk J_\bk} \,\cJ_2\right\rangle_J
& = & \delta_{\mu,0}\sum_{1,2,3}\left\langle \left(\lambda_1+\lambda_1^2 J_1\right)J_2J_3e^{\sum_\bk \lambda_\bk J_\bk} \right\rangle_J
 |L^{+,\sigma_2,\sigma_3}_{\bk_1,\bk_2,\bk_3}|^2 |\Delta(\sigma_2\omega_2+\sigma_3\omega_3-\omega_1)|^2
\delta_{\bk_1,\,\sigma_2\bk_2+\sigma_3\bk_3} \cr
& = & 9\delta_{\mu,0}\sum_{1,2,3}\left(\lambda_1+\lambda_1^2 \frac{\partial}{\partial\lambda_1}\right)\frac{\partial^2\cZ}{\partial\lambda_2\partial\lambda_3}
\times |H^{-,\sigma_2,\sigma_3}_{\bk_1,\bk_2,\bk_3}|^2 |\Delta(\sigma_2\omega_2+\sigma_3\omega_3-\omega_1)|^2
\delta_{\bk_1,\,\sigma_2\bk_2+\sigma_3\bk_3} \cr
&&
\label{J2A} \end{eqnarray}
Papers \cite{JakobsenNewell04,Choietal05b} retain {\it both} of the terms in the first factor in (\ref{J2}), i.e. 
both $\lambda_1$ and $\lambda_1^2J_1.$ However, with the proper scaling, $\lambda_1=O(1)$ while 
$\lambda_1^2J_1=\left(\frac{2\pi}{L}\right)^d \lambda_1\tJ_1=O(L^{-d}).$ Thus, the second term is even
smaller than the contribution from $\cJ_1$ in the limit as $L\rightarrow\infty$ and should be neglected. This can be seen also 
substituting $J_i= (\frac{2\pi}{L})^{d/2} \tJ_i$ and $\lambda_1=i\lambda(\bk_1)$ and taking the limit $L\rightarrow\infty$ using (\ref{largeL}):
\begin{eqnarray}
\left\langle e^{\sum_\bk \lambda_\bk J_\bk} \,\cJ_2\right\rangle_J &\sim &
-9i\delta_{\mu,0} \sum_{\ul{\sigma}=(-1,\sigma_2,\sigma_3)}\int d^dk_1 d^dk_2 d^dk_3 
     \left( \lambda(\bk_1) +\left(\frac{2\pi}{L}\right)^d \lambda(\bk_1)\frac{\delta}{\delta\lambda(\bk_1)}\right)\frac{\delta^2\cZ}{\delta\lambda(\bk_2)\delta\lambda(\bk_3)}\cr
  &&   \,\,\,\,\,\,\,\,\,\,\,\,\,\,\,\,\,\,\,\,\,\,\,\,\,\,\,\,\,\,\,\,\,\,\,\,\,\,\,\,\,\,\,\,\,\,\,\,\,\,\,\,\,\,\,\,\,\,\,\,\,\,\,\,\,\,\,\,\,\,\,\,
  \times |H^{-,\sigma_2,\sigma_3}_{\bk_1,\bk_2,\bk_3}|^2|\Delta(\ul{\sigma}\cdot\ul{\omega})|^2\delta^d(\ul{\sigma}\cdot\ul{\bk}).
\end{eqnarray}
The only surviving term in the limit as $L\rightarrow\infty$ is the one proportional to $\lambda(\bk_1).$ Verifying our initial 
estimation, the term proportional to $\lambda(\bk_1)\delta/\delta\lambda(\bk_1)$ is $O(L^{-d})$ and vanishes in the limit. 

A similar analysis may be carried through for the remaining contributions from $\cJ_3,\cJ_4$ and $\cJ_5.$ 
Detailed discussion of all the terms is given in \ref{appendix-spectral}, where it is shown that only $\cJ_2,\cJ_3,\cJ_5$ 
give $O(1)$ contributions in the large-box limit $L\rightarrow\infty$:
\begin{eqnarray}
\cX(T)-\cX(0) &\sim&
-9i\delta_{\mu,0} \epsilon^2\sum_{\ul{\sigma}=(-1,\sigma_2,\sigma_3)}\int d^dk_1 d^dk_2 d^dk_3 
\delta^d(\ul{\sigma}\cdot\ul{\bk})
\big|\Delta_T(\ul{\sigma}\cdot\omega(\ul{\bk}))\big|^2
     \cr
&& \,\,\,\,\,\,\,\,\,\,\,\,\,\,\,\,\,\,\,\,\,\,\,\,\,\,\,\,\,\,\,\,\,\,\,\,\,\,\,\,\,\,\,\,\,\,\,\,\,\,\,\,\,\,\,\,\,\,\,\,\,\,\,\,\,\,\,\,\,\,\,\,
  \times \lambda(\bk_1)  \big|H^{\ul{\sigma}}_{\ul{\bk}}\big|^2 \frac{\delta^2\cZ}{\delta\lambda(\bk_2)\delta\lambda(\bk_3)}
  \cr
&&+\epsilon^2\left[18i\delta_{\mu,0}\sum_{\ul{\sigma}=(-1,\sigma_2,\sigma_3)}\sigma_2\int_{(\mT^d)^3}d^dk_1d^dk_2d^dk_3 \delta^d(\ul{\sigma}\cdot\ul{\bk}) E_T(0,\ul{\sigma}\cdot\omega(\ul{\bk}))\right.  \cr
&&\,\,\,\,\,\,\,\,\,\,\,\,\,\,\,\,\,\,\,\,\,\,\,\,\,\,\,\,\,\,\,\,\,\,\,\,\,\,\,\,\,\,\,\,\,\,\,\,\,\,\,\,\,\,\,\,\,\,\,\,\,\,\,\,\times\lambda(\bk_1)\big|H^{\ul{\sigma}}_{\ul{\bk}}\big|^2\frac{\delta^2\cZ}{\delta\lambda(\bk_1)\delta\lambda(\bk_3)}        \cr
&&-9{\sum_1} \delta_{\mu_1,1}\delta_{\mu_{-1,1}}\prod_{\bk\neq\bk_1,-\bk_1}\delta_{\mu_\bk,0}\sum_{\ul{\sigma}=(-1,\sigma_2,\sigma_3)}\sigma_2\int d^dk_2d^dk_3 \delta^d(\ul{\sigma}\cdot\ul{\bk})
\cr
&&\,\,\,\,\,\,\,\,\,\,\,\,\left.E_T(\omega(\bk_1)+\omega(-\bk_1),\ul{\sigma}\cdot\omega(\ul{\bk})) 
H^{\ul{\sigma}}_{\ul{\bk}} H^{+,\sigma_2,\sigma_3}_{-\bk_1,\bk_2,\bk_3}\llangle e^{\sum_\bk \lambda_\bk J_\bk}\sqrt{\frac{\tJ_{-1}}{\tJ_1}}\tJ_3\rrangle_J
+(2\leftrightarrow 3)\right]
\cr
&&-\frac{9}{2}\epsilon^2{\sum_1} \delta_{\mu_1,1}\delta_{\mu_{-1},1}\prod_{\bk\neq\pm\bk_1}\delta_{\mu_\bk,0}\sum_{\ul{\sigma}=(-,\sigma_2,\sigma_3)}\int d^dk_2d^dk_3\delta^d(\ul{\sigma}\cdot\ul{\bk})
\cr
&&\,\,\,\,\,\,\,\,\,\,\left[\Delta_T(\ul{\sigma}\cdot\omega(\ul{\bk}))\Delta_T(-\ul{\sigma}\cdot\omega(\ul{\bk}'))\right]
H^{\ul{\sigma}}_{\ul{\bk}}H^{+,\sigma_2,\sigma_3}_{-\bk_1,\bk_2,\bk_3}\llangle\frac{\tJ_2\tJ_3}{\sqrt{\tJ_1\tJ_{-1}}}e^{\sum_\bk\lambda_\bk J_\bk}\rrangle_J.
\end{eqnarray}
The first term is that already found for $\cJ_2,$ the second bracketed term is from $\cJ_3,$ and the final term
is from $\cJ_5.$ Note that the expressions proportional to $ \delta_{\mu_1,1}\delta_{\mu_{-1,1}}$ were 
missed in \cite{JakobsenNewell04,Choietal05b} (although they are larger than many extra terms that those 
authors retained in their final equations which actually vanish as $L\rightarrow\infty$!) The terms 
proportional to $ \delta_{\mu_1,1}\delta_{\mu_{-1,1}}$ indeed do not appear in the final equations, 
not because of the large-$L$ limit but because they are nonsecular.  

We therefore consider the limit of weak nonlinearity. The limit is achieved by choosing the time variable 
$T$ in the expansion (\ref{expansion}) to lie between the wave period and nonlinear timescale $O(\epsilon^{-2})$ 
and by then taking $\epsilon\rightarrow 0.$ For this purpose we use the following standard asymptotic relations
for $T\rightarrow\infty$ \cite{BenneyNewell69}:
\begin{eqnarray}
&& \Delta_T(x)\sim \wt{\Delta}(x)=\pi\delta(x)+iP\left(\frac{1}{x}\right),\,\,\,\,E_T(x;y)\sim\Delta_T(x)\Delta_T(y)\sim\wt{\Delta}(x)\wt{\Delta}(y), \cr
&& |\Delta_T(x)|^2\sim 2\pi T\delta(x)+2P\left(\frac{1}{x}\right)\frac{\partial}{\partial x},\,\,\,\,
E_T(x;0)\sim\wt{\Delta}(x)\left(T-i\frac{\partial}{\partial x}\right),\,\,
\label{smallep}\end{eqnarray}
The terms multiplied by $\delta_{\mu,0}$ contain secular contributions proportional to $T,$ while the terms 
proportional to $ \delta_{\mu_1,1}\delta_{\mu_{-1,1}}$ are nonsecular. We now use (\ref{ZXX}) to calculate $\cZ(T)-\cZ(0),$
with the observation that it is changed by our replacement $\lambda_\bk=i\lambda(\bk)$ into 
\begin{equation}
\cZ[\lambda,\mu,T]=\cX\{\lambda,\mu,T\}+\cX^*\{-\lambda,-\mu,T\}. \label{ZXX2}
\end{equation}
Finally, replacing $(\cZ(T)-\cZ(0))/T$ by $\dot{\cZ}$ and using time variable $\tau=\epsilon^2 t$, one obtains 
\begin{eqnarray}
\frac{d}{d\tau}{\cZ}[\lambda,\mu,\tau]&=&-36i\pi\delta_{\mu,0}\sum_{\ul{\sigma}=(-1,\sigma_2,\sigma_3)}\int d^dk_1 d^dk_2 d^dk_3
\delta^d(\ul{\sigma}\cdot\ul{\bk})\delta(\ul{\sigma}\cdot\omega(\ul{\bk}))\lambda(\bk_1)|H^{\ul{\sigma}}_{\ul{\bk}}|^2
\cr
&&\left\{\frac{\delta^2\cZ}{\delta\lambda(\bk_2)\delta\lambda(\bk_3)}-\sigma_2\frac{\delta^2\cZ}{\delta\lambda(\bk_1)\delta\lambda(\bk_3)}-\sigma_3
\frac{\delta^2\cZ}{\delta\lambda(\bk_1)\delta\lambda(\bk_2)}\right\}. 
\label{Zspec-eq} \end{eqnarray}
Equation (\ref{Zspec-eq}) is the main result of this section.

The validity of (\ref{Zspec-eq}) does not, of course, require $\epsilon$ infinitesimally small or 
$L$ infinitely large, but just an $\epsilon$ sufficiently small and an $L$ sufficiently large (depending upon $\epsilon$).
However, it is worthwhile to stress the precise conditions, largely following the prior discussions of 
\cite{Connaughtonetal03} and \cite{Nazarenko11}, Chapter 11. In the first place, $\epsilon$ must be so 
small that there is a large separation between the wave period and the nonlinear time 
\be \frac{1}{\omega(\bk)} \ll \epsilon^{-2}\frac{n(\bk)}{dn(\bk)/d\tau}, \lb{nonuniform} \ee
with the latter of order $\epsilon^{-2}\gamma_\bk.$ This is required in order to be able to find an intermediate time
$T$ so that (\ref{smallep}) and $(\cZ(T)-\cZ(0))/T\doteq \dot{\cZ}$ are both well-satisfied, e.g. taking $T=O(\epsilon^{-1}).$
As emphasized in \cite{Connaughtonetal03}, condition (\ref{nonuniform}) is almost never uniformly valid for all 
wavenumbers $\bk,$ but is typically violated for either low or high $k$. This means that $\lambda(\bk)$ 
in (\ref{Zspec-eq}) must be restricted to be zero for $\bk$ outside the interval where (\ref{nonuniform}) holds and 
it is furthermore assumed that the wavenumber integrations are sufficient local that no wavenumbers outside that 
range give a substantial contribution. The size of $L$ is determined by the requirement that wavenumber summations 
over $[-\pi/a,\pi/a]^d$ can be approximated as continuous integrals. At the very least, it must be true that $L\gg a.$ 
The most stringent condition seems to arise from the requirement that the approximate delta functions of width $1/T$
in (\ref{smallep}) contain a large number of frequencies. If $\Delta\omega$ is the spacing of discrete frequencies,
then one must have $\Delta\omega\ll 1/T\sim O(\epsilon).$ Estimating $\Delta\omega=|\Delta\bk|\cdot|\grad_\bk\omega(\bk)|\sim
\frac{2\pi}{L}|\grad_\bk\omega(\bk)|,$ the essential requirement is that $L/|\grad_\bk\omega(\bk)|\gg T,$ i.e. the time required
for a wavepacket traveling at the group velocity to cross the box must be much larger than the time $T$ or, conservatively, 
the nonlinear interaction time. This requires extremely large boxes in practice, unless $\epsilon$ is only 
moderately small and (\ref{nonuniform}) only marginally satisfied.

\subsubsection{Properties}\lb{property-spect}

\noindent In this section we shall discuss the most basic properties of equation (\ref{Zspec-eq}) derived in the previous section.
As pointed out in \cite{Choietal05b}, the factor $\delta_{\mu,0}$ implies that the RP property of the initial conditions is preserved 
in time. Therefore, without loss of generality, we need only consider the characteristic functional for amplitudes, or 
${\cZ}[\lambda,\tau]\equiv \cZ[\lambda,\mu=0,\tau].$ Its evolution equation is 
\begin{eqnarray}
\dot{\cZ}[\lambda,\tau] &=&-36i\pi\sum_{\ul{\sigma}=(-1,\sigma_2,\sigma_3)}\int d^dk_1d^dk_2d^dk_3 \delta^d(\ul{\sigma}\cdot\ul{\bk})
\delta(\ul{\sigma}\cdot\omega(\ul{\bk}))\lambda(\bk_1)|H^{\ul{\sigma}}_{\ul{\bk}}|^2\cr
& & \,\,\,\,\,\,\,\,\,\,\,\,\,\,\,\,\,\,\,\,\,\,\,\,\,\,\,\, \times 
\left\{\frac{\delta^2\cZ}{\delta\lambda(\bk_2)\delta\lambda(\bk_3)}-\sigma_2\frac{\delta^2\cZ}{\delta\lambda(\bk_1)\delta\lambda(\bk_3)}
-\sigma_3\frac{\delta^2\cZ}{\delta\lambda(\bk_1)\delta\lambda(\bk_2)}\right\}
\lb{Zspec-eq2} \end{eqnarray}
Hereafter we  consider only this amplitude characteristic functional. 
Eq.(\ref{Zspec-eq2}) implies a hierarchy of evolution equations for the $M$-mode spectral correlation functions defined
in section \ref{RPA}, in the wave kinetic limit:
\be {\cN}^{(M)}(\bk_1,...,\bk_M,\tau)=\lim_{\epsilon\rightarrow 0}\lim_{L\rightarrow\infty} 
{\cN}^{(M)}_{L,\epsilon}(\bk_1,...,\bk_M,\epsilon^{-2}\tau). \ee
The hierarchy is easiest to derive by using the relation (\ref{Z-empspec}) between $\cZ[\lambda,\tau]$ and the empirical
spectrum, which implies that 
\be\cN^{(M)}(\bk_1,...,\bk_M,\tau)=\left.(-i)^M\frac{\delta^M \cZ[\lambda,\tau]}{\delta\lambda(\bk_1)\cdot\cdot\cdot\delta\lambda(\bk_M)}\right|_{\lambda=0}. \ee
By taking $M$ functional derivatives of (\ref{Zspec-eq2}) and setting $\lambda\equiv 0$, one derives for each integer
$M=1,2,3,...$ the following equation:
\begin{eqnarray}
&&\dot{\cN}^{(M)}(\bk_1,...,\bk_M,\tau)=36\pi\sum_{j=1}^M\sum_{\ul{\sigma}=(-1,\sigma_2,\sigma_3)}
\int d^d\ol{k}_2 d^d\ol{k}_3 \delta^d(\ul{\sigma}\cdot\ul{\bk}_j)\delta(\ul{\sigma}\cdot\omega(\ul{\bk}_j))|H^{\ul{\sigma}}_{\ul{\bk}_j}|^2\cr
&&\Big[\cN^{(M+1)}(\bk_1,...,\bk_{j-1},\bk_{j+1},...,\bk_M,\ol{\bk}_2,\ol{\bk}_3,\tau)
-\sigma_2\cN^{(M+1)}(\bk_1,...,\bk_M,\ol{\bk}_3,\tau)-\sigma_3\cN^{(M+1)}(\bk_1,...,\bk_M,\ol{\bk}_2,\tau)\Big].\cr
&&\,\,
\lb{spect-hier} \end{eqnarray}
which couples the $M$th-order correlation functions to the $(M+1)$st. We shall refer to the above collection of equations
for all $M=1,2,3...$ as the {\it spectral hierarchy} of kinetic wave turbulence. It is exactly analogous to the ``Boltzmann hierarchy''
derived by Lanford from the BBGKY hierarchy in the low-density limit \cite{Lanford75,Lanford76}. If the spectral correlation functions
satisfy bounds on their growth for large orders $M$ that allow them to uniquely characterize the distribution of the empirical 
spectrum, then the spectral hierarchy (\ref{spect-hier}) is not only a consequence of the equation (\ref{Zspec-eq2}) but is in
fact equivalent to that equation.  

An extremely important property of the equations (\ref{Zspec-eq2}) or (\ref{spect-hier}) is that they possess
certain exact solutions. In particular, if the initial functional $\cZ[\lambda,0]$ is of exponential form (\ref{Zexp0}), 
as follows for an initial  RP field with uncorrelated amplitudes, then an exact solution of  (\ref{Zspec-eq2}) is
\be \cZ[\lambda,\tau]=\exp\Big(i\int d^d\bk\,\lambda(\bk)n(\bk,\tau)\Big), \lb{Zexp-sol} \ee
where $n(\bk,\tau)$ satisfies the standard wave kinetic equation (\ref{wavekineq}) with initial condition 
$n(\bk,0)=n(\bk).$ This may be checked by direct substitution of (\ref{Zexp-sol}) into (\ref{Zspec-eq2}). 
Equivalently, with factorized $M$th-order correlation functions (\ref{spect-fac}) as initial data, there is a  
solution of the spectral hierarchy equations (\ref{spect-hier}) also of factorized form:
\be \cN^{(M)}(\bk_1,...,\bk_M,\tau)= \prod_{m=1}^M  n(\bk_m,\tau). \lb{spect-fac-sol} \ee
Note that such factorized solutions have only power-law growth for large orders $M,$ 
so that they uniquely characterize the exponential characteristic functional (\ref{Zexp-sol}). 
If it can be proved that solutions of the dynamical equations (\ref{Zspec-eq2}) or (\ref{spect-hier}) are unique for classes of initial data 
that include the forms (\ref{Zexp0}) and (\ref{spect-fac}), then the equations we have derived imply that spectral correlation 
functions initially factorized will remain so for $\tau>0.$ In this sense, therefore, the property of uncorrelated wave amplitudes 
is preserved in time by our equations.  This is an exact analogue of the ``propagation of chaos'' property for the Boltzmann hierarchy, 
which implies that the {\it Stosszahlansatz} is propagated in time \cite{Lanford75,Lanford76}. The uniqueness  of solutions of the 
linear functional equations (\ref{Zspec-eq2}) and (\ref{spect-hier}) is an interesting mathematical problem which we shall address in future work.

The results above have an important implication. As follows from our discussion in section \ref{RPA}, the conditions 
(\ref{Zexp-sol}) or (\ref{spect-fac-sol}) imply a law of large numbers for  the empirical spectrum at positive times. That is,
with probability going to 1 in the kinetic limit (first $L\rightarrow\infty,$ then $\epsilon\rightarrow 0$), it follows that 
\be \hn_L(\bk,\epsilon^{-2}\tau) \simeq n(\bk,\tau), \,\,\,\, \tau>0 \ee
where $n(\bk,\tau)$ is the solution of the wave kinetic equation (\ref{wavekineq}).  The interesting implication for laboratory 
and numerical experiments is that the wave kinetic equations will be valid for {\it typical} initial amplitudes and phases chosen 
from an RPA ensemble and not just for the spectrum averaged over the RPA ensemble.  That is, the empirical spectrum is 
``self-averaging.'' This is the exact analogue of the law of large numbers derived by Lanford for the empirical 1-particle 
distribution (Klimontovich density) in the low density limit for gases \cite{Lanford75,Lanford76}. 

In order to emphasize the close formal analogy of wave kinetic theory with the kinetic theory of gases, it is worthwhile 
to make here a few remarks about the role of entropy in both. It is well-known that the wave kinetic equation (\ref{wavekineq})
satisfies an ``H-theorem'' for the entropy defined, with Boltzmann's constant $k_B,$ by
\be S[n]= k_B \int_{\Lambda^*} d^dk\, \ln n(\bk).  \lb{entropy} \ee
That is, $dS/d\tau\geq 0$ for general solutions of (\ref{wavekineq}) and  $dS/d\tau=0$ for the thermal equilibrium 
solutions $n_{{\rm eq}}(\bk)=k_BT/\omega(\bk)$ at absolute temperature $T,$  which maximize the entropy 
(\ref{entropy}) for fixed energy $E[n]=\int_{\Lambda^*} d^dk\, \omega(\bk) n(\bk).$ See \cite{Zakharovetal92}, section 2.2.2. 
This is the exact analogue of the $H$-theorem originally derived by Boltzmann for his kinetic equation.  As pointed out more 
recently by Spohn \cite{Spohn06}, the entropy (\ref{entropy}) also follows from Boltzmann's prescription that  $S=k_B\log W$ 
\cite{Boltzmann72}, where $W$ is the Liouville measure of the set of microstates $(\tJ,\psi)$ consistent with the given 
``macrostate'' defined by the prescribed spectrum $n(\bk).$ More precisely, let $\Delta_i,$ $i=1,..,P$ be a partition of $\Lambda^*$ 
and let $\Gamma_{n,\Delta,\eta,L}$ be the set of microstates $\{(\tJ_\bk,\psi_\bk),\,\bk\in \Lambda_L^*\}$  
such that $ |\hn_L(\Delta_i)-n(\Delta_i)|<\eta,$ for $i=1,...,P$ where
\be \hn_L(\Delta_i)=\frac{1}{|\Delta_i|} \int_{\Delta_i} d^dk \,\,\hn_L(\bk), \,\,\,\,
      n(\Delta_i)=\frac{1}{|\Delta_i|} \int_{\Delta_i} d^dk \,\,n(\bk). \ee 
Then\footnote{
Note that $\Gamma_{n,\Delta,\eta,L}$ is a Cartesian product set of the form $\otimes_{i=1}^P\{ (\tJ_\bk,\psi_\bk),\,\bk\in 
\Lambda_L^*\cap \Delta_i:\,\,|\hn_L(\Delta_i)-n(\Delta_i)|<\eta\}$
and for each cell $\Delta_i$ of the partition, $ \hn_L(\Delta_i)\approx \frac{1}{2N_L(\Delta_i)}\sum_{\bk\in \Delta_i\cap\Lambda^*_L}  
(\tilde{p}_\bk^2+\tilde{q}_\bk^2).$ The stated result then follows using the formula $\frac{\pi^{D/2}}{(D/2)!}R^D$ for the volume 
of a ball in dimension $D=2 N(\Delta_i)$ of radius $R=(nD)^{1/2},$ dividing by $(2\pi)^{D/2}$ [eq.(\ref{liouville2})], and applying 
Stirling's approximation for the factorial $(D/2)!$ as $D\rightarrow\infty.$}\lb{logn}
\be \lim_{\eta\rightarrow 0}\lim_{L\rightarrow\infty} \left(\frac{2\pi}{L}\right)^d \ln |\Gamma_{n,\Delta,\eta,L}|=
       \sum_{i=1}^P |\Delta_i|\left( \ln n(\Delta_i)+1\right). \ee      
The result for each individual cell is the 
same as the microcanonical entropy $s(e)$ of a system of non-interacting harmonic oscillators or alternatively as the entropy 
of an ideal gas in a periodic box, thinking of $\tJ_\bk$ as kinetic energies and $\varphi_\bk$ as positions of the particles. 
Furthermore, defining $|\Delta|=\max_i|\Delta_i|,$ the partition may be refined by taking $|\Delta|\rightarrow 0,$ so that 
\be \lim_{|\Delta|\rightarrow 0}\lim_{\eta\rightarrow 0}\lim_{L\rightarrow\infty} \left(\frac{2\pi}{L}\right)^d \ln |\Gamma_{n,\Delta,\eta,L}|=
       \int_{\Lambda^*} d^dk\,\,(\ln n(\bk)+ 1).  \ee   
The result agrees, up to constants, with (\ref{entropy}).  The above argument introduces the ``microcanonical measure''
obtained by restricting Liouville measure to the set  $\Gamma_{n,\Delta,\eta,L}$ and normalizing by $ |\Gamma_{n,\Delta,\eta,L}|$
to yield a probability measure. By adapting the arguments of \cite{Lanford76} it follows that this ``microcanonical measure'' is 
a natural example which satisfies asymptotic factorization (\ref{spect-fac}) but {\it not} the RPA property. 

The previous results may appear somewhat  disappointing. Equations for the generating function $\cZ[\lambda,\tau]$ like our 
(\ref{Zspec-eq2}) have been proposed mainly in the hope of developing theories of intermittency of wave turbulence
and of higher-order statistics of the wave-amplitudes. It might be concluded from the discussion above that the 
correct equation for $\cZ[\lambda,\tau]$---i.e. our equation (\ref{Zspec-eq2})---is equivalent to the wave kinetic equation 
(\ref{wavekineq}) and has no more physical content. However, this is not correct. To clarify this point, we now 
classify all of  the realizable solutions of (\ref{Zspec-eq2}). That is, we characterize the solutions of the equation (\ref{Zspec-eq2}) 
for the most general possible initial conditions that can be physically attained in the kinetic limit, assuming RP  but 
{\it no} form of RPA. For this purpose, note that the characteristic functional $\cZ_L(\lambda,0)$ defined in (\ref{Z-empspec})
is a positive-definite functional, i.e. 
\be \sum_{i,j=1}^n c_ic_j^* \cZ_L(\lambda_i-\lambda_j,0)\geq 0\ee
for any set of $n$ fields $\lambda_1,...,\lambda_n$ and complex numbers $c_1,...,c_n.$ It is also normalized 
so that $\cZ_L(\lambda=0,0)=1.$ Since these properties are preserved under pointwise limits, the physical initial 
condition 
\be \cZ(\lambda,0)=\lim_{L\rightarrow\infty} \cZ_L(\lambda,0) \ee
for our eq.(\ref{Zspec-eq2}) also satisfies them. We must assume that the above limit exists, if eq.(\ref{Zspec-eq2}) 
is to have any validity at all, and we shall assume furthermore that the limiting $\cZ(\lambda,0)$ is a continuous 
functional of the $\lambda$ fields\footnote{We shall not attempt to identify here the precise topologies on the spaces of $\lambda$'s 
and $n$'s that would permit a rigorous formulation and proof of our results.}. By the Bochner-Minlos theorem, 
it therefore has the form
\be \cZ(\lambda,0)  = \int d\rho(n_0) \exp\Big(i\int d^d\bk\,\lambda(\bk)n_0(\bk)\Big) \ee
 for some probability measure $\rho.$ E.g. see \cite{Yamasaki85}. That is, the initial condition $\cZ(\lambda,0)$
 is a statistical superposition of exponential initial conditions of the form (\ref{Zexp0}) that arise from RPA fields. 
 Since equation (\ref{Zspec-eq2}) is linear in $\cZ$, a solution for such superposed initial data is 
\be \cZ(\lambda,\tau)  = \int d\rho(n_0) \exp\Big(i\int d^d\bk\,\lambda(\bk)n(\bk,\tau)\Big) \ee 
where $n(\bk,\tau)$ solves the kinetic equation with initial condition $n_0(\bk).$ This is the only solution if the 
uniqueness of solutions holds for (\ref{Zspec-eq2}). The conclusion of this argument is that {\it the most 
general, statistically realizable solutions\footnote{It should be noted that there may be non-realizable solutions 
of eq.(\ref{Zspec-eq2}) that do not have this form. In general, statistical moment equations may have ``parasitic'' 
solutions that do not correspond to realizable solutions of the underlying statistical problem. For an example of 
this phenomenon in the Kraichnan passive scalar model, see \cite{EyinkXin00}.} of eq.(\ref{Zspec-eq2}) correspond 
to ensembles of solutions of the wave kinetic equation (\ref{wavekineq}) with random initial conditions $n_0(\bk)$.}

The above argument is a formal analogue of a rigorous result of Spohn for the Boltzmann hierarchy in the 
kinetic theory of gases \cite{Spohn84}. Under appropriate technical assumptions, his argument can be 
carried over  to our spectral hierarchy (\ref{spect-hier}), with the conclusion that its general realizable solutions 
are statistical superpositions of factorized solutions, that is, 
\be \cN^{(M)}(\bk_1,...,\bk_M,\tau) = \int d\rho(n_0) \prod_{m=1}^M n(\bk_m,\tau), \ee
where again $n(\bk,\tau)$ solves the kinetic equation with initial condition $n_0(\bk)$. We shall refer to such 
solutions of the spectral hierarchy (\ref{spect-hier}) or of the equivalent equation (\ref{Zspec-eq2}) as 
``super-statistical solutions'', since they correspond to random ensembles of solutions of the spectral closure
equation. As we shall discuss at length in section \ref{intermittent}, such ``super-statistical solutions''
offer a possibility to explain intermittency and non-Gaussian statistics previously little discussed in the 
wave turbulence literature. However, we shall first consider the alternative approach based on closed 
equations for the PDF's of the wave amplitudes. 

\subsection{PDF Hierarchy}\lb{PDF}  

We now consider a second possible limit involving only a fixed number $M$ of modes $a_{\bk_m},$ $m=1,...,M$ 
as the total number $N\rightarrow\infty.$ As before, one must keep 
$\tJ_\bk=O(1)$ for all modes. We thus define the joint characteristic function:
\begin{align}
\cZ^{(M)}_L(\lambda_1,...,\lambda_M,\mu_1,...\mu_M,T;\bk_1,...,\bk_M)=\llangle\exp\left(i\sum_{m=1}^M 
\lambda_m \tJ_{\bk_m}(T)\right)\prod_{m=1}^M\psi_{\bk_m}^{\mu_m}(T)\rrangle, \lb{ZMfun}
\end{align}
This is the generating function (\ref{ChoiZ}) of \cite{Choietal05b} with 
\be \lambda_{\bk_m} =i\left(\frac{L}{2\pi}\right)^d\lambda_m,\,\,\,\, 
J_{\bk_m}=\left(\frac{2\pi}{L}\right)^d\tJ_{\bk_m},\,\,\,\,
m=1,...,M \lb{pdf-sub} \ee
and all other $\lambda_\bk=0.$ It also corresponds to the generating functional of 
\cite{Nazarenko11}, eq. (5.15), with essentially the same definitions of $\lambda_m$ and $J_m$ as ours, 
but inserting an imaginary $i$ in the exponent  and keeping only a finite number $M$ of nonzero arguments
\footnote{Nazarenko employed a different definition in \cite{Nazarenko11} than that used earlier 
in \cite{Choietal05a,Choietal05b} because he realized the latter definition to be inadequate to derive the 1-mode 
PDF equation (private communication).}. We use the shorthand $\cZ^{(M)}_L(\lambda,\mu,T)$ when there is no 
possibility of confusion. As we shall see,the limit $L\rightarrow\infty,$ $\epsilon\rightarrow 0$ of this 
object leads to a {\it hierarchy} of equations connecting different values of $M.$  We sketch the derivation 
of these equations, with more details in \ref{appendix-pdf}, and then analyze their basic properties.  

\subsubsection{Derivation}\lb{derive-PDF} 

The method is the same as before. We use the perturbation expansion in $\epsilon$ giving the formula
(\ref{ZXX}) for the generating functions, with the definitions (\ref{Xdef}) of $\cX_L(\lambda,\mu)$ and (\ref{J1})-(\ref{J5})
of the $\cJ$'s. As a consequence of (\ref{pdf-sub}) all separate terms in the prefactors of the $\cJ$'s 
are of the same order:
\be \lambda_{\bk_1}+\frac{\mu_{\bk_1}}{2J_{\bk_1}},\,\, \lambda_{\bk_1}+\lambda_{\bk_1}^2 J_{\bk_1}
-\frac{\mu_{\bk_1}^2}{4 J_{\bk_1}}=O(L^d) \ee
\be \frac{1}{2}\lambda_{\bk_1}^2+\frac{\mu_{\bk_1}}{4J_{\bk_1}^2}(\frac{\mu_{\bk_1}}{2}-1)
+\frac{\lambda_{\bk_1}\mu_{\bk_1}}{2J_{\bk_1}},\,\,
\lambda_{\bk_1}\lambda_{\bk_2},\,\,(\lambda_{\bk_1}+\frac{\mu_{\bk_1}}{4J_{\bk_1}})\frac{\mu_{\bk_2}}{J_{\bk_2}}=O(L^{2d}). \ee
Hence, we only need to calculate the leading-order graphical contributions. We already analyzed these in the previous section
and this discussion carries over here, except that now some wavenumbers are discrete and take on only $M$ values (mode 1
for $\cJ_1-\cJ_4$ and modes 1,2 for $\cJ_5$) whereas all others are continuous in the infinite-box limit. One must consider 
carefully whether free wavenumbers in graphical summations are discrete or continuous to see whether their contribution 
is $O(M)$ or $O(L^d).$ This analysis is carried out in \ref{appendix-pdf}. It is found that $\cX_L^{(M)}(\lambda,\mu,T)$ 
in the large-box limit gets no contributions from $\cJ_1,\cJ_4$ and $O(1)$ contributions from $\cJ_2,\cJ_3,\cJ_5.$ 
The result is
\begin{eqnarray}
\cX^{(M)}(\lambda,\mu,T)-\cX^{(M)}(\lambda,\mu,0) &\sim& -9\epsilon^2 i\delta_{\mu,0}\sum_{j=1}^M\sum_{\ul{\sigma}=(-1,\sigma_2,\sigma_3)}\int d^d\ol{k}_2d^d\ol{k}_3\, \delta^d(\ul{\sigma}\cdot\ul{\bk}_j)\big|\Delta_T(\ul{\sigma}\cdot\omega(\ul{\bk}_j))\big|^2
\cr
&&\,\,\,\,\,\,\,\,\,\,\,\,\,\,\,\,\,\,\,\,\,\,\,\,\,\,\,\,\,\,\times\big|H^{\ul{\sigma}}_{\ul{\bk}_j}\big|^2(\lambda_{\bk_j}+\lambda_{\bk_j}^2\frac{\partial}{\partial\lambda_{\bk_j}}) \frac{\partial ^2 \cZ^{(M+2)}}{\partial \ol{\lambda}_2\partial\ol{\lambda}_3}\Big|_{\ol{\lambda}_2=\ol{\lambda}_3=0}
\cr
&&+\epsilon^2\left[18 i\delta_{\mu,0}\sum_{j=1}^M\sum_{\ul{\sigma}=(-1,\sigma_2,\sigma_3)}\sigma_2\int d^dk_2d^dk_3\, \delta^d(\ul{\sigma}\cdot\ul{\bk}_j)E_T(0,\ul{\sigma}\cdot\omega(\ul{\bk}_j))\right.
\cr
&&\,\,\,\,\,\,\,\,\,\,\,\,\,\,\,\,\,\,\,\,\,\,\,\,\,\,\,\,\,\,\times \big|H^{\ul{\sigma}}_{\ul{\bk}_j}\big|^2\lambda_{\bk_j}\frac{\partial^2 \cZ^{(M+1)}}{\partial\ol{\lambda}_3\partial\lambda_j}\Big|_{\ol{\lambda}_3=0}+(2\leftrightarrow 3)
\cr
&&-9\sum_{j=1}^M\delta_{\mu_j,1}\delta_{\mu_{-j,1}}\prod_{m\neq j,-j}\delta_{\mu_m,0}\sum_{\ul{\sigma}=(-1,\sigma_2,\sigma_3)}
\sigma_2\int d^d\ol{k}_2\int d^d\ol{k}_3 \delta^d(\ul{\sigma}\cdot\ul{\bk}_j)   \cr
&& \left.\times E_T(\omega(\bk_j)+\omega(-\bk_j),\ul{\sigma}\cdot\omega(\ul{\bk}_j)) H^{-,\sigma_2,\sigma_3}_{\bk_j,\ol{\bk}_2,\ol{\bk}_3} H^{+,\sigma_2,\sigma_3}_{-\bk_j,\ol{\bk}_2,\ol{\bk}_3}\llangle 
e^{\sum_m i\lambda_m \tJ_m}\sqrt{\frac{\tJ_{-1}}{\tJ_1}}\tJ_3\rrangle_J +(2\leftrightarrow 3)\right]  \cr
&& +18\epsilon^2\sum_{j=1}^M\delta_{\mu_j,1}\delta_{\mu_{-j},1}\prod_{m\neq 1,-1}\delta_{\mu_m,0}
\left( \lambda_1\lambda_{-1}-i\frac{\lambda_1}{\tJ_{-1}}-\frac{1}{4\tJ_1\tJ_{-1}}\right)
 \sum_{\ul{\sigma}=(-,\sigma_2,\sigma_3)}\int d^d\ol{k}_2d^d\ol{k}_3\delta^d(\ul{\sigma}\cdot\ul{\bk}_j)
\cr
&& \times \Delta_T(\ul{\sigma}\cdot\omega(\ul{\bk}_j))\Delta_T(-\ul{\sigma}\cdot\omega(\ul{\bk}_j'))H^{-,\sigma_2,\sigma_3}_{\bk_j,\ol{\bk}_2,\ol{\bk}_3}
H^{+,\sigma_2,\sigma_3}_{-\bk_j,\ol{\bk}_2,\ol{\bk}_3}\llangle\sqrt{\tJ_1\tJ_{-1}}\tJ_2\tJ_3e^{i\sum_m\lambda_m \tJ_m}\rrangle_J.
\end{eqnarray}
Here $\ul{\bk}_j=(\bk_j,\ol{\bk}_2,\ol{\bk}_3)$ and $\ul{\bk}_j'=(-\bk_j,\ol{\bk}_2,\ol{\bk}_3).$
Taking the small-$\epsilon$ limit using the asymptotic formulas (\ref{smallep}), one finds 
the terms proportional to $\delta_{\mu,0}$ contain secular contributions while the terms proportional to $\delta_{\mu_j,1}\delta_{\mu_{-j},1}$
are nonsecular. Using (\ref{ZXX2}) to calculate $\cZ^{(M)}(T)-\cZ^{(M)}(0),$ replacing $(\cZ^{(M)}(T)-\cZ^{(M)}(0))/T$ by $\dot{\cZ}^{(M)}$ 
and using time variable $\tau=\epsilon^2 t$, one finally obtains
\begin{align}
\begin{split}
\frac{d}{d\tau}\cZ^{(M)}(\lambda,\mu,\tau)&=-36i\pi\delta_{\mu,0}\sum_{j=1}^M \sum_{\ul{\sigma}=(-1,\ol{\sigma}_2,\ol{\sigma}_3)}\int d^d\ol{k}_2 d^d\ol{k}_3\,\,\delta^d(\ul{\sigma}\cdot\ul{\bk}_j)\delta(\ul{\sigma}\cdot\omega(\ul{\bk}_j))\Big|H^{\ul{\sigma}}_{\ul{\bk}_j}\Big|^2
\\
&\Big\{(\lambda_j+\lambda_j^2\frac{\partial}{\partial\lambda_j})\frac{\partial^2\cZ^{(M+2)}}{\partial\ol{\lambda}_2\partial\ol{\lambda}_3}\Big|_{\ol{\lambda}_2=\ol{\lambda}_3=0}-\ol{\sigma}_2\lambda_j\frac{\partial\cZ^{(M+1)}}{\partial\ol{\lambda}_3\partial\lambda_j}\Big|_{\ol{\lambda}_3=0}-\ol{\sigma}_3\lambda_j\frac{\partial\cZ^{(M+1)}}{\partial\ol{\lambda}_2\partial\lambda_j}\Big|_{\ol{\lambda}_2=0}\Big\}.
\end{split}
\lb{Zpdf-hier} \end{align}
Equation (\ref{Zpdf-hier}) is the main result of this section. Its conditions of validity are similar to those of the spectral hierarchy 
discussed in section \ref{derive-spect}.  

\subsubsection{Properties}\lb{property-PDF}

\noindent We now consider the important properties of the multi-mode equations (\ref{Zpdf-hier}). As for the spectral 
characteristic functional in the previous section, the factors $\delta_{\mu,0}$ imply that the RP property is preserved in time. 
Therefore, we can consider the generating functions for the amplitudes alone, obtained by setting $\mu=0.$ These form 
a {\it hierarchy} of equations, for $M=1,2,3,....$
\begin{align}
\begin{split}
\frac{d}{d\tau}\cZ^{(M)}(\lambda,\tau)&=-36i\pi\sum_{j=1}^M \sum_{\ul{\sigma}=(-1,\ol{\sigma}_2,\ol{\sigma}_3)}\int d^d\ol{k}_2 d^d\ol{k}_3\,\,\delta^d(\ul{\sigma}\cdot\ul{\bk}_j)\delta(\ul{\sigma}\cdot\omega(\ul{\bk}_j))\Big|H^{\ul{\sigma}}_{\ul{\bk}_j}\Big|^2
\\
&\Big\{(\lambda_j+\lambda_j^2\frac{\partial}{\partial\lambda_j})\frac{\partial^2\cZ^{(M+2)}}{\partial\ol{\lambda}_2\partial\ol{\lambda}_3}\Big|_{\ol{\lambda}_2=\ol{\lambda}_3=0}-\ol{\sigma}_2\lambda_j\frac{\partial\cZ^{(M+1)}}{\partial\ol{\lambda}_3\partial\lambda_j}\Big|_{\ol{\lambda}_3=0}-\ol{\sigma}_3\lambda_j\frac{\partial\cZ^{(M+1)}}{\partial\ol{\lambda}_2\partial\lambda_j}\Big|_{\ol{\lambda}_2=0}\Big\}.
\end{split}
\lb{Zpdf-hier2} \end{align}
By straightforward Fourier transformation in the $\lambda$ variables,  one can obtain an equivalent\footnote{Although
naively equivalent, the formulation in terms of amplitude $s$ variables may be preferable to the $\lambda$-variable formulation if 
the ideas of \cite{Choietal05a,Nazarenkoetal10} are correct that the equations become invalid at sufficiently large amplitudes.
In that case, only the $s$-formulation would permit the incorporation of suitable boundary conditions. See section \ref{amplitude}.}
hierarchy of equations 
for the joint PDF's $\cP^{(M)}(s_1,...,s_M;\bk_1,...,\bk_M)$. For each $M=1,2,3,...$ these are equations for conservation of probability 
\be
\dot{\cP}^{(M)}+\sum_{m=1}^M\frac{\partial}{\partial s_m}\cF_m^{(M)}=0,
\lb{pdf-hier1} \ee
with the probability flux 
\begin{eqnarray} 
\cF_m^{(M)}&=& -36\pi s_m\sum_{\ul{\sigma}=(-1,\sigma_2,\sigma_3)}\int d^d\ol{k}_2 d^d\ol{k}_3 \,\,\delta^d(\ul{\sigma}\cdot\ul{\bk}_m)
\delta(\ul{\sigma}\cdot\omega(\ul{\bk}_m))|H^{\ul{\sigma}}_{\ul{\bk}_m}|^2\cr
&&\Big[\int d\ol{s}_2 d\ol{s}_3\frac{\partial\cP^{(M+2)}}{\partial s_m}(s_1,...,s_M,\ol{s}_2,\ol{s}_3)\ol{s}_2\ol{s}_3 \cr
&&+\sigma_2\int d\ol{s}_3\cP^{(M+1)}(s_1,...,s_M,\ol{s}_3)\ol{s}_3+\sigma_3\int d\ol{s}_2\cP^{(M+1)}(s_1,...,s_M,\ol{s}_2)\ol{s}_2\Big].
\lb{pdf-hier2} 
\end{eqnarray}
We obtain no closed equations for $P^{(M)}$ with any choice of $M$ but instead an infinite hierarchy, in which the evolution equation
for $\cP^{(M)}$ contains $\cP^{(M+1)}$ and $\cP^{(M+2)},$ for $M=1,2,3,....$ Our results are therefore not in agreement with previously 
proposed closed equations for $\cP^{(2)}$ with $M=2$ \cite{Choietal09}\footnote{Their 2-mode equations (26)-(33) 
reduce to the asymptotically exact 1-mode equation if one neglects some additional terms that vanish in the limit $L\rightarrow\infty$
and if one assumes that $\cP^{(2)}=\cP^{(1)}\cP^{(1)}.$ In this limit in which the 2-mode equation of \cite{Choietal09} becomes exact, 
therefore, it contains no more information than the 1-mode equation.}. 

We do, however, recover the equations for $\cP^{(1)}$ and $\cZ^{(1)} $ with $M=1$ which were previously obtained
\cite{Choietal05a,Choietal05b,JakobsenNewell04}, under appropriate conditions. Namely, assume that the initial data 
for the hierarchy (\ref{Zpdf-hier2}) are factorized
\be \cZ^{(M)}(\lambda_1,...,\lambda_M,0;\bk_1,...,\bk_M)=\prod_{m=1}^M Z(\lambda_m,0;\bk_m), \ee
as would follow from RPA initial conditions, for example. It is then easy to show by substitution into (\ref{Zpdf-hier2}) that 
there are solutions which remain factorized
\be \cZ^{(M)}(\lambda_1,...,\lambda_M,\tau;\bk_1,...,\bk_M)=\prod_{m=1}^M Z(\lambda_m,\tau;\bk_m), 
\,\,\,\, \tau>0,\ee
where the factors $Z(\lambda,\tau;\bk)$ satisfy the closed equations
\be \frac{\partial}{\partial\tau} Z(\lambda;\bk) =
i\eta_\bk \lambda\left(1+\lambda\frac{\partial}{\partial\lambda}\right)Z(\lambda;\bk)
             -\gamma_\bk\lambda\frac{\partial Z}{\partial\lambda}(\lambda;\bk) \ee
with           
\be \eta_\bk = 36\pi\sum_{\ul{\sigma}=(-1,\sigma_2,\sigma_3)}\int d^dk_2 d^dk_3 \,\,
\delta^d(\ul{\sigma}\cdot\ul{\bk})\delta(\ul{\sigma}\cdot\omega(\ul{\bk}))|H^{\ul{\sigma}}_{\ul{\bk}}|^2 n(\bk_2)n(\bk_3)\geq 0,
\lb{eta} \ee
\be \gamma_\bk = 36\pi\sum_{\ul{\sigma}=(-1,\sigma_2,\sigma_3)}\int d^dk_2 d^dk_3 \,\,
\delta^d(\ul{\sigma}\cdot\ul{\bk})\delta(\ul{\sigma}\cdot\omega(\ul{\bk}))|H^{\ul{\sigma}}_{\ul{\bk}}|^2 
[\sigma_3n(\bk_2)+\sigma_2n(\bk_3)]. \lb{gamma} \ee
These results are equivalent to the existence of solutions of the PDF hierarchy (\ref{pdf-hier1}),(\ref{pdf-hier2})
that remain factorized
\be \cP^{(M)}(s_1,...,s_M,\tau;\bk_1,...,\bk_M)=\prod_{m=1}^M P(s_m,\tau;\bk_m),\,\,\,\,\tau>0 \ee
where the 1-mode pdfs $P(s,\tau;\bk)$ satisfy
\be  \frac{\partial}{\partial\tau}P =\frac{\partial}{\partial s}\left[ s\left(\eta_\bk\frac{\partial P}{\partial s}
      +\gamma_\bk P\right)\right]. \lb{1pdf-eq} \ee
Preservation of RPA therefore follows if one can prove uniqueness of solutions of the hierarchy (\ref{Zpdf-hier2}) 
or, equivalently, (\ref{pdf-hier1}),(\ref{pdf-hier2}). The equations for the factors agree with previous results 
for the 1-mode equations \cite{Choietal05a,Choietal05b,JakobsenNewell04}. 

The equations (\ref{1pdf-eq}) are not simple linear Fokker-Planck equations, however, but are instead {\it nonlinear} 
Markov  evolution equations in the sense of McKean \cite{McKean66}. That is, the solutions must satisfy a set of 
self-consistency conditions,
\be n(\bk,\tau) = \int ds\,\, s P(s,\tau;\bk) \lb{self} \ee
where $n(\bk,\tau)$ is the same spectrum that appears in the formulas for the coefficients (\ref{eta}),(\ref{gamma}). 
These equations are the exact solutions of a model of ``self-consistent Langevin equations'', like those for the
DIA turbulence closure \cite{Kraichnan70,Leith71}. Here the model equations take the form  of the stochastic differential
equations 
\be  ds_\bk = (\eta_\bk-\gamma_\bk s_\bk)d\tau + \sqrt{2\eta_\bk s_\bk} \,\,dW_\bk, \lb{langevin} \ee
interpreted in the Ito sense and with self-consistent determination of $n(\bk,\tau)$ via (\ref{self}). 
Solutions of this stochastic model can be realized by a Monte Carlo procedure of McKean \cite{McKean66}, 
with (\ref{langevin}) generalized to 
\be ds_\bk^{(n)} = (\eta_\bk-\gamma_\bk s_\bk^{(n)})dt + \sqrt{2\eta_\bk s^{(n)}_\bk} \,\,dW_\bk^{(n)}, 
\,\,\,\,\bk\in \Lambda_L^*,n=1,...,N,\ee
with the spectrum obtained by an $N$-sample average
\be n_\bk = \frac{1}{N}\sum_{n=1}^N s_\bk^{(n)}, \ee
and with $\gamma_\bk,\eta_\bk$ given by formulas (\ref{eta}),(\ref{gamma}) in which wavenumber integrals are
discretized as sums. Taking first $N>>1$ and then $L>>a$ yields a solution of the PDF equations (\ref{1pdf-eq}). 
This procedure works as well to solve the general hierarchy equations (\ref{pdf-hier1}),(\ref{pdf-hier2}), 
without assuming factorized initial data. For the factorized case, a far simpler procedure is to solve first 
the wave kinetic equation (\ref{wavekineq}) for $n(\bk,\tau)$ and then, using this as input, to solve 
the 1-mode equation (\ref{1pdf-eq}) in order to obtain  $P(s,\tau;\bk)$ for any wavenumber mode 
$\bk$ of interest.  As initial condition one may  take any $P(s,0;\bk)$ which satisfies the consistency 
condition (\ref{self}) at $\tau=0$, since this condition is preserved in time by the 1-mode equation.  

It is interesting that, independent of the initial condition $P(s,0;\bk),$ the solution 
$P(s,\tau;\bk)$ relaxes as $\tau$ increases to a Rayleigh distribution 
\be Q(s,\tau;\bk) = \frac{1}{n(\bk,\tau)} \exp(-s/n(\bk,\tau)), \lb{rayleigh}  \ee
which corresponds to a Gaussian distribution of the Fourier coefficient $\widetilde{a}_\bk(\tau).$ 
It is easy to check that, for any solution $n(\bk,\tau)$ of the wave kinetic equation (\ref{wavekineq}), 
$Q(s,\tau;\bk)$ defined above solves the 1-mode PDF equation (\ref{1pdf-eq}). (Since wavenumber 
$\bk$ appears only as a parameter in our argument, we suppress its appearance for  the rest of this paragraph.)  
The relaxation of a general solution $P$ to $Q$ is indicated by an $H$-theorem for the {\it relative entropy}
\footnote{A related set of observations were made in \cite{JakobsenNewell04}, section 9.}
\be H(P|Q)=\int ds\,\, P(s)\ln\left(\frac{P(s)}{Q(s)}\right)
 = \int ds\,\, P(s)\ln P(s) +\ln n+1. \ee
 This is a convex function of $P,$ non-negative, and vanishing only for $P=Q$ 
 \cite{CoverThomas91}. Taking the time-derivative using (\ref{1pdf-eq}), it is straightforward to derive
\be \frac{d}{d\tau} H(P(\tau)|Q(\tau))= -\eta \int ds\,\,\frac{s|\partial_s P(s,\tau)|^2}{P(s,\tau)}
   +\frac{\eta}{n(\tau)},\ee
where 
\be \int -s\partial_s P(s,\tau)\,ds=\int P(s,\tau)\,ds=1 \lb{pdfid} \ee 
was used to cancel terms involving the $\gamma$ coefficient. But note the self-consistency condition 
$n(\tau)=\int s\,P(s,\tau)\,ds$ implies 
\be \frac{d}{d\tau} H(P(\tau)|Q(\tau))= -\eta \left(
\int ds\,\,\frac{s|\partial_s P(s,\tau)|^2}{P(s,\tau)}
   -\frac{1}{\int s\,P(s,\tau)\,ds}\right)\leq 0. \ee
The inequality follows from the Cauchy-Schwartz inequality applied to (\ref{pdfid})
\be 1=\int \sqrt{sP}\cdot \sqrt{\frac{s}{P}}(-\partial_s P)\,ds 
      \leq \sqrt{\int sP(s)\,ds\cdot\int \frac{s|\partial_s P|^2}{P}\,ds}. \ee   
Equality holds and relative entropy production vanishes if and only if $\sqrt{sP}=c \sqrt{\frac{s}{P}}(-\partial_s P)$
for some constant $c$, or $P=-c\partial_sP.$ The solution of this latter equation
gives $P=Q$ with $n=1/c.$ We thus see that $P(\tau)$ should relax to $Q(\tau)$
as $\tau$ increases. For the purpose of later discussion, we emphasize that this argument assumes 
that the kinetic theory is valid over the entire range of amplitudes $s\in (0,\infty)$ and it could 
otherwise fail. 

Let us remark that the relative entropy has a simple probabilistic meaning, similar to that discovered 
by Boltzmann for his entropy function, which involves the empirical PDF $\hP_L(s;\Delta)$ defined 
in (\ref{emp-PDF-2} ). For a single cell $\Delta$ in wavenumber space $\Lambda^*,$
let $\mu_{\Delta,n,\eta}$ be the microcanonical measure on the set of microstates $\Gamma_{\Delta,n,\eta}=
\{ (\tJ,\psi): \,\, |\hn_L(\Delta)-n(\Delta)|<\eta\}.$ Then, 
with respect to this microcanonical measure, the most probable value of $\hP_L(s;\Delta)$ as first 
$L\rightarrow 0$, then $\eta\rightarrow 0$ is  $Q(s;\Delta)=e^{-s/n(\Delta)}/n(\Delta)$. However, the 
probability of  observing another PDF $P$ as a rare fluctuation is
\be \lim_{\eta\rightarrow 0}\lim_{L\rightarrow\infty} \frac{1}{N_L(\Delta)}
\ln\mu_{\Delta,n,\eta}(\{\hP_L\approx P\}) = -H(P|Q). \lb{sanov} \ee 
This can be shown heuristically by adapting the original argument of Boltzmann \cite{Boltzmann72}. 
Let $P(s)$ be a probability density function over possible values of $s,$ let $\Sigma=\{\Sigma_j,\,\,\,j=1,...,R\}$ 
be a finite partition of the positive reals into intervals with $s_j$ the midpoint values and define the integers
$ N_j =\left\lfloor N \int_{\Sigma_j} P(s)\,ds \right\rfloor $ 
where $\lfloor\cdot\rfloor$ denotes integer part and $N=N_L(\Delta).$ (Since $\Delta$ is fixed in this argument,
we hereafter omit explicit reference to that quantity in the remainder of this paragraph.) Note that $\sum_{j=1}^R N_j=N$ for large 
enough $L.$ Now let $\hat{N}_j$ denote the number of modes $\bk\in \Lambda_L^*\cap\Delta$ such that 
$\tJ_\bk\in \Sigma_j$ and define the set $\Gamma_{P,\Sigma,L}=\{ (\tJ,\psi):\, \hat{N}_j=N_j,\,\,j=1,...,R\}.$
The Liouville measure of this set is
$ |\Gamma_{P,\Sigma,L}| = \frac{N!}{N_1!\cdots N_R!} |\Sigma_1|^{N_1}\cdots |\Sigma_R|^{N_R} 
\sim \exp\left(-N \sum_j |\Sigma_j| P_j\ln P_j\right)$ with $P_j=N_j/N|\Sigma_j|$ and using Stirling's
approximation for the factorials, exactly as in the original argument of Boltzmann. Then
\be \lim_{|\Sigma|\rightarrow 0} \lim_{L\rightarrow\infty }\frac{1}{N}\ln |\Gamma_{P,\Sigma,L}| =-\int ds\,\,P(s)\ln P(s).\ee
Now consider the set $\Gamma_{n,\eta,L}=\cup_{\{P: \left|n-\sum_j |\Sigma_j|s_j P_j\right|<\eta\}} \Gamma_{P,\Sigma,L}.$
Its Liouville measure $|\Gamma_{n,\eta,L}|$ is dominated in the limits $L\rightarrow \infty,$ $\eta\rightarrow 0$,  
$|\Sigma|\rightarrow 0$ by $|\Gamma_{P,\Sigma,L}|$ for the pdf $P$ satisfying 
$\int ds\,\, sP(s)=n$ with the largest entropy $-\int ds\,\,P(s)\ln P(s).$
As is well-known,  this maximum entropy distribution is the Rayleigh pdf $Q(s)=e^{-s/n}/n.$ Thus,
\be \lim_{|\Sigma|\rightarrow 0}\lim_{\eta\rightarrow 0} \lim_{L\rightarrow\infty }\frac{1}{N}\ln |\Gamma_{n,\eta,L}| 
=-\int ds\,\,Q(s)\ln Q(s)=\ln n+1.\ee
Taking $\mu_{n,\eta}(\{\hP_L\approx P\})=\frac{|\Gamma_{P,\Sigma,L}|}{|\Gamma_{n,\eta,L}|},$ then
$ \lim_{|\Sigma|\rightarrow 0}\lim_{\eta\rightarrow 0} \lim_{L\rightarrow\infty }\frac{1}{N}\ln
\frac{|\Gamma_{P,\Sigma,L}|}{|\Gamma_{n,\eta,L}|}=-\left(\int ds\, P(s)\ln P(s) + \ln n +1\right),$ as claimed. The result 
(\ref{sanov}) becomes the standard Sanov theorem \cite{CoverThomas91}, if the measure $\mu_{\Delta,n,\eta}$
is replaced by the RPA measure for which each mode $\tJ_\bk,\,\,\bk\in \Lambda_L^*\cap\Delta$ has the 
independent density  $Q(s;\Delta).$ 

A simple picture emerges of entropy production in kinetic wave turbulence. For the macroscopic 
state specified by the 1-mode PDF's $P(s;\bk)$ of all wavenumbers $\bk$, the Boltzmann entropy obtained by counting 
microstates is   
\be S(P) = -\int  ds \int d^dk \,\, P(s;\bk) \ln P(s;\bk), \ee
which satisfies the 2nd-law $dS/d\tau\geq 0$ under the evolution equations (\ref{1pdf-eq}). Associated to the 
decomposition 
\be S(P) = - H(P|Q) + S(Q), \ee
there are two distinct sources of entropy production. The increase in the usual entropy of wave turbulence 
$S(Q)=\int d^dk \,\, (\ln n(\bk)+1)$ corresponds to the approach of the wave spectrum  $n(\bk)$ to its thermal 
equilibrium form $n_{eq}(\bk)=T/\omega(\bk).$ The increase in $-H(P|Q)$ corresponds to the approach of the
PDF's $P(s;\bk)$ to the Rayleigh distributions $Q(s;\bk)$ with the same spectrum $n(\bk)=\int ds\,\, sP(s;\bk).$ 
It is interesting that this approach to Gaussian statistics is quite slow according to the equation (\ref{1pdf-eq}),
taking place on the nonlinear time-scale $O(\epsilon^{-2}).$ This is in contrast to the physical space statistics
which---even if non-Gaussian initially---are expected to approach Gaussian on the fast wave time-scale.
See footnote \#\ref{fast}. The physical-space statistics are furthermore expected to remain close to Gaussian 
under the influence of the weakly nonlinear dynamics.  There is no contradiction with our $H$-theorem, 
since preservation of RPA in the weak senses (\ref{spect-fac}), (\ref{fact-pdf2}) and the results in section 
\ref{RPA} imply that the physical space statistics of the random field $u(\bx,t)$ indeed remain Gaussian 
to leading orders in $\epsilon$ and $1/L$ \cite{Eyink12}.

The above argument motivates a microcanonical measure $\mu_{P,\Delta,\Sigma,\eta,L}$ defined
for a given $P(s;\bk)$ by normalizing Liouville measure on the set 
\be \Gamma_{P,\Delta,\Sigma,\eta,L}=\left\{ (\tJ,\psi):\,\, \left|\int_{\Sigma_j}ds\, \hP_L(s,\Delta_i)
        -\int_{\Sigma_j} ds \, P(s,\Delta_i)\right|<\eta, \,\, i=1,...,P,j=1,...,R\right\} \ee 
for partitions $\Delta$ of $\Lambda^*$ and $\Sigma$ of $\mR^+.$ This measure satisfies the RP
property, as it contains no dependence on the phases. One can define for this 
measure the $M$-mode correlation functions of the empirical PDF,  $\hP_L(\varphi) = 
\left(\frac{2\pi}{L}\right)^d \sum_{\bk\in\Lambda_L^*}\varphi(\tJ_\bk,\bk)$:
\be \cP^{(M)}_{P,\Delta,\Sigma,\eta,L}(\varphi_1,...,\varphi_M) = 
\mu_{P,\Delta,\Sigma,\eta,L}\left( \hP_L(\varphi_1) \cdots \hP_L(\varphi_M)\right) 
\ee
for any choice of continuous functions $\varphi_1,...,\varphi_M$ on $\mR^+\times\Lambda^*$ vanishing at infinity.
The arguments of \cite{Lanford76} can be adapted to show that 
\be  \lim_{|\Delta|,|\Sigma|\rightarrow 0}\lim_{\eta\rightarrow 0}\lim_{L\rightarrow\infty} 
\cP^{(M)}_{P,\Delta,\Sigma,\eta,L}(\varphi_1,...,\varphi_M) =
P(\varphi_1) \cdots P(\varphi_M) \ee
where
$P(\varphi)=\int_0^\infty  ds\int_{\Lambda^*} d^dk\,\,\varphi(s,\bk) P(s;\bk).$ Thus the factorization property 
of the $M$-mode PDFs holds asymptotically for the microcanonical measure $\mu_{P,\Delta,\Sigma,\eta,L},$
although it does not satisfy the strict RPA property. Our derivation of the PDF hierarchy equations  
(\ref{pdf-hier1}),(\ref{pdf-hier2}) assumed existence of the $M$-mode PDF's in a somewhat stronger 
sense in the limit as $L\rightarrow\infty$ (i.e. for fixed values of $\bk_1,..,\bk_M$),  but it is reasonable to expect 
that they remain valid for such ``generalized RPA'' initial conditions. 

Lastly, we can ask what are the solutions of the PDF hierarchy for general initial data which are RP
but which are not factorized even asymptotically. One must assume at least that \newpage
\begin{eqnarray} 
\lim_{L\rightarrow\infty} \langle \hP_L(\varphi_1,0) \cdots \hP_L(\varphi_M,0)\rangle 
      & = &  \int_0^\infty ds_1\cdots \int_0^\infty ds_M \int d^dk_1\cdots d^dk_M \cr
      & & \,\,\,\,\,\,\,\,\,\,\,\,\,\,\,\,\,\,\,\,\,\,\,\,\,\,\,\,\,\,\,\,\,\,\,\,\,\,\,\,\times\,
      \varphi_1(s_1,\bk_1)\cdots \varphi_M(s_M,\bk_M) \cP_0^{(M)}(s_1,..,s_M;\bk_1,..,\bk_M), 
\end{eqnarray} 
in order to provide suitable initial data $\cP_0^{(M)}(s_1,..,s_M;\bk_1,..,\bk_M)$ for the hierarchy.       
However, with appropriate technical assumptions, it follows by the argument of Spohn \cite{Spohn84} that
\be \cP_0^{(M)}(s_1,..,s_M;\bk_1,..,\bk_M) =\int d\rho_0(P) \,\, P(s_1; \bk_1)\cdots P(s_M;\bk_M), \ee
where $\rho_0$ is a probability measure on the PDF's. Since the PDF hierarchy equations  
(\ref{pdf-hier1}),(\ref{pdf-hier2}) are linear in the $\cP^{(M)}$'s,  a solution will be provided by 
\be \cP^{(M)}(s_1,..,s_M,\tau;\bk_1,..,\bk_M) =\int d\rho_0(P) \,\, P(s_1,\tau; \bk_1)\cdots P(s_M,\tau;\bk_M). \ee
This will be the only solution if existence and uniqueness of solutions holds for the PDF hierarchy. 
Thus, the most general realizable  solutions of the PDF hierarchy equations are again expected to be 
``super-statistical solutions'' that correspond to ensembles of solutions of the 1-mode PDF equations 
(\ref{1pdf-eq}) with random initial conditions. 

\section{Intermittency in Kinetic Wave Turbulence}\lb{intermittent}

One of the important potential applications of multimode equations in wave kinetics is the explanation of 
observed intermittency and anomalous scaling in wave turbulence \cite{Falconetal07,Nazarenkoetal10,Yokoyama04}.
On the face of it,  wave kinetic theory appears to have few resources to explain such phenomena and instead 
has all the attributes of a ``mean-field theory'' \cite{Goldenfeld92}. As we have seen in previous sections, 
it is a theory which ignores fluctuations and in which important quantities in fact become deterministic. 
All interactions of the infinite collection of modes is through a mean-field $n(\bk,\tau).$ Furthermore, scaling exponents 
in the Kolmogorov solutions of the wave kinetic equations are given by dimensional analysis \cite{Connaughtonetal03}.
These are the hallmarks of a mean-field theory, which generally cannot predict anomalous scaling.

Nevertheless, there are at least two approaches based on wave kinetics which seem to have some promise to 
explain intermittency and non-Gaussian statistics. One is the idea of a ``cascade in amplitude space'' that 
was proposed by Choi et al. \cite{Choietal05a,Nazarenkoetal10}, based on the 1-mode PDF equation (\ref{1pdf-eq}).
Another is the idea of a ``super-turbulence'' in wave-kinetics, advanced in the present work. We shall now 
discuss both of these possibilities.

\subsection{Cascade in Amplitude Space?}\lb{amplitude}

As already mentioned at the end of section \ref{derive-spect},
wave kinetic equations generally cannot be uniformly valid over the whole range
of wavenumbers, but must break down in either low or high wavenumbers where nonlinearity become 
strong \cite{Connaughtonetal03}. Choi et al.  \cite{Choietal05a} (also \cite{Nazarenkoetal10}) have 
proposed that there is a similar non-uniformity in amplitude space, with equation (\ref{1pdf-eq}) for 
the 1-mode PDF  $P(s,\tau;\bk)$ restricted in validity to $s<s_{nl},$ where $s_{nl}$ is the amplitude for 
which nonlinear interactions become strong at wavenumber $\bk.$ [For the remainder of this section we 
shall assume that the wavenumber $\bk$ is fixed and drop it as an explicit label.] For amplitudes $s>s_{nl}$
strong-interaction processes such as ``wave-breaking'', ``cresting'', etc. are assumed to occur which are 
beyond the description of wave kinetics. The hope is that the effects of such nonlinear processes can 
be modeled by supplementing the equation (\ref{1pdf-eq}) with suitable boundary conditions and 
forcing terms. We find this a very intriguing suggestion but, as we now argue, it seems to us to have as
yet no successful formulation, either analytically or physically. 

Let us review the specific proposals of \cite{Choietal05a,Nazarenkoetal10}. They note that eq.(\ref{1pdf-eq})
is a probability conservation law
\be  \partial_t P + \partial_s F=0 \lb{1pdf-eq2} \ee
with probability flux in amplitude space given by 
\be F(s)=-s(\gamma P+\eta\partial_sP). \lb{1flux} \ee
The general time-independent solution of (\ref{1pdf-eq2}) with constant flux $F(s)=F_*$
was observed in \cite{Choietal05a} to be 
\be P(s) = C e^{-s/\nu} - \frac{F_*}{\eta} {\rm Ei}(s/\nu) e^{-s/\nu} \ee
with $\nu\equiv \eta/\gamma$ and with ${\rm Ei}(x)={\rm P}\int_{-\infty}^x e^t \frac{dt}{t}$ the standard exponential integral. 
The first term has zero flux $F(s)\equiv 0$ while the second term has non-vanishing flux $F(s)\equiv F_*.$ 
Since ${\rm Ei}(x)>0$ for $x>x_*\doteq 0.3725,$ positivity of $P(s)$ for $s\gg \nu$ requires $F_*\leq 0.$  
When $F_*=0$ then $P(s)=Q(s),$ the Rayleigh distribution with spectral density $n=\nu.$ Not noted in 
\cite{Choietal05a} is the fact that one must more generally make a distinction between $\nu$ and $n$ 
given by the self-consistency condition
\be n= \int ds\,\, s P(s), \ee
which are only the same when $P=Q.$ The above integral diverges for $F_*\neq 0, $ if the upper limit extends 
to infinity, because $P(s)\sim \frac{|F_*|}{\gamma s}$ due to the asymptotics 
${\rm Ei}(x)\sim \frac{e^x}{x}\left(1 + \frac{1!}{x} + \frac{2!}{x^2}+\cdots \right)$ for $x\gg 1.$ The integral is 
finite if it is cut off at the upper limit $s_{nl}.$ This assumes that whatever nonlinear processes such as 
``wave-breaking'' occur at $s>s_{nl}$ lead to a PDF which decays very rapidly for $s\gtrsim s_{nl}.$

The papers \cite{Choietal05a,Nazarenkoetal10} propose to model the nonlinear effects at $s>s_{nl}$
by a negative flux $F(s_{nl})=F_*<0,$ which represents a flow of samples back into the kinetic range $s<s_{nl},$
at least in some range of wavenumbers. For example, this might occur at low-wavenumbers as 
wavecaps and cusps in the nonlinear range break up and feed back into the weak, incoherent background. 
However, the equation (\ref{1pdf-eq2}) will then no longer conserve probability but instead satisfies
$ \frac{d}{dt} \int_0^{s_{nl}} ds P(s) = |F_*|>0. $ 
Paper \cite{Choietal05a} proposes to add to the equation a constant ``drag'' $D$ 
\be  \partial_t P + \partial_s F= -D,\,\,\,\, s<s_{nl}  \lb{1pdf-mod} \ee
to represent the dilution in the weight of samples at $s<s_{nl}$ by the flux $F_*$ of new 
samples into the ensemble at $s=s_{nl}.$  It is easy to see that one must choose $D=|F_*|/s_{nl}$ 
to conserve total probability\footnote{In fact, \cite{Choietal05a} proposed to take $D=-\gamma P(s_{nl})/s_{nl},$ 
which does not conserve probability and which is not even dimensionally correct. 
This equation is a typographical error not caught by the authors of \cite{Choietal05a} 
(S. Nazarenko, private communication).}. Paper \cite{Choietal05a} then claims that the general solution to the 
modified equation is $P(s)=Ce^{-s/n} +\frac{|F_*|}{\eta} {\rm Ei}(s/n-\log(s/n)) e^{-s/n}.$ The
picture proposed in \cite{Choietal05a,Nazarenkoetal10} is one of an ``inverse probability cascade'' with 
flux $F_*<0$ in amplitude space, coexisting with the usual forward energy cascade in wavenumber space. 
The power-law tails $P(s)\sim \frac{|F_*|}{\gamma s}$ for $s\gg \nu$ are a source of non-Gaussianity 
and intermittency in wave turbulence if $F_*\neq 0.$ A ``critical balance'' argument is used in 
\cite{Nazarenkoetal10} to estimate that $|F_*|\sim \gamma n/s_{nl}$. 

It is easy to see, however, that the proposed cascade picture cannot be correct for the specific model 
in \cite{Choietal05a}, on both physical and mathematical grounds. A constant ``drag'' 
$D$ is not localized at small and large $s$ but has effects felt over the whole range of $s.$ Thus, one 
would not expect a constant-flux ``inertial range'' to exist for such a drag. In fact, for any choice of $D,$ 
a particular solution of the inhomogeneous equation (\ref{1pdf-mod}) is 
\be P_{part}(s) = D/\gamma. \ee
The general solution $P(s)$ of (\ref{1pdf-mod}) is a superposition of $P_{part}(s)$ with any solution of the homogeneous 
equation (\ref{1pdf-eq2}):
\be P(s) =\frac{D}{\gamma} + \frac{C_1}{\nu} e^{-s/\nu} +  \frac{C_2}{\nu} e^{-s/\nu}{\rm Ei}\left(\frac{s}{\nu}\right). \ee
This has flux $F(s)=-Ds-C_2\gamma.$ With the b.c. $F(s_{nl})=F_*$ and the unique probability-conserving choice
$D=|F_*|/s_{nl}$ one finds $C_2=0$ and thus
\be  P(s) = \frac{|F_*|}{\gamma} \cdot \frac{1}{s_{nl}} + \left(1- \frac{|F_*|}{\gamma}\right) 
                  \frac{e^{-s/\nu}}{\nu(1-e^{-s_{nl}/\nu})}, \lb{our-pdf} \ee
using normalization. Note the requirement $|F_*|<\gamma$ for positivity of this solution, which is then a 
superposition of a Rayleigh distribution and a uniform distribution. As expected, it has not constant 
flux but instead $F(s)=F_* s/s_{nl}.$

We claim that (\ref{our-pdf}) is the correct solution of the model formulated in \cite{Choietal05a}.
It gives an alternative possibility to explain intermittency and non-Rayleigh distributions, with 
constant tails $P(s)\propto s^0$ for $s\gg \nu$ rather than $P(s)\propto s^{-1}.$ Using the estimate 
$|F_*|\sim \gamma n/s_{nl}$ of \cite{Nazarenkoetal10} one finds that the realizability inequality 
$|F_*|<\gamma$ is satisfied whenever $n\lesssim s_{nl}.$ However, it  is unclear to us that the flux b.c. 
and constant drag $D$ is a physically correct model of wave-breaking. Constant-flux solutions 
as originally proposed in \cite{Choietal05a} could be relevant with a more general  model of strong 
nonlinear effects, in which the constant $D$ is replaced with a function $D(s)$ such that  
$\int_0^{s_{nl}} ds \, D(s)=|F_*|$ but for which $D(s)$ nearly vanishes except at the upper and lower limits. 
This would provide a ``transparency window'' that would allow constant probability-flux solutions to exist.  

Whatever may be the correct model of the strong nonlinear effects at $s>s_{nl}$, an important general point 
is that any modification of the 1-mode PDF eq.(\ref{1pdf-eq2}) requires that the wave kinetic equation must 
also be modified. This can be illustrated for the specific model in (\ref{1pdf-mod}) with $D=|F_*|/s_{nl}.$
Taking the time-derivative of $n= \int_0^{s_{nl}} ds\,\, sP(s)$ gives
\be  \dot{n} = -\gamma n+\eta -\eta s_{nl} P(s_{nl}) + \frac{1}{2}|F_*|s_{nl}, \lb{kineq-mod} \ee
which differs from the standard kinetic equation $\dot{n} = -\gamma n+\eta.$ The stationary form of the
``modified kinetic equation'' (\ref{kineq-mod}) is easily checked to be valid for our solution (\ref{our-pdf}) 
by calculating $n= \int_0^{s_{nl}} ds\,\, sP(s)$ to be
\begin{eqnarray}
n &=& \nu - \nu s_{nl} P(s_{nl}) + \frac{|F_*|}{2\gamma} s_{nl} \cr
   &\simeq & \nu\left(1-\frac{|F_*|}{\gamma}\right) + \frac{|F_*|}{2\gamma} s_{nl},
\lb{n-mod} \end{eqnarray}   
with the latter approximation valid for $s_{nl}\gg \nu.$ The correction to the usual value $n=\nu$ predicted by the kinetic
equation need not be small. If we use the estimate $|F_*|\sim \gamma n/s_{snl}$ from ``critical balance'' 
\cite{Nazarenkoetal10} then the second term in (\ref{n-mod}) is comparable to the first term, or even much larger 
if $\nu\ll n$. If such changes in $n(\bk)$ occur for a large set of wavenumbers $\bk$, then the coefficients
$\gamma$ and $\eta$ defined through the integrals (\ref{eta}),(\ref{gamma}) may also be strongly 
modified. These remarks make clear the nontriviality of constructing a self-consistent hybrid model of wave 
kinetics and of the strong nonlinear effects. 

\subsection{Super-Turbulence of Wave Kinetics?}\lb{super}

A second possibility to explain intermittency and anomalous scaling entirely within the framework of wave 
kinetics is by  ``super-turbulence''. We shall discuss this mechanism here, mainly at a general, 
qualitative level. 

It is useful to begin with a review of the work of Spohn \cite{Spohn84}, who already discussed turbulence
in gas dynamics as a natural motivation for ensembles of Boltzmann solutions. This is easiest to understand 
in the hydrodynamic regime where length-scale of variations in local equilibrium parameters (temperature, 
density, velocity) is much larger than the mean-free-path length.  The Boltzmann equation reduces then
to a hydrodynamic description and one  can appeal to the extensive literature on turbulent solutions of 
hydrodynamic equations. The specific example discussed in \cite{Spohn84} is the Rayleigh-Bernard system, 
considered as a many-particle system subject to thermal boundary conditions. In such a situation driven by the 
boundary conditions, Boltzmann's original $H$-theorem is no longer valid and the stationary solution of 
the Boltzmann equation for the pure conducting state is no longer purely Maxwellian but has a small 
correction corresponding to thermal non-equilibrium \cite{Espositoetal98}. In the turbulent regime 
at sufficiently high Reynolds and Rayleigh numbers, the laminar purely conducting state is unstable 
and turbulent convection develops. The temporal dynamics is chaotic so that long-time-averages,
for example, are described naturally by invariant measures correspondng to ensembles of Boltzmann
solutions. 

Such a statistical description is natural also for turbulent situations without driving by body 
forces or boundaries and with time-dependent statistics. For example, a single realization of decaying,
statistically homogeneous turbulence is very spatially complex and heterogeneous. Averages over large 
volumes ---by space-ergodicity---are again described by ensembles of solutions. Note in this example
that Boltzmann's $H$-theorem applies, with a monotonic increase of entropy due to heating of the fluid.
There is eventual approach to global thermodynamic equilibrium and a space-independent Maxwellian 
1-particle distribution, but the turbulent state with strong spatial variations exists  as a long transient 
for an intermediate range of times. The previous examples are in the hydrodynamic regime, but there
should be similar turbulence in the fully kinetic regime. For example, in compressible turbulence in gases 
at sufficiently high Mach numbers the thickness of shocks should be  of the order of the mean-free-path length 
\cite{Griffith81} and, for a quantitative description, the Boltzmann equation should be used rather than 
hydrodynamic equations. 

These ideas can carry over from the Boltzmann equation to the wave kinetic equation.  As noted
in the introduction, Zakharov et al. (\cite{Zakharovetal92}, section 4.2.2) had already pointed out that 
strong instability of the Kolmogorov solutions to the wave kinetic equation could lead to such a ``secondary 
turbulence'' . The statistical distribution of wave amplitudes  that would be obtained in a long time-series from 
experiment or simulation would then correspond to a random ensemble of solutions of the 1-mode 
PDF equation (\ref{1pdf-eq}):
\be \cP^{(1)}(s;\bk) =\int d\rho(P) \,\, P(s;\bk). \ee
Even if the individual solutions $P(s;\bk)$ were close to Rayleigh distributions, the measurable distributions
$\cP^{(1)}(s;\bk)$ could be arbitrarily far from Rayleigh, depending upon the ``super-statistical'' measure $\rho.$  
To determine this measure becomes a very difficult problem, as daunting as the corresponding problem for 
Navier-Stokes turbulence! Here we can only identify some of the sources of randomness, including instability, 
random forcing,  and boundary conditions. We consider these in turn.

Instability of the Kolmogorov solution is a natural origin for ``super-turbulence''. As a related example, consider
the GOY shell model of turbulence, a dynamical system with a stationary Kolmogorov solution supporting a constant 
energy flux with dimensional scaling \cite{Biferale03}. This solution, however, is subject to a rich array of 
instabilities \cite{Biferaleetal95,Kadanoffetal97} and the statistical behavior of the GOY model shows strong 
intermittency and anomalous scaling. There is a detailed theory of linear stability of the Kolmogorov
solutions of wave kinetic theory \cite{FalkovichShafarenko87,BalkZakharov88a,BalkZakharov88b,BalkZakharov98},
reviewed in \cite{Zakharovetal92}, Ch. 4. This theory studies perturbations to a scale-homogeneous
Kolmogorov solution which, strictly speaking, requires an infinitely long inertial interval. The conclusion
of this theory is that, for most common cases,  the Kolmogorov solution is linearly stable.  However, the 
theory,  although sophisticated, is not quite definitive. Quoting from \cite{Zakharovetal92}:
\begin{quote}
``It should be noted that a more consistent formulation of the 
problem of the stability of the Kolmogorov spectrum should be as follows. First of all, the kinetic 
equation should be supplemented by terms describing the isotropic pumping and damping regions 
and a stationary solution of this equation should be found that is close to the Kolmogorov spectrum 
in an interval $(k_1, k_2)$; outside this interval the solution may strongly differ from the Kolmogorov 
spectrum. Then the kinetic equation must be linearized in the vicinity of this stationary solution 
and expanded in angular harmonics. $\cdots$ 
Having examined the behavior of the solutions 
equation, one should clarify the changes that occur when the ranges of the source and sink 
in $k$-space go to zero or to infinity and examine the behavior of the perturbations established in the 
interval $(k_1, k_2)$. Finally, one should analyze in which situations this behavior is independent of the 
specific type of the source and sink. 

This program for examining the stability of Kolmogorov spectra turns out to be too complex. 
Currently there exists no strict proof of the fact that in general the kinetic equation with a source 
and sink has a stationary solution close to the Kolmogorov spectrum in some interval.'' 
\end{quote}
In fact, the situation is even less certain than this statement implies. It has recently been 
found that the space-independent Kolmogorov-Zakharov solutions can be linearly unstable to spatially 
inhomogeneous perturbations \cite{Newelletal12}. In addition, there are hydrodynamic flows
such as plane Couette and pipe Poiseuille which,  despite being linearly stable for all Reynolds numbers, 
become turbulent in laboratory experiments and in numerical simulations at moderate Reynolds number
\cite{SchmidHenningson01}.  The usual understanding is that these flows are unstable to finite amplitude 
perturbations, e.g. \cite{DubrulleZahn91}. Thus, linear stability does not rule out transition to turbulence.
For these reasons, we must regard ``super-turbulence'' as a viable possibility in wave kinetics. 

As emphasized in the previous quote, external forcings are required to maintain a stationary 
energy cascade state. It is important to consider more deeply the origin and role of such forces.
The monograph \cite{Zakharovetal92} considers the necessary conditions on the force for the 
existence of a stationary distribution (section 2.2.3) and how to match the Kolmogorov solutions
to the wavenumber regions of the force (section 3.4). This discussion assumes a particular idealized
model for the forcing, in which to the collision integral $I(\bk,\tau)$ of the wave kinetic equation
there is added an additional term:
\be \dot{n}(\bk,\tau)= \Gamma(\bk,\tau)n(\bk,\tau)+ I(\bk,\tau). \ee
This model describes ``pumping'' for wavenumbers at which $\Gamma(\bk,\tau)>0$ and ``damping'' 
for $\Gamma(\bk,\tau)<0.$ Validity of this model requires a suitably weak, slowly-changing force in the 
equations of motion (\ref{A-eq}):
\be\frac{d}{dt} A_\bk^\sigma = \frac{1}{2} \epsilon^2 \Gamma(\bk,\epsilon^2 t)A_\bk^\sigma +\cdots \ee
Here $\cdots$ denotes the original terms in (\ref{A-eq}). The addition of this term to the dynamics 
can be easily accommodated into our derivations, with a new term appearing in eq.(\ref{a2})
\be a_\bk^{(2)}(T) = \frac{1}{2} T \cdot \Gamma_\bk a_\bk^{(0)} +\cdots, \ee
in eq.(\ref{J3})
\be \cJ_3 =\frac{1}{2} T \delta_{\mu,0} \sum_1 \Gamma_1 \lambda_1 J_1 +\cdots,  \ee
in eq.(\ref{Zspec-eq2})
\be \frac{d\cZ}{d\tau} = \int d^dk \,\,\Gamma(\bk,\tau) \lambda(\bk) \frac{\delta\cZ}{\delta\lambda(\bk)} 
      + \cdots, \ee
in eq.(\ref{Zpdf-hier}) 
\be \frac{d\cZ^{(M)}}{d\tau}= \sum_{m=1}^M \Gamma(\bk_m,\tau) \lambda_m \frac{\partial\cZ^{(M)}}{\partial\lambda_m}
      +\cdots \ee      
and in eq.(\ref{pdf-hier2})      
\be \cF^{(M)}_m = \Gamma(\bk_m,\tau) s_m +\cdots. \ee      
Thus, the standard forcing model from \cite{Zakharovetal92} is obtained. All of our previous results, e.g. on 
preservation of factorized solutions, law of large numbers, etc., still carry over. In general, forced wave turbulence 
can be described by kinetic equations only for suitable assumptions on the forcing, like those above. In the above 
derivation, the only randomness was in the phases and amplitudes of the waves. However, the forcing amplitudes 
$\Gamma(\bk,\tau)$ may be chosen to be a realization of a stationary and time-ergodic random  process.  
In that case, long-time averages in the statistical steady-state can be described by averages over ensembles 
of solutions of the wave kinetic equations with different realizations of the force.  In fact, \cite{Falconetal08}
find that in their experiments the forcing is well modelled by an Ornstein-Uhlenbeck process with large fluctuations
in the energy input rate. Even without chaos or turbulence in the wave kinetic equations, this will yield
``super-randomness'', since the Kolmogorov solution is neutrally stable to change in the energy flux rate
\cite{Zakharovetal92}. 

In addition to pumping by an external force, hydrodynamic turbulence may also be (in fact, more commonly, is)
driven by boundary conditions. Such a situation can also occur in wave kinetics, as we have discussed 
in the previous section \ref{amplitude}, where strongly nonlinear effects such as wave-breaking can provide 
both sinks and sources to kinetic wave turbulence. These effects restrict the validity of wave kinetics to a 
``bounded domain'' in wavenumbers and amplitudes and could provide suitable forcings at the boundaries 
to represent those nonlinear effects. In this setting, Kolmogorov cascade solutions which are stable when 
considered without restriction, might become unstable and ``super-turbulence'' develop. The two scenarios 
in this section and the previous one are thus not completely exclusive.

\section{Conclusions}\lb{conclusion}

The major contributions of this paper are the following:
\begin{enumerate}
\item We have derived by formal calculations the leading-order, asymptotically correct multimode equations 
for wave kinetics in Hamiltonian systems with 3-wave resonances. 
\item We have shown that these equations possess factorized solutions for factorized initial conditions, corresponding 
to ``propagation of chaos'' or preservation of ``random phases \& amplitudes''.  
\item We introduced the ``empirical spectrum'' and ``empirical  1-mode PDF'' and showed that the above factorization
implies a law of large numbers, so that these quantities satisfy the wave-kinetic closure equations for nearly every 
initial realization of random phases and amplitudes, without necessity of averaging.  
\item We have demonstrated the close formal relations of wave kinetic theory with the kinetic theory of gases, 
especially regarding the role of entropy and the 2nd law of thermodynamics in both theories. 
\item We have completely classified all realizable solutions of our multimode equations and have shown that they correspond
to ``super-statistical solutions'', or ensembles of the wave-kinetic closure equations with either random initial conditions 
or random forces. 
\item We have exploited our results to discuss the possibilities to explain intermittency and non-Gaussian statistics of 
wave turbulence within the kinetic description, in particular by a ``super-turbulence'' corresponding to chaotic or 
stochastic solutions of the wave-kinetic closure equations. 
\end{enumerate}
Our work solves the problem posed in \cite{NewellRumpf11}, section 5.2.5 on the continuum limit of finite-dimensional 
wave turbulence to show ``how the natural closure arises in taking the limit $L\rightarrow\infty.$'' Although we have focused 
here on 3-wave resonant models, we expect that analogous results should hold also for other closely related wave systems, 
e.g. those whose first nontrivial resonances are 4-wave. 

It is perhaps important to emphasize what we have {\it not} accomplished in this paper: we have not  provided 
a rigorous mathematical proof of the wave kinetic theory. At this time, no set of sufficient conditions are 
known that would imply the validity of wave-kinetics for any general class of systems (\cite{NewellRumpf11},
section 5.2.6). Our derivations fail to constitute a proof in particular because we have made no 
attempt to rigorously bound the $O(\epsilon^3)$ remainder term and higher order terms in the perturbation expansion 
(\ref{Xdef}). In principle, there exist methods to calculate the collision integral in the wave kinetic equation formally 
to any order in the small parameter $\epsilon$ \cite{BenneyNewell69,BenneyNewell67,ErofeevMalkin89}. As a matter 
of fact, there are reasons to expect that the collision integral is non-analytic in $\epsilon,$ if the kinetic theory of gases 
is any guide. The multiscale asymptotic formalism devised by Bogolyubov \cite{Bogolyubov46} to calculate systematic 
corrections to the Boltzmann equation in powers of gas density is closely analogous to the methods employed 
to derive the wave kinetic equation perturbatively in $\epsilon.$ However, it was discovered in the mid-1960's 
that the 4th- and higher-order terms in Bogolyubov's density expansion of the collision operator for gases are 
divergent and the true dependence appears to be non-analytic in density \cite{Cohen93}. Even for gas kinetics, 
many fundamental issues regarding higher-order corrections remain unresolved. The rigorous works of Lanford 
\cite{Lanford75,Lanford76} imply that these higher-order terms are asymptotically negligible in the low-density 
limit for gases, but similar mathematical study of wave kinetics is only just begun. For a recent rigorous proof of 
validity of wave-kinetic predictions in a 4-wave model problem at  thermal equilibrium, see \cite{LukkarinenSpohn11}. 

Our goal here was the more directly physical one of formulating the limiting equations of wave kinetics and 
understanding their content. We have found that the equation (\ref{Peierls}) originally obtained by Peierls 
\cite{Peierls29} and many subsequent authors \cite{BroutPrigogine56,ZaslavskiiSagdeev67,Choietal05a,Choietal05b,
JakobsenNewell04} is not the leading-order, asymptotically valid equation in the wave kinetic limit. The Peierls 
equation contains additional terms which vanish as $L\rightarrow\infty$ and which do not appear in our 
limit equations.  It should be stressed that the derivation of the Peierls equation in \cite{Choietal05a,Choietal05b,
JakobsenNewell04} and also earlier derivations of Peierls himself \cite{Peierls29}, Brout \& Prigogine \cite{BroutPrigogine56}, 
Zaslavskii \& Sagdeev \cite{ZaslavskiiSagdeev67}, etc. are {\it not} systematic in the limit $L\rightarrow\infty.$ 
At present, the Peierls equation has no established validity (or even a precise meaning) in the large-volume limit. 
Although the Peierls equation does not have any obvious asymptotic validity, it does serve as a ``generating 
equation''\footnote{This terminology was suggested to us by S. Nazarenko (private communication).} 
from which all of our simpler equations can be obtained in the limit
$L\rightarrow\infty.$ It is an interesting issue whether the additional terms retained in the Peierls equation describe
correctly sub-leading phenomena in powers of $1/L.$ This question shall be addressed in our subsequent 
paper \cite{Eyink12}, where we shall derive the limiting equation for the generating functional (\ref{ZMfun}) 
with all $N$ modes retained and use our techniques to evaluate the next-to-leading order corrections in $1/L$ 
to the asymptotic limit. In any case, the virtue of our new equations (\ref{Zspec-eq}) and (\ref{Zpdf-hier})
is that they contain all and only the leading-order terms and thus allow us to clarify the structure of that limit. 
We have completely characterized the solutions of the limiting hierarchies and shown them to consist of 
random ensembles of solutions of the wave kinetic equation (\ref{wavekineq}) and 1-mode PDF equation 
(\ref{1pdf-eq}). This should permit a better evaluation of the theory by simulation and experiment. 
In particular, our most subversive conclusion is that the ``mean-field'' scaling of the Kolmogorov-Zakharov solutions 
may not be the true prediction of kinetic theory for scaling exponents of spectra and higher-order statistics in wave 
turbulence and that wave kinetics may instead allow for intermittency and anomalous scaling. 

\vspace{.25in}

\noindent {\bf Acknowledgements.} We are grateful to U. Frisch, S. Nazarenko, A. Newell, and 
H. Spohn for many useful comments on our work. We thank also two anonymous referees for their suggestions
which have helped to improve the paper.  

\newpage

\appendix\lb{appendix}

\section{Derivation of the Spectral Hierarchy}\lb{appendix-spectral}

As discussed in the text, we have verified the results of \cite{Choietal05b} for the $\cJ$-terms in eqs.(\ref{J1})-(\ref{J2}),
up to minor corrections. The contributions to the spectral generating functional are obtained by making the substitutions  
$J_i= (2\pi/L)^{d} \tJ_i$ and $\lambda_1=i\lambda(\bk_1)$ and taking the limit $L\rightarrow\infty.$ The asymptotics of 
the various terms can be summarized in the following general rule: each action variable $J$ changes the order by $L^{-d}$ 
and each free sum over wavenumbers by $L^d$. We illustrate this rule in our detailed calculations below. 

{\it{Calculation of $\cJ_1$:}} Substituting the definition (\ref{expansion}) of $a^{(1)}_1$, one obtains 
\begin{eqnarray}
\cJ_1
= \sum_{1,2,3} \Big(\lambda_1+\frac{\mu_1}{2J_1}\Big)L^{+,\sigma_2,\sigma_3}_{\bk_1,\bk_2,\bk_3} \sqrt{J_1J_2J_3} \Big\langle\psi_1^{-1}\psi_2^{\sigma_2}\psi_3^{\sigma_3}\prod_{\bk}\psi_{\bk}^{\mu_\bk}\Big\rangle_{\psi}
                   \Delta_T(\sigma_2\omega_2+\sigma_3\omega_3-\omega_1)\delta_{\bk_1,\sigma_2\bk_2+\sigma_3\bk_3},
\end{eqnarray} 
which can be represented by the following diagram before averaging over phases:
\begin{center}
\includegraphics[width=150pt,height=100pt]{1.pdf}
\end{center}
After phase averaging, there are two contributions represented by the following diagrams:
\begin{center}
\includegraphics[width=150pt,height=100pt]{1_1.pdf}\,\,\,\,\,\,\,\,\,\,\,\,\,\,\,\,\,\,\,\,\,\,\,\,\,\,\,\,\,
\includegraphics[width=200pt,height=100pt]{1_2.pdf}
\end{center}
or explicitly written as
\begin{eqnarray}
\cJ_1&= &}{\sum_{1,2,3}}' \Big(\lambda_1+\frac{\mu_1}{2J_1}\Big)L^{+,\sigma_2,\sigma_3}_{\bk_1,\bk_2,\bk_3}\sqrt{J_1J_2J_3} 
                   \Delta_T(\sigma_2\omega_2+\sigma_3\omega_3-\omega_1)\delta_{\bk_1,\sigma_2\bk_2+\sigma_3\bk_3}\delta_{\mu_1,1}\delta_{\mu_2,-\sigma_2}\delta_{\mu_3,-\sigma_3}\prod_{m\neq 1,2,3}\delta_{\mu_m,0
                   \cr
        &&+{\sum_{1,2}}' \Big(\lambda_1+\frac{\mu_1}{2J_1}\Big)L^{+,\sigma_2,\sigma_2}_{\bk_1,\bk_2,\bk_2}  \sqrt{J_1} J_2             
          \Delta_T(2\sigma_2\omega_2-\omega_1)\delta_{\bk_1,2\sigma_2\bk_2} \delta_{\mu_1,1}\delta_{\mu_2,-2\sigma_2}\prod_{m\neq 1,2}\delta_{\mu_m,0}.
\end{eqnarray} 
Note that a prime $\prime$ on a sum indicates that all wavenumbers must be distinct. Here all of the summations are ``pinned.'' 
Now making the substitutions  $J_i= (2\pi/L)^{d} \tJ_i$ and $\lambda_1=i\lambda(\bk_1)$ into the above and taking the limit 
$L\rightarrow\infty$ gives
\begin{eqnarray}
\cJ_1
& =&  \frac{1}{2}\left(\frac{2\pi}{L}\right)^{d/2}\sum_{\ul{\sigma}=(-1,\sigma_2,\sigma_3)}{\sum_{\ul{\bk}}}' L^{+,\sigma_2,\sigma_3}_{\bk_1,\bk_2,\bk_3} 
\sqrt{\frac{\tJ_2\tJ_3}{\tJ_1}} 
                    \Delta_T(\ul{\sigma}\cdot\omega(\ul{\bk}))\delta_{\ul{\sigma}\cdot\ul{\bk},{\bf 0}}\delta_{\ul{\mu},-\ul{\sigma}}\prod_{m\neq 1,2,3}\delta_{\mu_m,0} \cr
 && + \frac{1}{2} \left(\frac{2\pi}{L}\right)^{d/2}\sum_{\ul{\sigma}=(-1,2\sigma_2)}{\sum_{\ul{\bk}'}}' L^{+,\sigma_2,\sigma_2}_{\bk_1,\bk_2,\bk_2}  
 \frac{\tJ_2}{\sqrt{\tJ_1}}
                    \Delta_T(\ul{\sigma}\cdot\omega(\ul{\bk}'))\delta_{\ul{\sigma}\cdot\ul{\bk}',{\bf 0}}
                    \delta_{\ul{\mu}',-\ul{\sigma}}\prod_{m\neq 1,2}\delta_{\mu_m,0},
\end{eqnarray}   
where $\ul{\mu}'=(\mu_1,\mu_2)$ and $\ul{\bk}'=(\bk_1,\bk_2)$. Hence $\cJ_1$ is $O(L^{-d/2})$ in the large box limit. In fact, the result easily follows from the aforementioned asymptotic rules. The amplitude factor $\sqrt{J_1J_2J_3}$ changes the order of $\cJ_1$ by $L^{-3d/2}$. Since there is no free sum here, the contribution of the term proportional to $\lambda_1$ is at most $O(L^{-3d/2})$. The term proportional to $\mu_1$ introduces an additional action variable $1/J_1$ and thus gives the leading contribution $O(L^{-d/2})$.

{\it{Calculation of $\cJ_2$:}} Again substituting the definition (\ref{a1}) of $a_1^{(1)}$, one obtains
\begin{equation}
\begin{split}
\cJ_2=&\frac{1}{2}\sum_{1,2,3,4,5} \Big(\lambda_1+\lambda_1^2 J_1-\frac{\mu_1^2}{4 J_1}\Big)L^{+,\sigma_2,\sigma_3}_{\bk_1,\bk_2,\bk_3}L^{-,\sigma_4,\sigma_5}_{\bk_1,\bk_4,\bk_5} \sqrt{J_2 J_3 J_4 J_5}\llangle\psi_2^{\sigma_2}\psi_3^{\sigma_3}\psi_4^{\sigma_4}\psi_5^{\sigma_5}\prod_{\bk}\psi_{\bk}^{\mu_k}\rrangle_\psi \\
&\Delta_T(\sigma_2\omega_2+\sigma_3\omega_3-\omega_1)
\Delta_T(\sigma_4\omega_4+\sigma_5\omega_5+\omega_1)\delta_{\bk_1,\,\sigma_2\bk_2+\sigma_3\bk_3}\delta_{-\bk_1,\,\sigma_4\bk_4+\sigma_5\bk_5},
\end{split}
\label{J2phase}\end{equation}
which can be represented by the following diagram before phase-averaging:
\begin{center}
\includegraphics[width=150pt,height=100pt]{2.pdf}
\end{center}
After phase averaging, $\cJ_2$ has three types of contributions. The leading contributions are two type I diagrams with no external couplings.
We show one such diagram here (the other one is obtained by the exchange $2\leftrightarrow 3$).
\begin{center}
\includegraphics[width=120pt,height=100pt]{2_1.pdf}
\end{center}
These contain three distinct wavenumbers and two vertices, but the wavenumber delta-functions at the two vertices turn out to give the same constraint.
Thus the type I diagrams contain two free wavenumbers corresponding to unconstrained sums. 
Subleading contributions are given by type II diagrams with two wavenumbers externally coupled. We show one diagram here with the 
wavenumbers 2 and 4 coupled to distinct external blobs. The three others can be obtained 
by exchanging $4\leftrightarrow 5$ and $2\leftrightarrow 3$.
\begin{center}
\includegraphics[width=150pt,height=100pt]{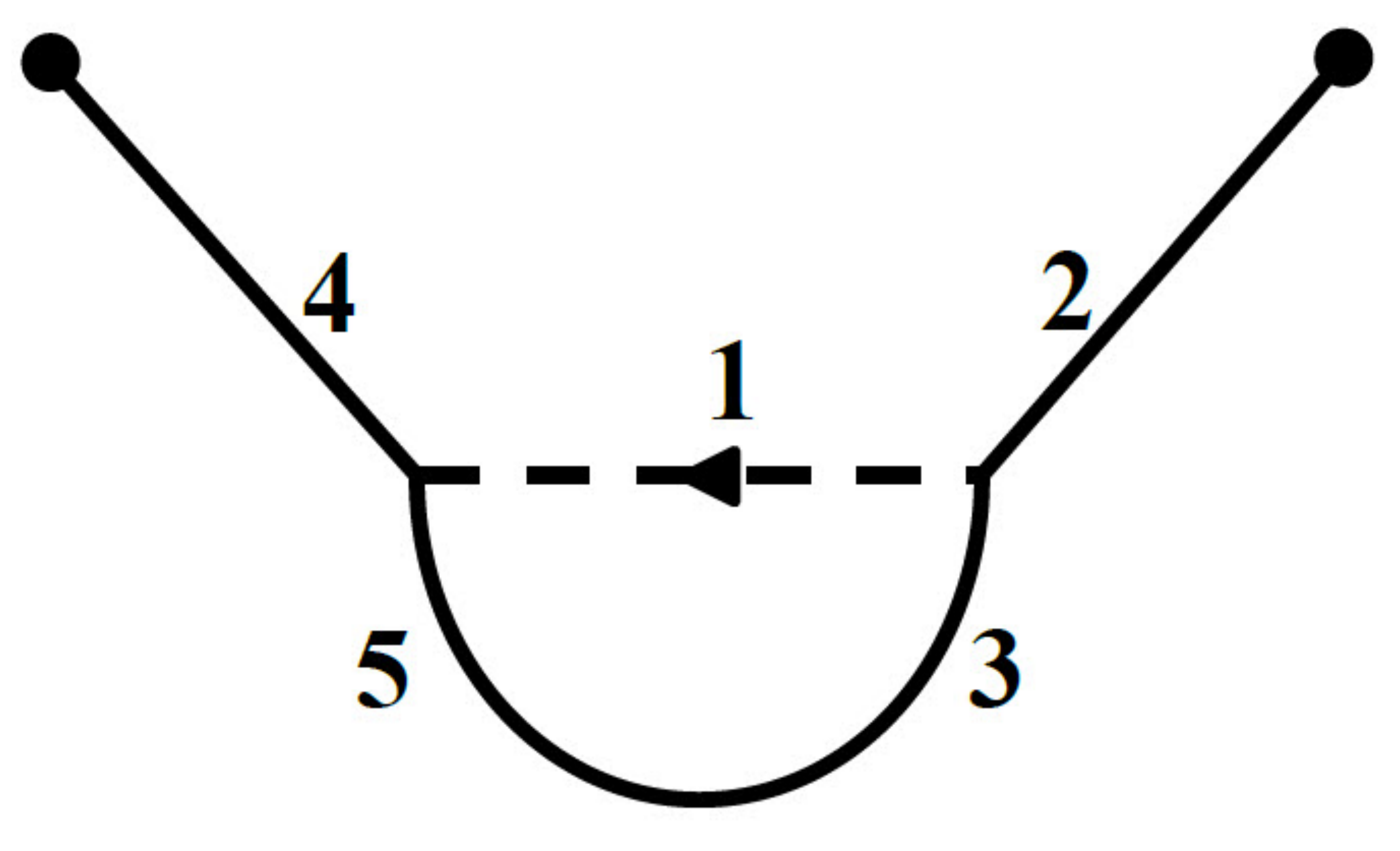}
\end{center}
These subleading contributions contain only one free wavenumber. The two delta functions $\delta_{\bk_1,\,\sigma_2\bk_2+\sigma_3\bk_3}$ 
and $\delta_{-\bk_1,\,\sigma_4\bk_4+\sigma_5\bk_5}$ along with the internal connection $\delta_{\bk_3,\bk_5}\delta_{\sigma_3,-\sigma_5}$ 
require $\sigma_2\bk_2=-\sigma_4\bk_4$.  Therefore, $\bk_2=-\bk_4$ (since $\bk_1$ and $\bk_4$ must be distinct) and $\sigma_2=\sigma_4$ 
in this diagram.  In principle, another type II contribution could exist with two wavenumbers coupled to the same external blob, as follows:
\begin{center}
\includegraphics[width=120pt,height=100pt]{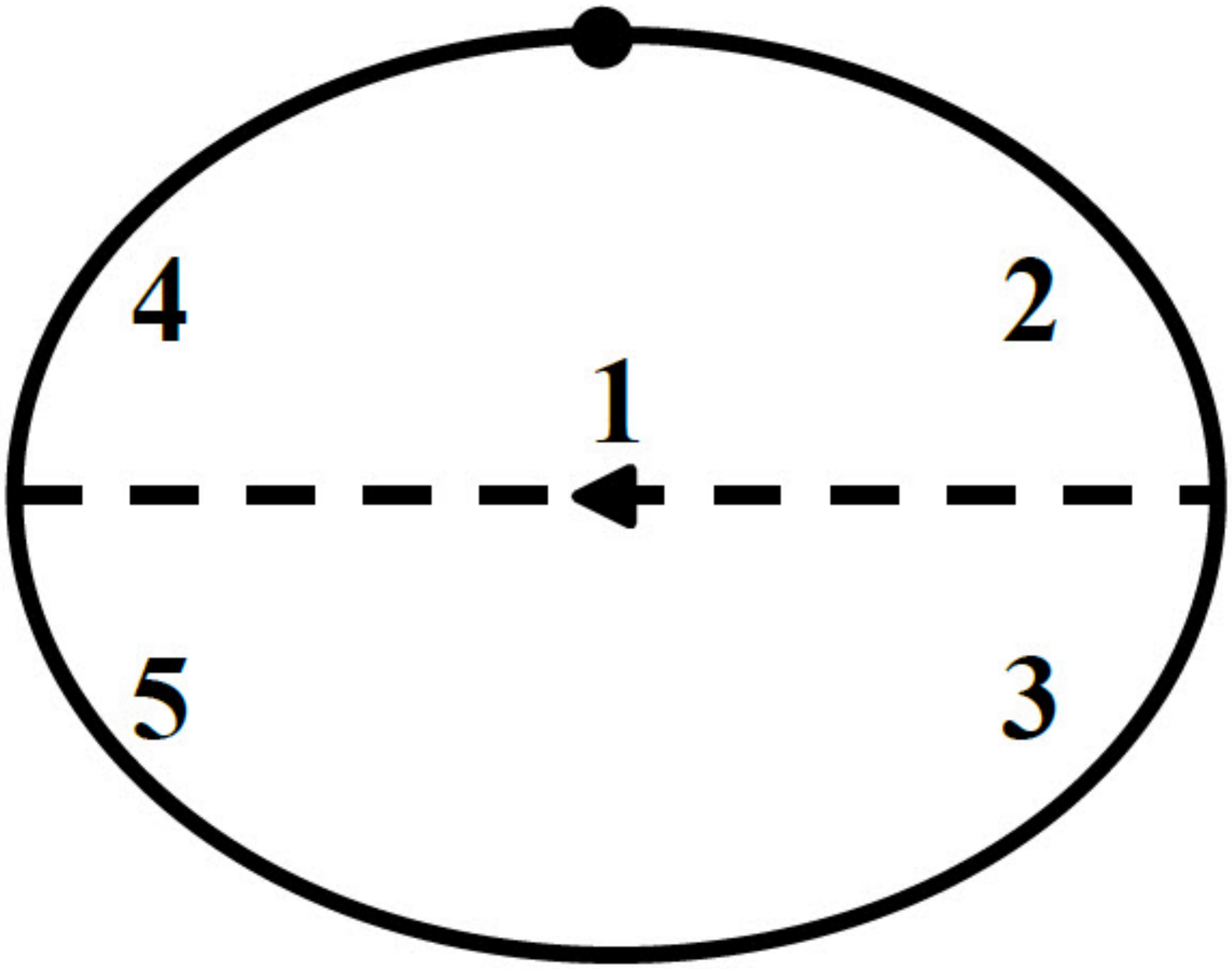}
\end{center}
together with diagrams resulting from $4\leftrightarrow 5$ and $2\leftrightarrow 3$. However, the condition $\sigma_2\bk_2=-\sigma_4\bk_4$ 
in this case requires that $\bk_2=\bk_4$, $\sigma_2=-\sigma_4,$ and $\mu_2=0,$ which coincides with the type I diagram.

All other contributions are type III where all wavenumbers are pinned, for example,
\begin{center}
\includegraphics[width=150pt,height=100pt]{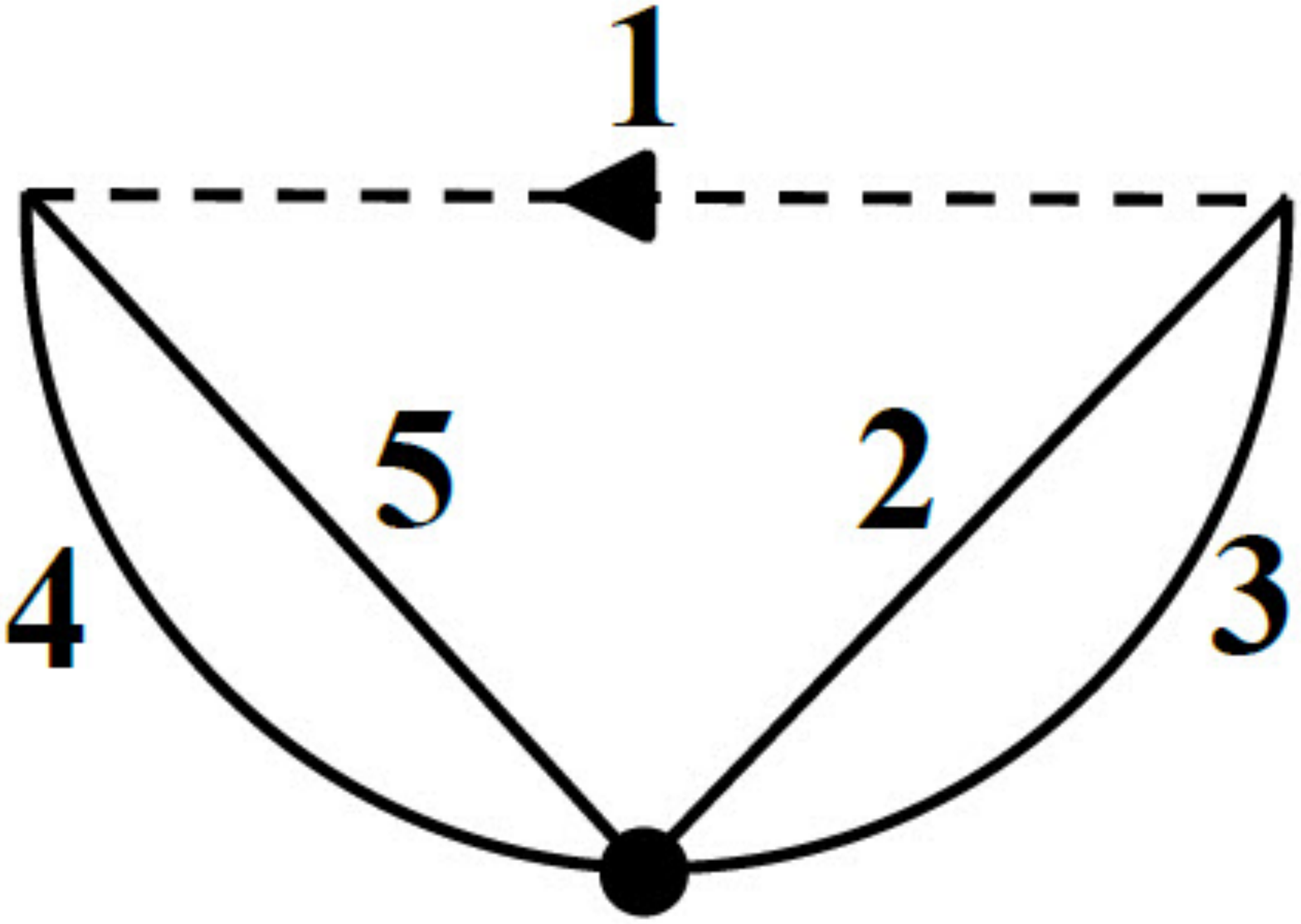}
\end{center}
Now consider large $L.$
The amplitude factor $\sqrt{J_2J_3J_4J_5}$ changes the order by $L^{-2d}$. Now consider the contributions 
proportional to the three terms in the prefactor $\lambda_1+\lambda_1^2 J_1-\frac{\mu_1^2}{4J_1}.$ In the term proportional to $\lambda_1$ the 
type I diagrams provide two free sums that increase the order by $L^{2d}$, giving an $O(1)$ contribution. The next order contributions from type II
diagrams have only one free wavenumber and thus are at most $O(L^{-d})$. The term proportional to $\lambda_1^2 J_1$ is at most $O(L^{-d})$ 
because of the additional action variable $J_1$. For the term proportional to $\frac{\mu_1^2}{4J_1}$, we note for both type I and type II diagrams 
$\bk_1$ is not pinned, which requires $\mu_1$ to be zero. Thus the only nonzero contribution here must come from type III graphs, which is at 
most $O(L^{-d})$. Therefore we can conclude that the leading contributions in $\cJ_1$ are $O(1)$, or explicitly
\begin{eqnarray} 
\cJ_2 &\sim &9\delta_{\mu,0}\sum_{1,2,3} \lambda_1J_2J_3 
 |H^{-,\sigma_2,\sigma_3}_{\bk_1,\bk_2,\bk_3}|^2 |\Delta_T(\sigma_2\omega_2+\sigma_3\omega_3-\omega_1)|^2
\end{eqnarray}
Substituting $J_i= (2\pi/L)^{d} \tJ_i$ and $\lambda_1=i\lambda(\bk_1)$ and taking the limit $L\rightarrow\infty$ using (\ref{largeL}), one finds
\begin{eqnarray}
\left\langle e^{\sum_\bk \lambda_\bk J_\bk} \,\cJ_2\right\rangle_J &\sim &
-9i\delta_{\mu,0} \sum_{\ul{\sigma}=(-1,\sigma_2,\sigma_3)}\int d^dk_1 d^dk_2 d^dk_3 
\delta^d(\ul{\sigma}\cdot\ul{\bk})
\big|\Delta_T(\ul{\sigma}\cdot\omega(\ul{\bk}))\big|^2
     \cr
&& \,\,\,\,\,\,\,\,\,\,\,\,\,\,\,\,\,\,\,\,\,\,\,\,\,\,\,\,\,\,\,\,\,\,\,\,\,\,\,\,\,\,\,\,\,\,\,\,\,\,\,\,\,\,\,\,\,\,\,\,\,\,\,\,\,\,\,\,\,\,\,\,
  \times \lambda(\bk_1)  \big|H^{-,\sigma_2,\sigma_3}_{\bk_1,\bk_2,\bk_3}\big|^2 \frac{\delta^2\cZ}{\delta\lambda(\bk_2)\delta\lambda(\bk_3)}
\end{eqnarray}
It is worth mentioning here that the term proportional to $\lambda_1^2$ kept in \cite{Choietal05b} gives only an $O(L^{-d})$ contribution in the large box limit, which can be calculated exactly as
\begin{eqnarray} 
9i\left(\frac{2\pi}{L}\right)^d\delta_{\mu,0}\sum_{\ul{\sigma}=(-1,\sigma_2,\sigma_3)}\int d^dk_1d^dk_2d^dk_3\lambda^2(\bk_1)\frac{\delta^3\cZ}{\delta\lambda(\bk_1)\delta\lambda(\bk_2)\delta\lambda(\bk_3)}|H^{-,\sigma_2,\sigma_3}_{\bk_1,\bk_2,\bk_3}|^2|\Delta_T(\ul{\sigma}\cdot\omega(\ul{\bk}))|^2\delta^d(\ul{\sigma}\cdot\ul{\bk}),
\end{eqnarray}
which is even smaller than the $O(L^{-d/2})$ correction from $\cJ_1$ which we calculated before.


{\it{Calculation of $\cJ_3$:}} Substituting the definition (\ref{a2}) of $a_1^{(2)}$ we have
\begin{eqnarray}
\cJ_3=&\sum_{1,2,3,4,5}\Big(\lambda_1+\frac{\mu_1}{2J_1}\Big)L^{+,\sigma_2,\sigma_3}_{\bk_1,\bk_2,\bk_3}L^{\sigma_2,\sigma_4,\sigma_5}_{\bk_2,\bk_4,\bk_5}
\sqrt{J_1J_3J_4J_5}\llangle\psi_1^{-1} \psi_3^{\sigma_3}\psi_4^{\sigma_4}\psi_5^{\sigma_5}\prod_{\bk}\psi_{\bk}^{\mu_\bk}\rrangle_{\psi} \\
&\left.E_T(\omega^1_{345},\omega^1_{23})\delta^1_{23}\delta^{2}_{45}\right|_{\sigma_1=1}+(2\leftrightarrow 3).
\end{eqnarray}
By symmetry, we only need to consider the first part, which has the following diagram pre-phase-averaged:
\begin{center}
\includegraphics[width=140pt,height=100pt]{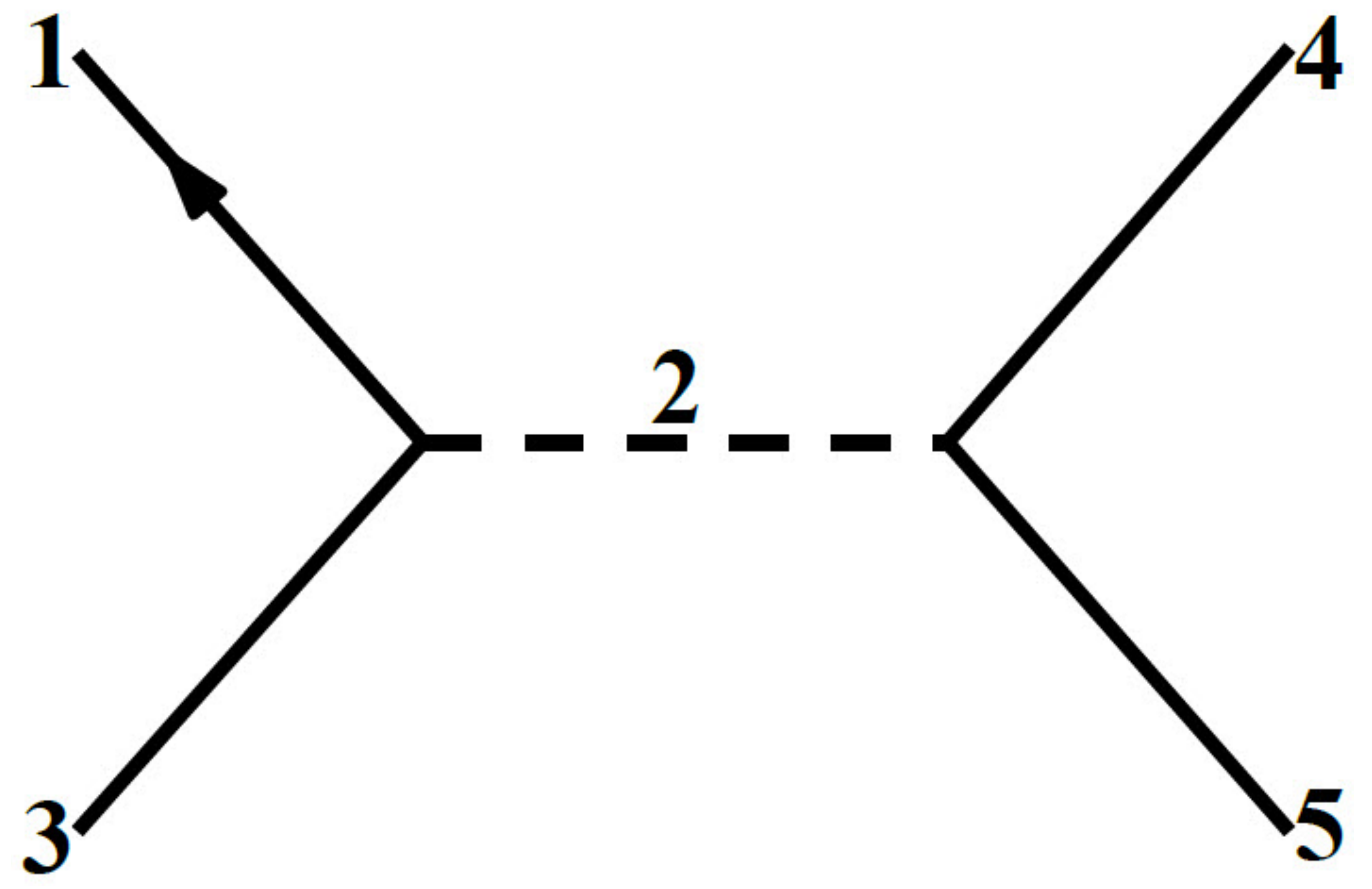}.
\end{center}
After phase averaging, there are three types of contributions. The type I diagrams are two leading contributions which contain two free wavenumbers. We show only one diagram here, the other obtained by exchanging $4\leftrightarrow 5$ 
\begin{center}
\includegraphics[width=120pt,height=100pt]{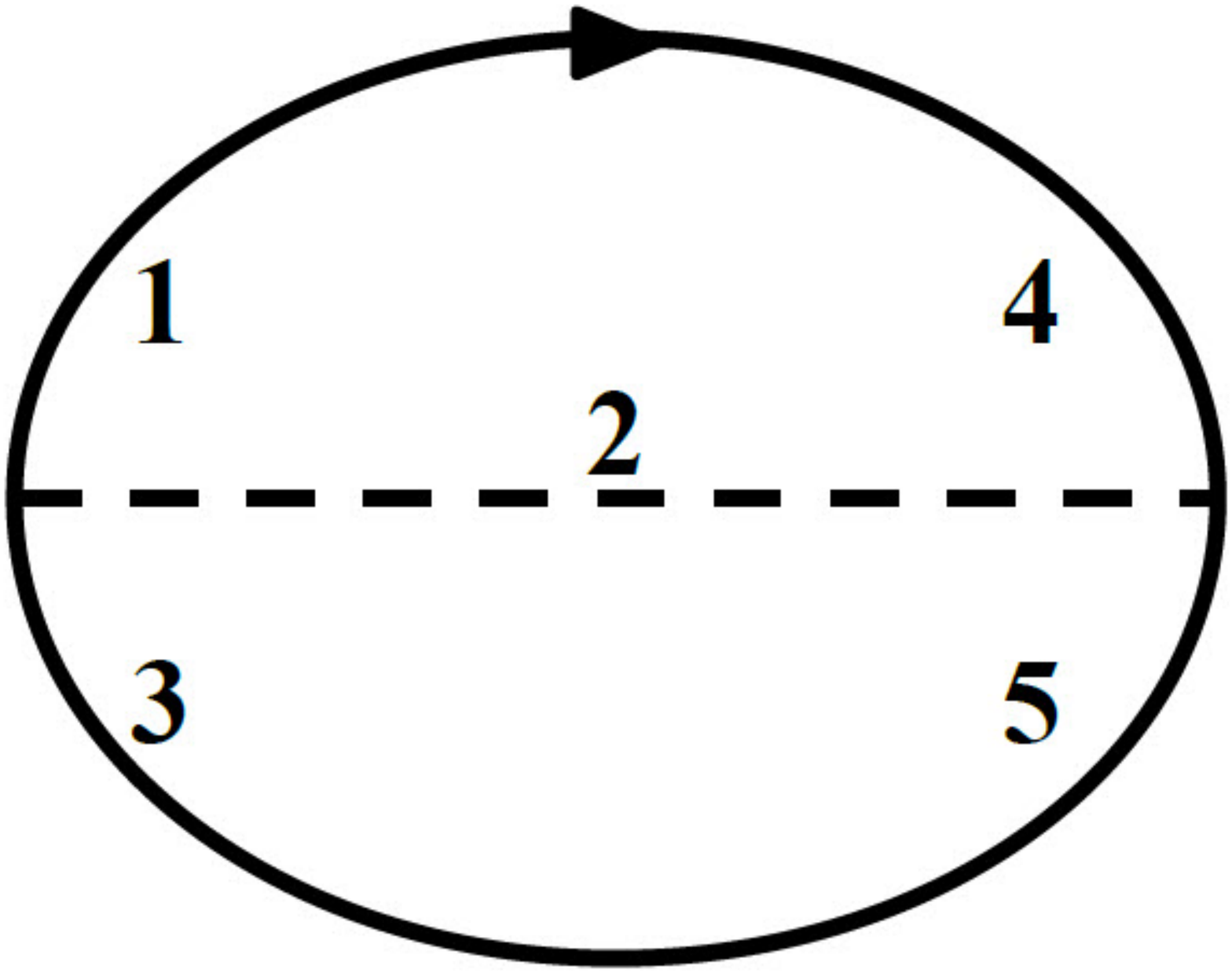}
\end{center}
The type II diagrams are four subleading contributions which contain one free wavenumber. Again we only show one diagram here, with mode 1 externally coupled:
\begin{center}
 \includegraphics[width=150pt,height=100pt]{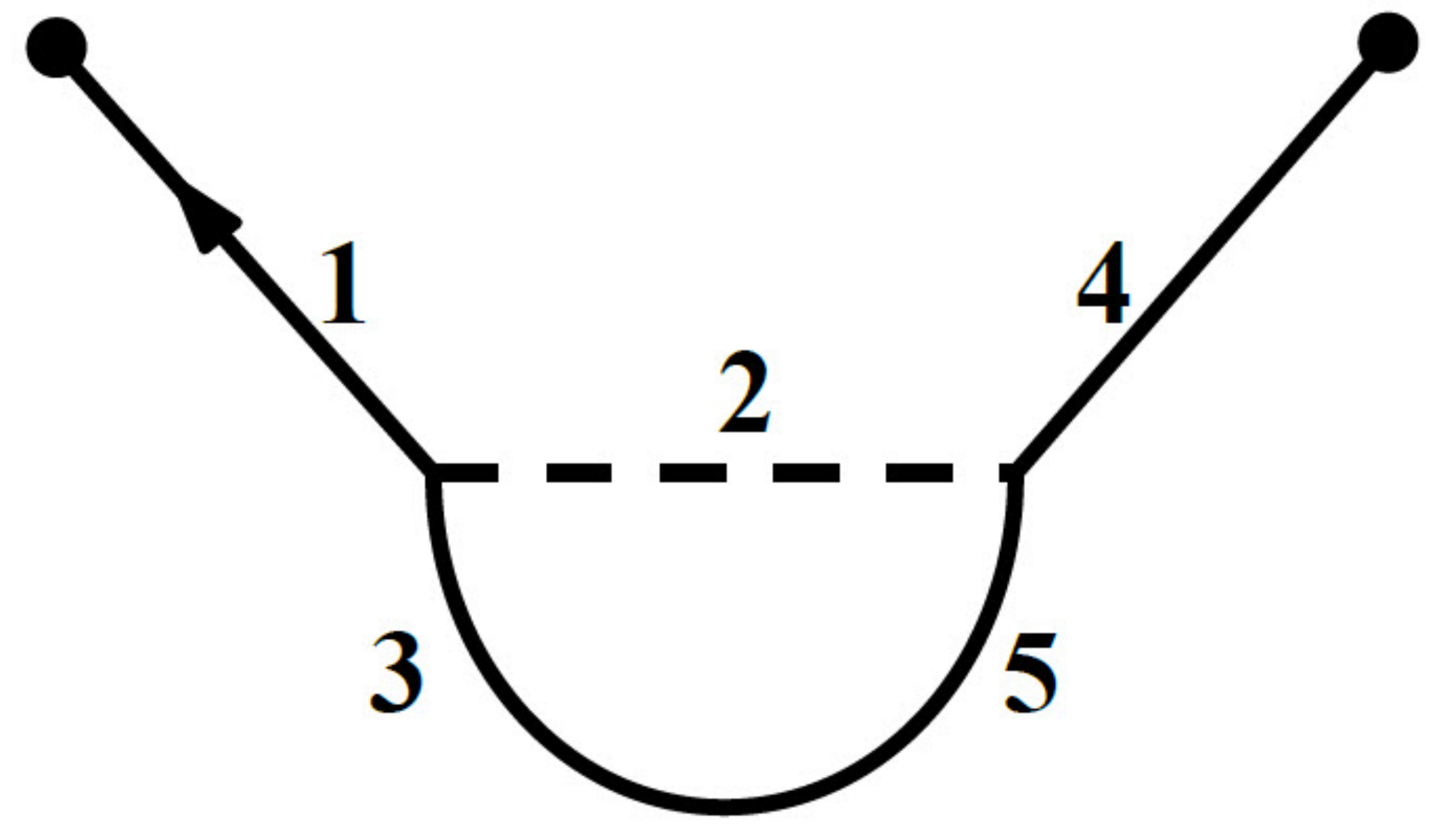}
 \end{center}
 It turns out that the graphs with 1 externally coupled are the most important type II contributions to $\cJ_3.$ The delta functions require $\bk_4=-\bk_1$ 
 and $\sigma_4=-1$ in this diagram. Additional type II graphs can be obtained by symmetry. All other contributions are type III where all wavenumbers are pinned. 

Now consider large $L$. The amplitude factor $\sqrt{J_1J_3J_4J_5}$ changes the order of $\cJ_3$ by $L^{-2d}$. In the term proportional to $\lambda_1$, 
the type I diagrams increase the order by $L^{2d}$ since they contain two free sums, giving an $O(1)$ contribution to $\cJ_1$. In the term proportional to $\frac{\mu_1}{2J_1}$, the type I diagrams give zero contribution because they require $\mu_1$ to vanish. The leading $O(1)$ contribution now comes from the type II diagrams 
with mode 1 externally coupled and $\mu_1=1.$ The type III diagrams contain no free wavenumber and give contributions at most $O(L^{-d})$.  The leading 
contribution to $\cJ_3$ in the large-box limit is thus:
\begin{eqnarray}
\cJ_3 & \sim & 2\sum_{\sigma_2,\sigma_3}\prod_{\bk}\delta_{\mu_\bk,0}\sum_{1,2,3}\lambda_1L^{+,\sigma_2,\sigma_3}_{\bk_1,\bk_2,\bk_3} L^{\sigma_2,-1,-\sigma_3}_{\bk_2,\bk_1,\bk_3} J_1J_3 E_T(0,\sigma_2\omega_2+\sigma_3\omega_3-\omega_1) \delta_{\bk_1,\sigma_2\bk_2+\sigma_3\bk_3} \cr
&&+\sum_{\sigma_2,\sigma_3}\sum_{1,2,3}\delta_{\mu_1,1}\delta_{\mu_{-1},1}\prod_{\bk\neq \bk_1,-\bk_1}\delta_{\mu_\bk,0} \frac{\mu_1}{2J_1} L^{+,\sigma_2,\sigma_3}_{\bk_1,\bk_2,\bk_3}L^{\sigma_2,-1,-\sigma_3}_{\bk_2, -\bk_1,\bk_3} \sqrt{J_1J_{-1}}J_3 \cr
&&\,\,\,\,\,\,\,\times E_T(\omega_1+\omega_{-1},\omega_1-\sigma_2\omega_2-\sigma_3\omega_3) \delta_{\bk_1,\sigma_2\bk_2+\sigma_3\bk_3}  +(2\leftrightarrow 3)
\end{eqnarray}
Substituting $J_i=(2\pi/L)^d\tJ_i$ and $\lambda_1=i\lambda(\bk_1)$ and taking the limit $L\to\infty$ using (\ref{largeL}), one finds
\begin{eqnarray}
\left\langle e^{\sum_\bk \lambda_\bk J_\bk} \,\cJ_3\right\rangle_J&\sim&18i\delta_{\mu,0}\sum_{\ul{\sigma}=(-1,\sigma_2,\sigma_3)}\sigma_2
\int d^dk_1\int d^dk_2 \int d^dk_3 \delta^d(\ul{\sigma}\cdot\ul{\bk})E_T(\ul{\sigma}\cdot\omega(\ul{\bk}),0)  \cr
&&\,\,\,\,\,\,\,\,\,\,\,\,\,\,\,\,\,\,\,\,\,\,\,\,\lambda(\bk_1)\big|H^{-,\sigma_{2},\sigma_{3}}_{\bk_1,\bk_2,\bk_3}\big|^2\frac{\delta^2\cZ}{\delta\lambda(\bk_1)\delta\lambda(\bk_3)}        \cr
&&-9{\sum_1}\delta_{\mu_1,1}\delta_{\mu_{-1,1}}\prod_{\bk\neq\bk_1,-\bk_1}\delta_{\mu_\bk,0}\sum_{\ul{\sigma}=(-1,\sigma_2,\sigma_3)}\sigma_2\int d^dk_2
\int d^dk_3 \delta^d(\ul{\sigma}\cdot\ul{\bk})E_T(\omega(\bk_1)+\omega(-\bk_1),\ul{\sigma}\cdot\omega(\ul{\bk}))     \cr
&&\,\,\,\,\,\,\,\,\,\,\,\,\,\,\,\,\,\,\,\,\,\,\,\, H^{-,\sigma_2,\sigma_3}_{\bk_1,\bk_2,\bk_3} H^{+,\sigma_2,\sigma_3}_{-\bk_1,\bk_2,\bk_3}\llangle e^{\sum_\bk \lambda_\bk J_\bk}\sqrt{\frac{\tJ_{-1}}{\tJ_1}}\tJ_3\rrangle_J+(2\leftrightarrow 3).
\end{eqnarray}

{{\it Calculation of $\cJ_4$:}} Similarly, substituting (\ref{a1}) for $a_1^{(1)}$
\begin{eqnarray}
\cJ_4&=&\sum_{1,2,3,4,5}\Big(\frac{1}{2}\lambda_1^2+\frac{\mu_1}{4J_1^2}(\frac{\mu_1}{2}-1)+
\frac{\lambda_1\mu_1}{2J_1}\Big)L^{+,\sigma_2,\sigma_3}_{\bk_1,\bk_2,\bk_3}
L^{+,\sigma_4,\sigma_5}_{\bk_1,\bk_4,\bk_5}J_1\sqrt{J_2J_3J_4J_5}\llangle\psi_1^{-2}\psi_2^{\sigma_2}\psi_3^{\sigma_3}\psi_4^{\sigma_4}\psi_5^{\sigma_5}\prod_{\bk}\psi_{\bk}^{\mu_\bk}\rrangle_\psi     \cr
&&\,\,\,\,\,\, \times\Delta_T(\sigma_2\omega_2+\sigma_3\omega_3-\omega_1)\Delta_T(\sigma_4\omega_4+\sigma_5\omega_5-\omega_1)\delta_{\bk_1,\sigma_2\bk_2+\sigma_3\bk_3}\delta_{\bk_1,\sigma_4\bk_4+\sigma_5\bk_5},
\end{eqnarray}
which can be represented by the following diagram before phase-averaging:
\begin{center}
\includegraphics[width=150pt,height=100pt]{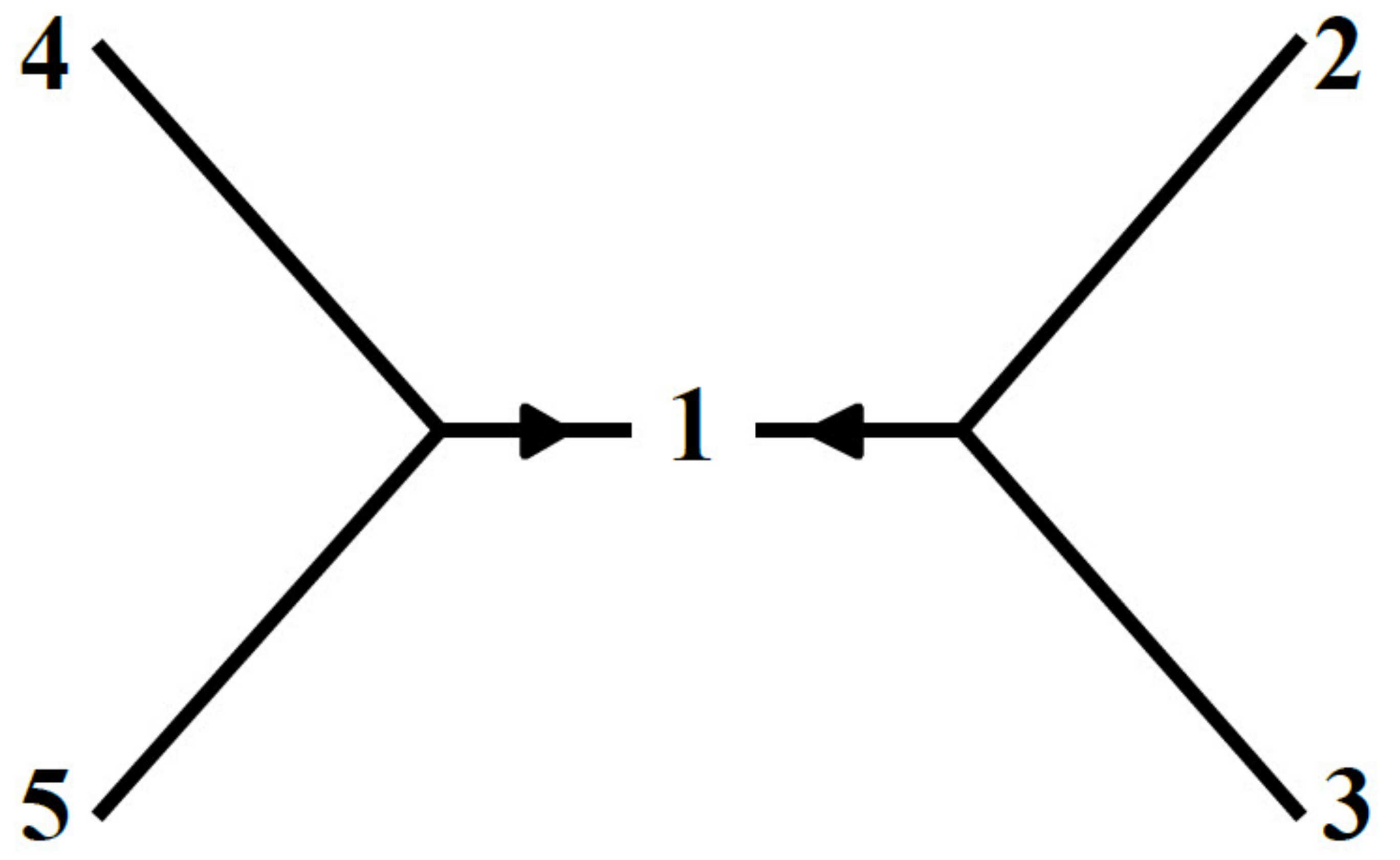}
\end{center}
After phase averaging, there are two types of contributions. Type I diagrams have only the 1 mode externally coupled,
corresponding to the following diagram and its partner with $2\leftrightarrow 3$:
\begin{center}
 \includegraphics[width=120pt,height=100pt]{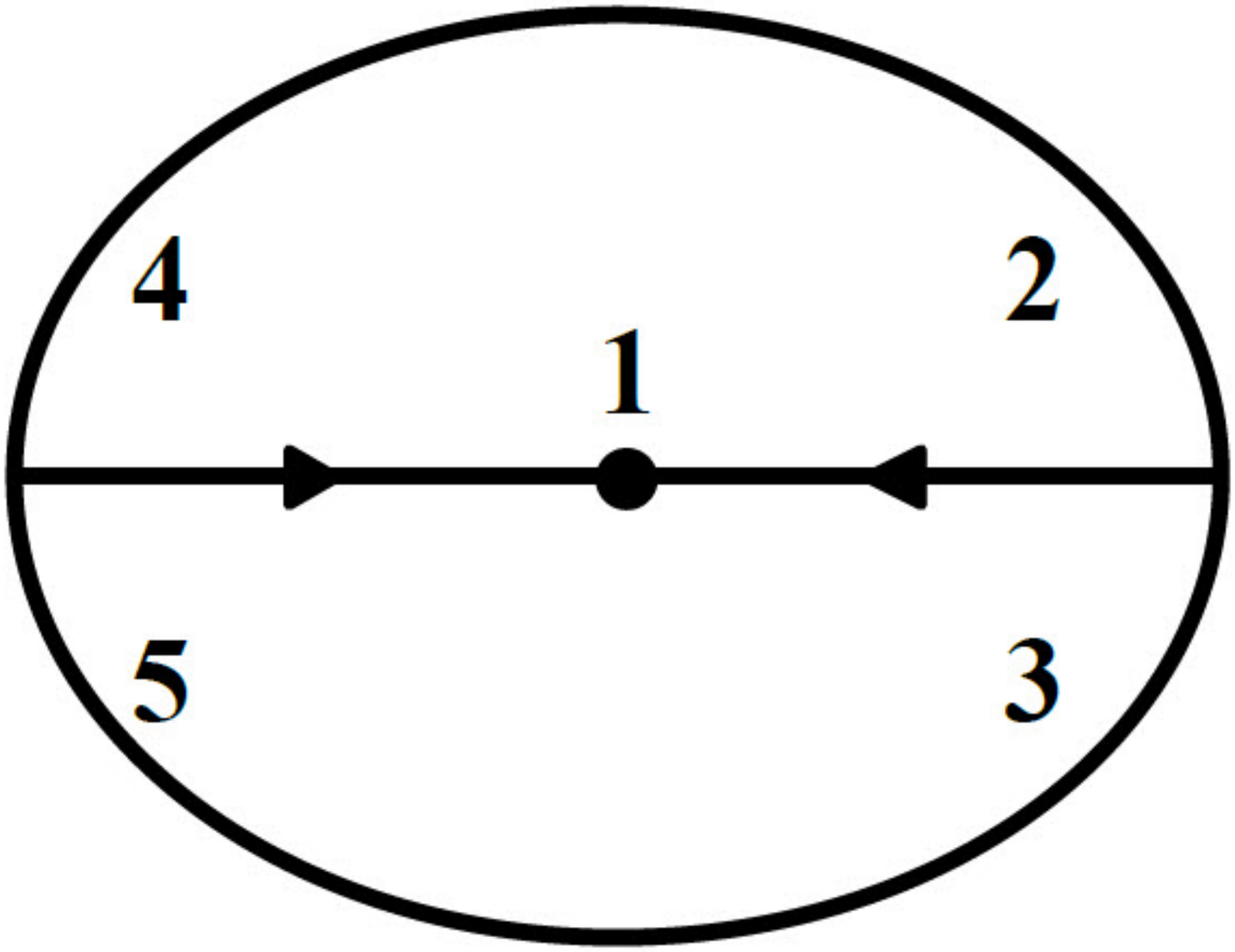}
\end{center} 
These are formally  the leading contributions, with one free wavenumber. However, the wavenumber delta functions require $\bk_1=\bzed$ 
and thus this graph vanishes identically. All other nonzero contributions are type II where there are no free wavenumbers.

Now consider large $L$. The amplitude factor $J_1\sqrt{J_2J_3J_4J_5}$ in $\cJ_4$ changes the order by $L^{-3d}$. 
The largest possible contribution comes from the term proportional to $\frac{\mu_1}{4J_1^2}(\frac{\mu_1}{2}-1),$ which is a  
large factor of $O(L^{2d}).$ However, such contributions are still only $O(L^{-d})$. Hence  $\cJ_4$ is negligible in the large box limit.

{{\it Calculation of $\cJ_5$:}} Write $\cJ_5\equiv\sum_{1\neq 2}\left\{\lambda_1\lambda_2(\cB_1+\cB_2)
+(\lambda_1+\frac{\mu_1}{4J_1})\frac{\mu_2}{J_2}(\cB_1-\cB_3)\right\}$, where (\ref{a1}) gives
\begin{eqnarray}
\cB_1&\equiv&\llangle\prod_{\bk}\psi_\bk^{(0)\mu_\bk} a_1^{(1)}a_1^{(0)*}a_2^{(1)}a_2^{(0)*}\rrangle_\psi=\sum_{3,4,5,6}L^{+,\sigma_3,\sigma_4}_{\bk_1,\bk_3,\bk_4}L^{+,\sigma_5,\sigma_6}_{\bk_2,\bk_5,\bk_6}\sqrt{J_1J_2J_3J_4J_5J_6}\llangle\psi_1^{-1}\psi_2^{-1}\psi_3^{\sigma_3}\psi_4^{\sigma_4}\psi_5^{\sigma_5}\psi_6^{\sigma_6}\prod_{\bk}\psi_\bk^{\mu_\bk}\rrangle_\psi
\cr
&&\,\,\,\,\,\, \times\Delta(\sigma_3\omega_3+\sigma_4\omega_4-\omega_1)\Delta(\sigma_5\omega_5+\sigma_6\omega_6-\omega_2)\delta_{\bk_1,\sigma_3\bk_3+\sigma_4\bk_4}\delta_{\bk_2,\sigma_5\bk_5+\sigma_6\bk_6}
\cr
\cB_2 &\equiv&\llangle\prod_{\bk}\psi_\bk^{\mu_\bk}a_1^{(1)*}a_1^{(0)}a_2^{(1)}a_2^{(0)*}\rrangle_\psi
=\sum_{3,4,5,6}L^{-,\sigma_3,\sigma_4}_{\bk_1,\bk_3,\bk_4}L^{+,\sigma_5,\sigma_6}_{\bk_2,\bk_5,\bk_6}\sqrt{J_1J_2J_3J_4J_5J_6}\llangle\psi_1^{+1}\psi_2^{-1}\psi_3^{\sigma_3}\psi_4^{\sigma_4}\psi_5^{\sigma_5}\psi_6^{\sigma_6}\prod_{\bk}\psi_\bk^{\mu_\bk}\rrangle_\psi
\cr
&&\,\,\,\,\,\, \times\Delta(\sigma_3\omega_3+\sigma_4\omega_4+\omega_1)\Delta(\sigma_5\omega_5+\sigma_6\omega_6-\omega_2)\delta_{-\bk_1,\sigma_3\bk_3+\sigma_4\bk_4}\delta_{\bk_2,\sigma_5\bk_5+\sigma_6\bk_6}
\cr
\cB_3 &\equiv&\llangle\prod_{\bk}\psi_\bk^{\mu_\bk}a_1^{(1)}a_1^{(0)*}a_2^{(1)*}a_2^{(0)}\rrangle_\psi
=\sum_{3,4,5,6}L^{+,\sigma_3,\sigma_4}_{\bk_1,\bk_3,\bk_4}L^{-,\sigma_5,\sigma_6}_{\bk_2,\bk_5,\bk_6}\sqrt{J_1J_2J_3J_4J_5J_6}\llangle\psi_1^{-1}\psi_2^{+1}\psi_3^{\sigma_3}\psi_4^{\sigma_4}\psi_5^{\sigma_5}\psi_6^{\sigma_6}\prod_{\bk}\psi_\bk^{\mu_\bk}\rrangle_\psi
\cr
&&\,\,\,\,\,\, \times\Delta(\sigma_3\omega_3+\sigma_4\omega_4-\omega_1)\Delta(\sigma_5\omega_5+\sigma_6\omega_6+\omega_2)\delta_{\bk_1,\sigma_3\bk_3+\sigma_4\bk_4}\delta_{-\bk_2,\sigma_5\bk_5+\sigma_6\bk_6}.
\end{eqnarray}
Note under the interchanges $1\leftrightarrow 2,$ $3\leftrightarrow 5,$ $4\leftrightarrow 6,$ that $\cB_2\leftrightarrow \cB_3.$ 
All three of the $\cB$'s contain the same amplitude factor $\sqrt{J_1J_2J_3J_4J_5J_6}$ that changes the order by $L^{-3d}$. 
We list their diagrams as follows,
\begin{center}
\includegraphics[width=150pt,height=100pt]{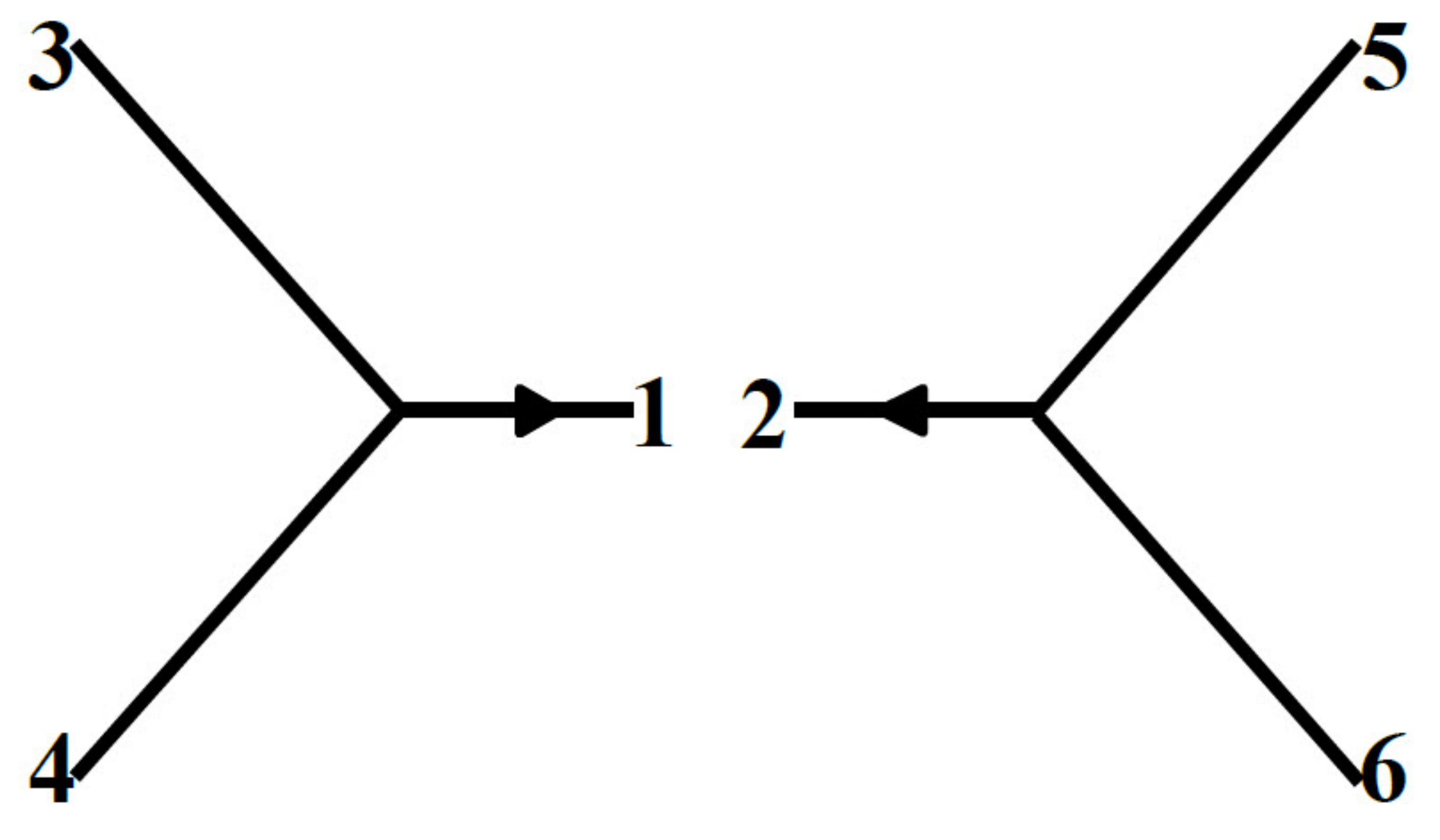}
\includegraphics[width=150pt,height=100pt]{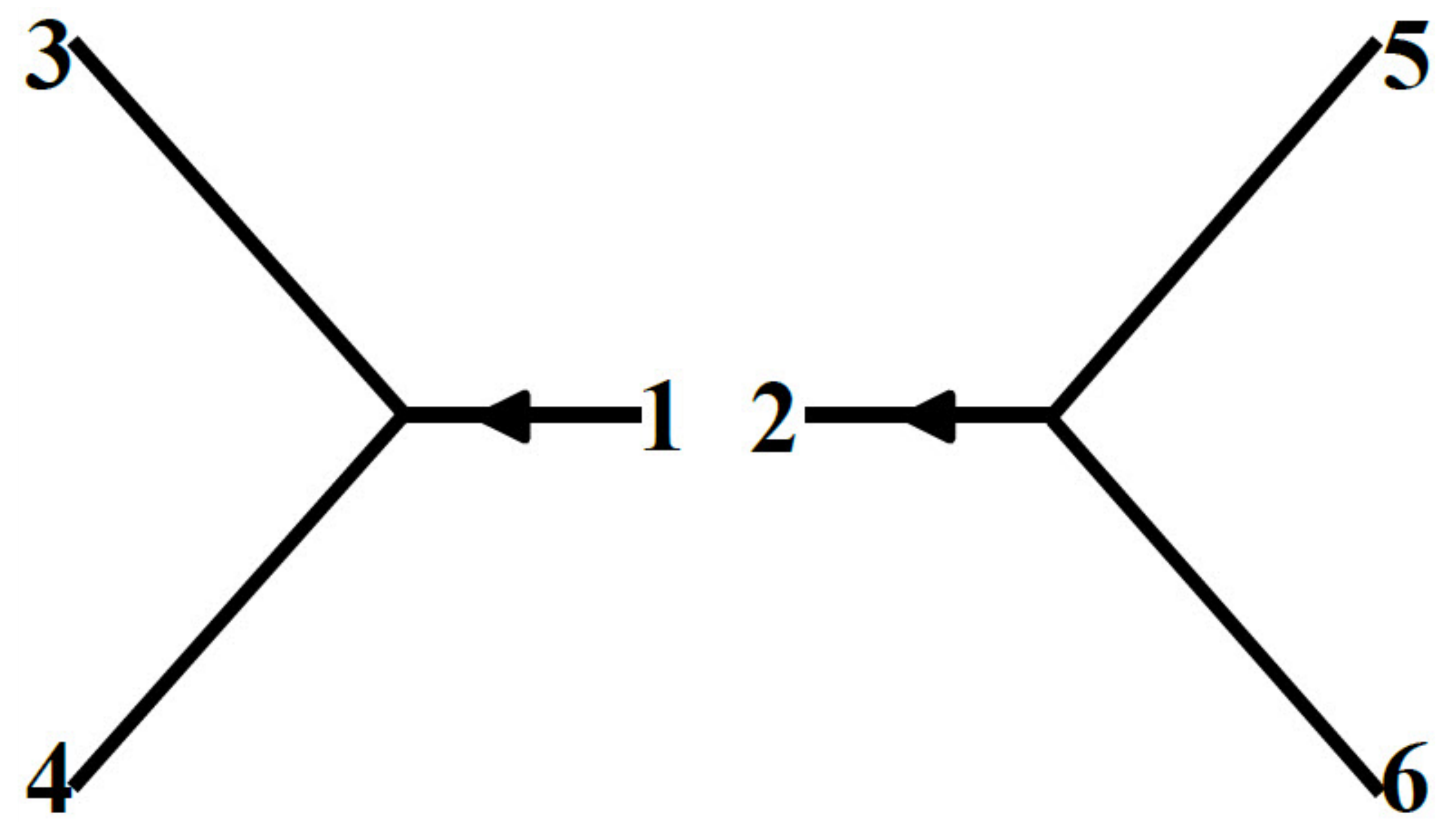}
\includegraphics[width=150pt,height=100pt]{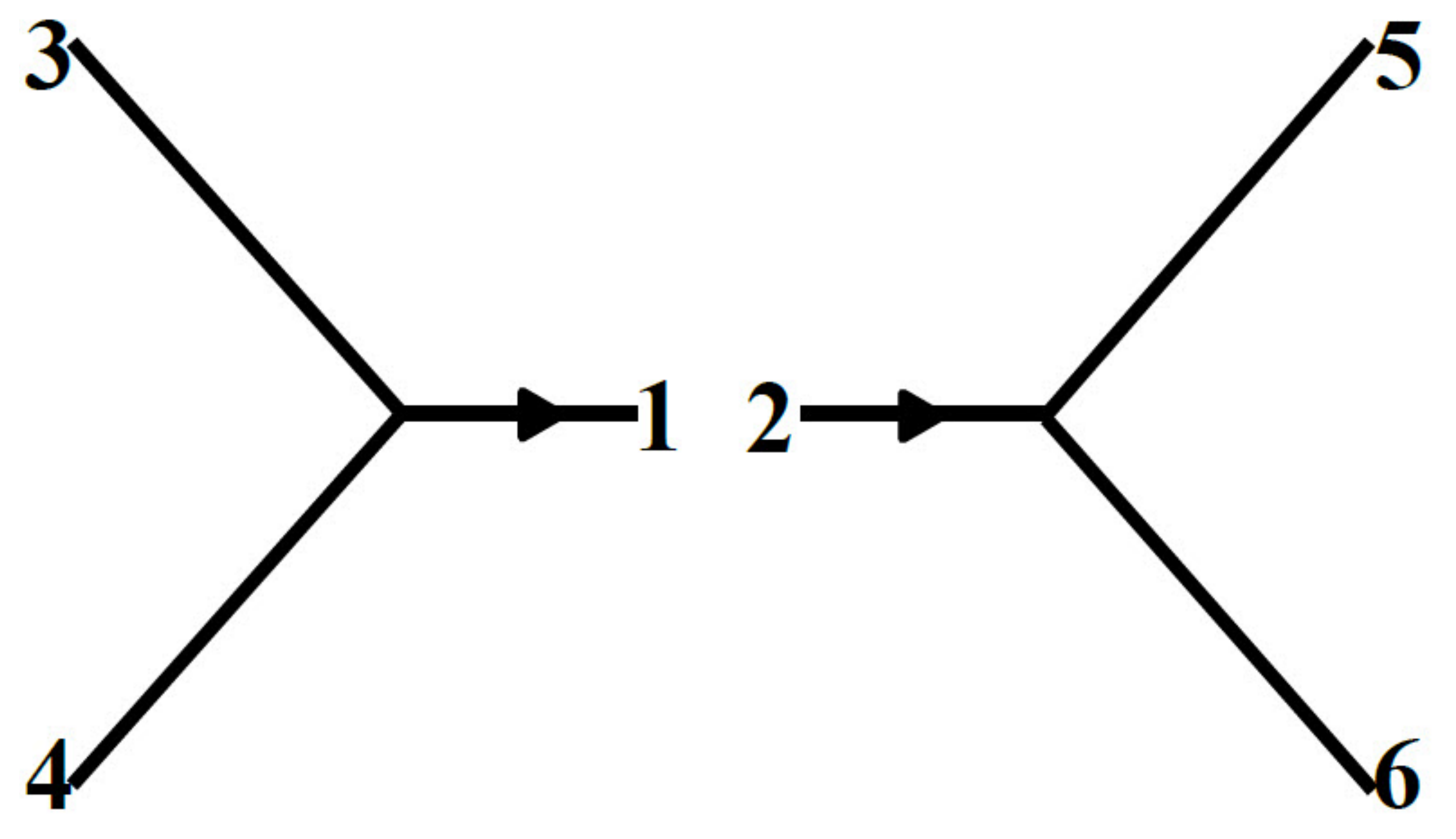}
\end{center}
After phase averaging, there are three types of diagrams for the $\cB$'s. The type I diagrams are leading contributions with no external 
couplings. Shown here is one type I diagram for $\cB_1:$   
\begin{center}
\includegraphics[width=120pt,height=100pt]{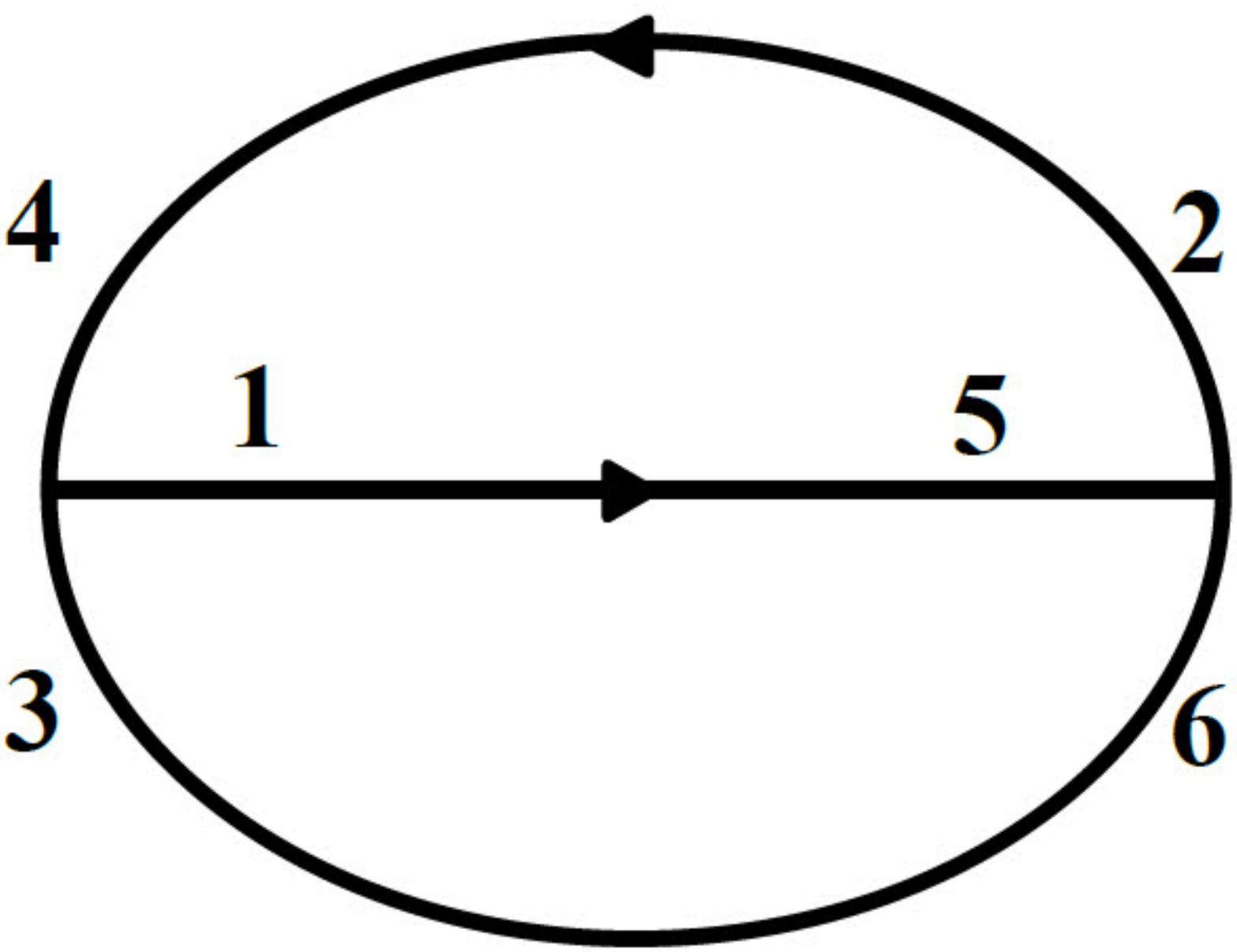}
\end{center}
with others obtained by $3\leftrightarrow 4$ and $5\leftrightarrow 6.$ These contain two free wavenumbers. For  $\cB_2,\cB_3$ there are 
additional type I contributions with 1 and 2 coupled, but these do not contribute to $\cJ_5$ because of the restriction that $1\neq 2.$
The type II diagrams are subleading contributions which contain one free wavenumber. Shown here is one such contribution for $\cB_1$
with both 1 and 2 externally coupled:
\begin{center}
\includegraphics[width=150pt,height=100pt]{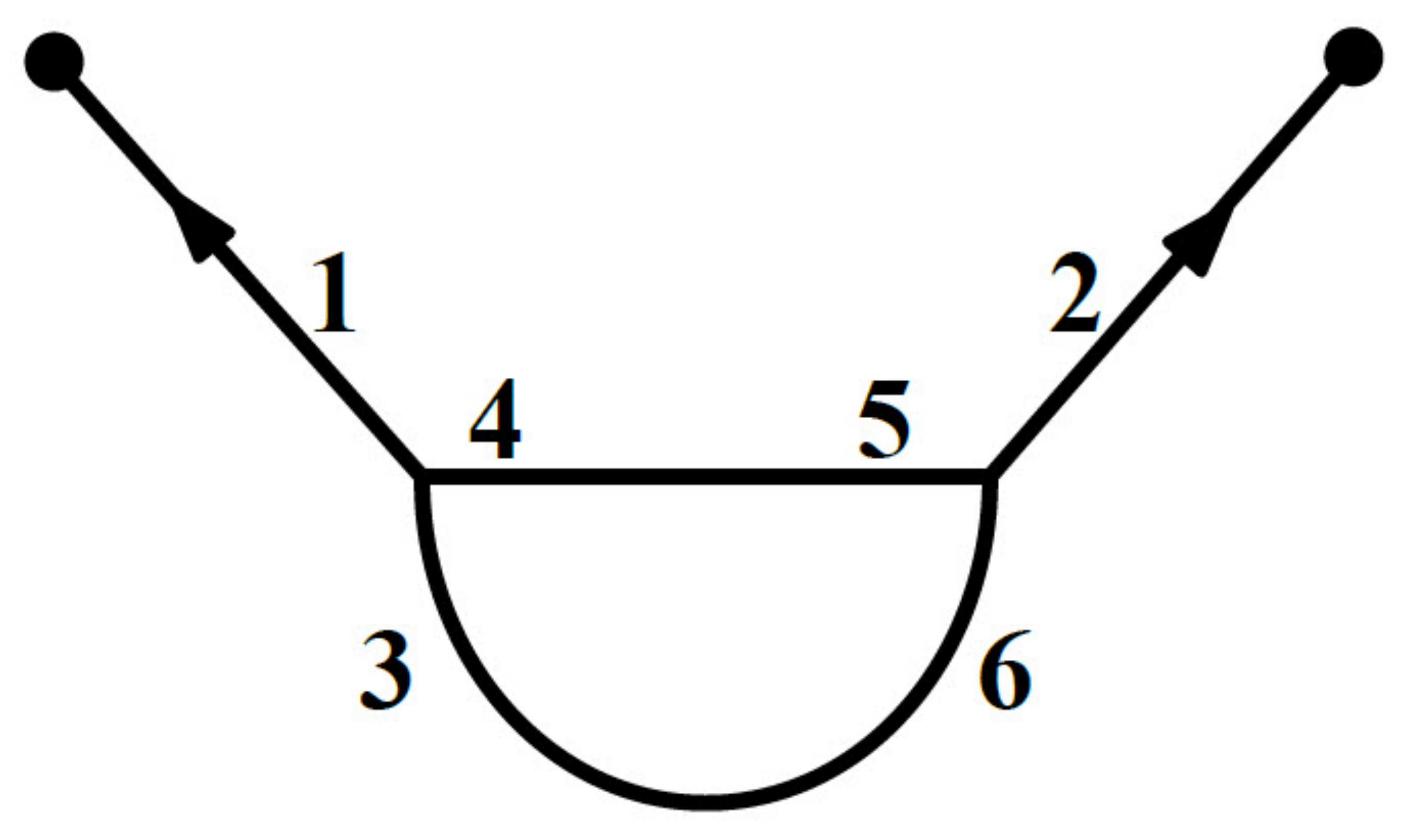}
\end{center}
The delta functions here require $\bk_2=-\bk_1$. In the corresponding diagrams for $\cB_2,\cB_3$ the delta functions instead require 
$\bk_1=\bk_2,$ so these do not exist. (Contributions to $\cB_2,\cB_3$ with 1 and 2 coupled to the {\it same} external blob coincide  
with the type I diagrams that do not contribute to $\cJ_5.$) There are also type II contributions for the $\cB$'s with only one of 1 or 2 externally 
coupled. For example, one such contribution is obtained for $\cB_1$ by exchanging $1\leftrightarrow 4$ in the above graph and then the delta
functions require that $\bk_2=\sigma_4\bk_4$ so that $\bk_2=-\bk_4,\,\sigma_4=-1.$ 
Similar type II contributions exist for $\cB_2,\cB_3.$
Other type II contributions with neither 1 nor 2 externally coupled must have both $\mu_1=\mu_2=0$ and do not contribute at non-vanishing 
order to $\cJ_5.$ Type III contributions with additional external couplings have no free wavenumbers.

Now consider the size of the various terms for large $L$.  For $\sum_{1\neq 2}\lambda_1\lambda_2(\cB_1+\cB_2)$, the leading contribution comes from type I diagrams 
and is $O(L^{-d})$. For $\sum_{1\neq2}\frac{\lambda_1\mu_2}{J_2}(\cB_1-\cB_3)$, the type I diagrams give zero contribution since $\mu_2=0$ and type II diagrams are at most $O(L^{-d})$. For $\sum_{1\neq 2}\frac{\mu_1\mu_2}{4J_1J_2}\cB_1$, type I diagrams give zero contribution and type II diagrams with both 1 and 2 externally coupled give 
an $O(1)$ contribution.  The corresponding term $\sum_{1\neq 2}\frac{\mu_1\mu_2}{4J_1J_2}\cB_3=O(L^{-d})$ from type III diagrams, since, as noted above, there do not 
exist type II diagrams for $\cB_3$ with $1$ and $2$ externally coupled to distinct blobs. In large box limit, the term $\cJ_5$ is given, to leading order, by
\begin{eqnarray}
\cJ_5 &\sim& -\frac{9}{2}{\sum_1}\delta_{\mu_1,1}\delta_{\mu_{-1},1}\prod_{\bk\neq\pm\bk_1}\delta_{\mu_\bk,0}\sum_{\ul{\sigma}=(-1,\sigma_2,\sigma_3)}\int d^dk_2\int d^dk_3\delta^d(\ul{\sigma}\cdot\ul{\bk})\Delta_T(\ul{\sigma}\cdot\omega(\ul{\bk}))\Delta_T(\ul{\sigma}\cdot\omega(\ul{\bk}'))
\cr
&&\,\,\,\,\,\,\,\,\,\,\,\,\,\,\,\,\,\,\,\,\,\,\,\,\,\,\,\,\,\,\,\,\,\,\,\,\,\,\,\times H^{-,\sigma_2,\sigma_3}_{\bk_1,\bk_2,\bk_3}H^{+,\sigma_2,\sigma_3}_{-\bk_1,\bk_2,\bk_3}\llangle\frac{\tJ_2\tJ_3}{\sqrt{\tJ_1\tJ_{-1}}}e^{\sum_\bk\lambda_\bk J_\bk}\rrangle_J,
\end{eqnarray}
where $\ul{\bk}=(\bk_1,\bk_2,\bk_3)$ and $\ul{\bk}'=(-\bk_1,\bk_2,\bk_3)$. Now making the substitution $\lambda_\bk=i\lambda(\bk)$ and taking the limit, one obtains,
\begin{eqnarray}
\left\langle e^{\sum_\bk \lambda_\bk J_\bk} \,\cJ_5\right\rangle_J&\sim&
 -\frac{9}{2}{\sum_1}\delta_{\mu_1,1}\delta_{\mu_{-1},1}\prod_{\bk\neq\pm\bk_1}\delta_{\mu_\bk,0}\sum_{\ul{\sigma}=(-,\sigma_2,\sigma_3)}\int d^dk_2d^dk_3\delta^d(\ul{\sigma}\cdot\ul{\bk})\Delta_T(\ul{\sigma}\cdot\omega(\ul{\bk}))\Delta_T(-\ul{\sigma}\cdot\omega(\ul{\bk}'))
\cr
&&\,\,\,\,\,\,\,\,\,\,\,\,\,\,\,\,\,\,\,\,\,\,\,\,\,\,\,\,\,\,\,\,\,\,\,\,\,\,\,\times H^{-,\sigma_2,\sigma_3}_{\bk_1,\bk_2,\bk_3}H^{+,\sigma_2,\sigma_3}_{-\bk_1,\bk_2,\bk_3}\llangle\frac{\tJ_2\tJ_3}{\sqrt{\tJ_1\tJ_{-1}}}e^{\sum_\bk\lambda_\bk J_\bk}\rrangle_J.
\end{eqnarray}

\section{Derivation of the PDF Hierarchy}\lb{appendix-pdf}  

We here evaluate the $\cJ$-terms that contribute to $\cX^{(M)}(\lambda,\mu,T).$ Most of the essential work has already been
done in \ref{appendix-spectral}. As discussed in the text,  the main difference is that mode 1 is now discrete for $\cJ_1-\cJ_4$ 
and assumes only $M$ values, while for $\cJ_5$ modes 1,2 are both discrete. All other modes are continuous in the 
infinite-box limit. One must consider carefully whether free wavenumbers in graphical summations are discrete or 
continuous to see whether their contribution is $O(M)$ or $O(L^d).$  

$\cJ_1$: The graphs contain no free wavenumbers, so that $\cJ_1$ is still $O(L^{-d/2})$.

$\cJ_2$: The leading order $O(1)$ contribution comes again from the type I diagrams, which contain two free wavenumbers.
One of these free wavenumbers is mode 1 which is discrete, but this is compensated by the extra factor of $L^d$ from the 
prefactors $\lambda_{\bk_1}$ and $\lambda_{\bk_1}^2 J_{\bk_1},$ which now contribute equally.  
One thus obtains by substituting $\lambda_{\bk_m}=i(L/2\pi)^d\lambda_m$ and $J_m=(2\pi/L)^d\tJ_m$ the result
\begin{eqnarray}
\left\langle e^{\sum_\bk \lambda_\bk J_\bk} \,\cJ_2\right\rangle_J
&\sim&9\delta_{\mu,0}\sum_{j=1}^M\sum_{\ul{\sigma}=(-1,\sigma_2,\sigma_3)}
\int d^d\ol{k}_2d^d\ol{k}_3\, \delta^d(\ul{\sigma}\cdot\ul{\bk}_j)\big|\Delta_T(\ul{\sigma}\cdot\omega(\ul{\bk}_j))\big|^2\big|H^{\ul{\sigma}}_{\ul{\bk}_j}\big|^2
\cr
&&\,\,\,\,\,\,\,\,\,\,\,\,\,\,\,\,\,\,\,\,\,\,\,\,\,\,\,\,\,\,\times\llangle(i\lambda_j-\lambda_j^2\tJ_j) \tJ_{\ol{2}}\tJ_{\ol{3}}\,e^{\sum_m i\lambda_m \tJ_m}\rrangle_J
\cr
&=&-9i\delta_{\mu,0}\sum_{j=1}^M\sum_{\ul{\sigma}=(-1,\sigma_2,\sigma_3)}\int d^d\ol{k}_2d^d\ol{k}_3\, \delta^d(\ul{\sigma}\cdot\ul{\bk}_j)\big|\Delta_T(\ul{\sigma}\cdot\omega(\ul{\bk}_j))\big|^2\big|H^{\ul{\sigma}}_{\ul{\bk}_j}\big|^2
\cr
&&\,\,\,\,\,\,\,\,\,\,\,\,\,\,\,\,\,\,\,\,\,\,\,\,\,\,\,\,\,\,\times(\lambda_j+\lambda_j^2\frac{\partial}{\partial\lambda_j}) \frac{\partial ^2 \cZ^{(M+2)}}{\partial \ol{\lambda}_2\partial\ol{\lambda}_3}\Big|_{\ol{\lambda}_2=\ol{\lambda}_3=0}.
\end{eqnarray}
Here $\ul{\bk}_j=(\bk_j,\ol{\bk}_2,\ol{\bk}_3)$.


$\cJ_3$: There are two contributions, from the two terms in the prefactor, $\lambda_{\bk_1}$ and $\mu_{\bk_1}/2J_{\bk_1}.$ For the 
$\lambda_{\bk_1}$ term,  the leading $O(1)$ contribution comes from the type I diagrams, just as for $\cJ_2.$ One of the two free wavenumbers
is discrete mode 1, but this is compensated by the $O(L^d)$ prefactor $\lambda_{\bk_1}.$ For the $\mu_{\bk_1}/2J_{\bk_1}$ term, 
the leading $O(1)$ contribution comes from the type II diagrams with modes $1$ and $-1$ externally coupled, so that $\mu_{\bk_1}=\mu_{-\bk_1}=1.$
The type II diagram has one free continuous wavenumber. Thus, its contribution is also $O(1)$ if the set 
of $M$ wavenumbers includes at least one pair of opposite wavenumbers. Assuming this is not so, the final result is:
\begin{eqnarray}
\left\langle e^{\sum_\bk \lambda_\bk J_\bk} \,\cJ_3\right\rangle_J^{(I)}
&\sim&-18\delta_{\mu,0}\sum_{j=1}^M \sum_{\ul{\sigma}=(-1,\sigma_2,\sigma_3)}\sigma_2\int d^d\ol{k}_2d^d\ol{k}_3\, \delta^d(\ul{\sigma}\cdot\ul{\bk}_j)
E_T(0,\ul{\sigma}\cdot\omega(\ul{\bk}_j))\big|H^{\ul{\sigma}}_{\ul{\bk}_j}\big|^2
\cr
&&\,\,\,\,\,\,\,\,\,\,\,\,\,\,\,\,\,\,\,\,\,\,\,\,\,\,\,\,\,\,\times\llangle i\lambda_j \tJ_j\tJ_{\ol{3}}\,e^{\sum_m i\lambda_m \tJ_m}\rrangle_J +(2\leftrightarrow 3). \cr
&=&18i\delta_{\mu,0}\sum_{j=1}^M\sum_{\ul{\sigma}=(-1,\sigma_2,\sigma_3)}\sigma_2
\int d^d\ol{k}_2d^d\ol{k}_3\, \delta^d(\ul{\sigma}\cdot\ul{\bk}_j)E_T(0,\ul{\sigma}\cdot\omega(\ul{\bk}_j))
\cr
&&\,\,\,\,\,\,\,\,\,\,\,\,\,\,\,\,\,\,\,\,\,\,\,\,\,\,\,\,\,\,\times \lambda_j\big|H^{\ul{\sigma}}_{\ul{\bk}_j}\big|^2\frac{\partial^2 \cZ^{(M+1)}}{\partial\ol{\lambda}_3\partial\lambda_j}\Big|_{\ol{\lambda}_3=0}+(2\leftrightarrow 3).
\end{eqnarray}
However, if the $M$ wavenumbers include pairs of opposites, then there is an additional contribution
\begin{eqnarray}
\left\langle e^{\sum_\bk \lambda_\bk J_\bk} \,\cJ_3\right\rangle_J^{(II)} &\sim &-9\sum_{j=1}^M\delta_{\mu_j,1}\delta_{\mu_{-j,1}}\prod_{m\neq j,-j}\delta_{\mu_m,0}\sum_{\ul{\sigma}=(-1,\sigma_2,\sigma_3)}
\sigma_2\int d^d\ol{k}_2\int d^d\ol{k}_3 \delta^d(\ul{\sigma}\cdot\ul{\bk}_j)   \cr
&& \times E_T(\omega(\bk_j)+\omega(-\bk_j),\ul{\sigma}\cdot\omega(\ul{\bk}_j)) H^{-,\sigma_2,\sigma_3}_{\bk_j,\ol{\bk}_2,\ol{\bk}_3} H^{+,\sigma_2,\sigma_3}_{-\bk_j,\ol{\bk}_2,\ol{\bk}_3}\llangle 
e^{\sum_m i\lambda_m \tJ_m}\sqrt{\frac{\tJ_{-1}}{\tJ_1}}\tJ_3\rrangle_J \cr
&& +(2\leftrightarrow 3).  
\end{eqnarray}

$\cJ_4$: The type I contribution vanishes and the leading contribution from type II diagrams with no free sums remains $O(L^{-d}).$ 

$\cJ_5$: The amplitude factors contribute $O(L^{-3d})$ while the prefactors are $O(L^{2d}).$ Thus, to give an overall $O(1)$ contribution, the $\cB$ factors
must contain a free continuous wavenumber. For $\cB_1$ the type I diagrams have two free discrete wavenumbers and type II diagrams with mode 2 only pinned 
have one free discrete wavenumber, but type II diagrams with both 1 and 2 pinned have one free continuous wavenumber. For $\cB_2,\cB_3,$ there 
are no graphs with modes 1 and 2 distinct, contributing to $\cJ_5,$ that have a free continuous wavenumber. Type I diagrams have two free wavenumbers, 
both discrete if 1 and 2 are uncoupled. Type II diagrams with only one of the modes 1 or 2 externally coupled have one discrete free wavenumber.  

Thus, the only $O(1)$ contribution to $\cJ_5$ is from $\sum_{1\neq 2} \left\{\lambda_{\bk_1}\lambda_{\bk_2}+\frac{\lambda_{\bk_1}\mu_{\bk_2}}{J_{\bk_2}}
+\frac{\mu_{\bk_1}\mu_{\bk_2}}{4J_{\bk_1}J_{\bk_2}}\right\}\cB_1$
and type II diagrams for $\cB_1$ with both 1 and 2 pinned. Corrections are at most $O(M/L^d).$ The $O(1)$ contribution is thus like that obtained before 
for $\cJ_5,$ except that the mode-1 sum is over just $M$ discrete values and all terms in the prefactor contribute equally. This gives 
\begin{eqnarray}
\left\langle e^{\sum_\bk \lambda_\bk J_\bk} \,\cJ_5\right\rangle_J&\sim&
 18\sum_{j=1}^M\delta_{\mu_j,1}\delta_{\mu_{-j},1}\prod_{m\neq 1,-1}\delta_{\mu_m,0}
\left( \lambda_1\lambda_{-1}-i\frac{\lambda_1}{\tJ_{-1}}-\frac{1}{4\tJ_1\tJ_{-1}}\right)
 \sum_{\ul{\sigma}=(-,\sigma_2,\sigma_3)}\int d^d\ol{k}_2d^d\ol{k}_3\delta^d(\ul{\sigma}\cdot\ul{\bk}_j)
\cr
&& \times \Delta_T(\ul{\sigma}\cdot\omega(\ul{\bk}_j))\Delta_T(-\ul{\sigma}\cdot\omega(\ul{\bk}_j'))H^{-,\sigma_2,\sigma_3}_{\bk_j,\ol{\bk}_2,\ol{\bk}_3}
H^{+,\sigma_2,\sigma_3}_{-\bk_j,\ol{\bk}_2,\ol{\bk}_3}\llangle\sqrt{\tJ_1\tJ_{-1}}\tJ_2\tJ_3e^{i\sum_m\lambda_m \tJ_m}\rrangle_J.
\end{eqnarray}

\bibliographystyle{model1-num-names}
\bibliography{reference}

\end{document}